\documentclass[prd,nofootinbib,showpacs,showkeys,twocolumn]{revtex4}
\pdfoutput=1

\usepackage{graphicx}
\usepackage{amsmath,amssymb}
\usepackage{mathrsfs}
\usepackage{color}
\usepackage{slashed}
\usepackage{hyperref}

\allowdisplaybreaks

\definecolor{blue}{rgb}{0,0,0.5}
\definecolor{lightblue}{rgb}{0,0,1}
\definecolor{red}{rgb}{0.5,0,0}
\definecolor{lightred}{rgb}{1,0,0}
\definecolor{green}{rgb}{0,0.5,0}
\definecolor{darkgreen}{rgb}{0.0,0.3,0.0}
\definecolor{orange}{rgb}{1,0.4,0}
\definecolor{grey}{rgb}{0.5,0.5,0.5}
 
\providecommand{\eqn}{Eq.~}
\providecommand{\eqns}{Eqs.~}
\providecommand{\fig}{Fig.~}
\providecommand{\figs}{Figs.~}
\providecommand{\sect}{Sec.~}

\providecommand{\refs}{Refs.~}
\providecommand{\reference}{Ref.~}

\DeclareMathOperator\arctanh{arctanh}
\DeclareMathOperator\BR{BR}

\newcommand{\ev}[1]{\ensuremath{\left\langle #1 %
                     \right\rangle}} 

\begin{document}

\title{Loopy Constraints on Leptophilic Dark Matter and Internal Bremsstrahlung}

\author{Joachim Kopp}
\email[\,]{jkopp@mpi-hd.mpg.de}    

\author{Lisa Michaels}
\email[\,]{lisa.michaels@mpi-hd.mpg.de}    

\author{Juri Smirnov}
\email[\,]{juri.smirnov@mpi-hd.mpg.de}

\affiliation{
Max-Planck-Institut f\"ur Kernphysik, Saupfercheckweg 1, 69117 Heidelberg, Germany}



\begin{abstract}
  A sharp and spatially extended peak in an astrophysical gamma ray spectrum
  would provide very strong evidence for the existence of dark matter (DM),
  given that there are no known astrophysical processes that could mimic such a
  signal. From the particle physics perspective, perhaps the simplest
  explanation for a gamma ray peak is internal bremsstrahlung in DM
  annihilation through a charged $t$-channel mediator $\eta$ close in mass to
  the DM particle $\chi$.  Since DM annihilation to quarks is already tightly
  constrained in this scenario, we focus here on the leptophilic case.  We
  compute the electromagnetic anapole and dipole moments that DM acquires at
  1-loop, and we find an interesting enhancement of these moments if the DM particle
  and the mediator are close in mass.  We constrain the DM anapole and dipole moments
  using direct detection data, and then translate these limits into bounds on
  the DM annihilation cross section. Our bounds are highly competitive with those from
  astrophysical gamma ray searches.  In the second part of the paper, we derive
  complementary constraints on internal bremsstrahlung in DM annihilation using
  LEP mono-photon data, measurements of the anomalous magnetic moments of the
  electron and the muon, and searches for lepton flavor violation. We also
  comment on the impact of the internal bremsstrahlung scenario on the
  hyperfine splitting of true muonium.
\end{abstract}

\maketitle

\section{\label{sec:introduction}Introduction}

One of the cleanest signatures in indirect dark matter (DM) searches are peaks
in the cosmic gamma ray spectrum from the Galactic Center or other regions of
high DM density.  On the one hand, there are no known astrophysical sources
that could mimic such a signal.\footnote{The authors of
\reference\cite{Aharonian:2012cs} show that a particular composition of a pulsar
wind nebula could generate a peaked gamma ray signal, but an observation of a
peak at the same energy in different regions of the galaxy would rule out this
possibility.} On the other hand, gamma ray observatories are making tremendous
progress in terms of statistics, resolution and control of systematic
uncertainties.

\begin{figure}
  \begin{center}
    \includegraphics[width=0.5\textwidth]{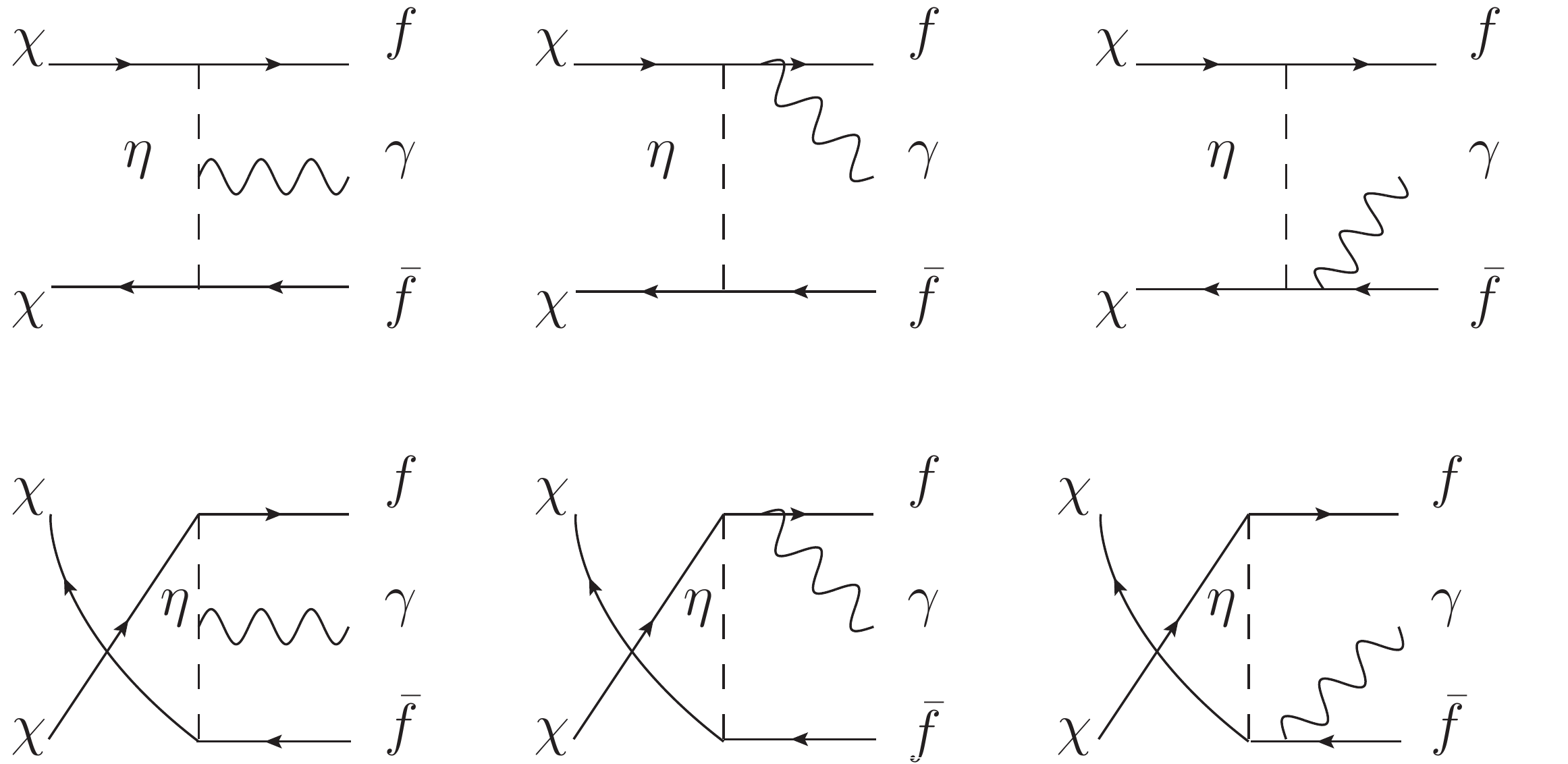}
  \end{center}
  \caption{The Feynman diagrams contributing to virtual internal bremsstrahlung (VIB)
    and to final state radiation in the case of Majorana DM annihilating
    through a scalar $t$-channel mediator. For Dirac DM, the second row of
    diagrams is absent.  Note that only the sum of VIB and final state
    radiation diagrams is gauge invariant.}
  \label{fig:vib-diagrams}
\end{figure}

From the particle physics point of view, peaks in the gamma ray spectrum can
originate from DM annihilation or decay to two photons, a photon and a $Z$ boson,
or a photon and a Higgs boson.  However, since DM is electrically
neutral, these processes can only happen at the 1-loop level, making it likely
that DM is first discovered in other annihilation or decay channels.
There is, however, a class of models where the first experimental hint for DM
is a gamma ray peak. Namely, this can happen in models where DM annihilates
via a charged $t$-channel mediator, so that a photon can be emitted from the
mediator, see \fig\ref{fig:vib-diagrams}.  This process is called virtual internal bremsstrahlung (VIB)~\cite{Bringmann:2007nk, Bell:2010ei, Bringmann:2012vr}.
If the mediator mass $m_\eta$ and the
DM mass $m_\chi$ are close to each other, the resulting photon energy is
strongly peaked (see~\fig\ref{fig:vdsigmadx}) and can yield a line-like gamma
ray signal if the width of the peak is below the detector resolution.

Of particular interest in this context are models in which the DM couples
preferentially to leptons. These leptophilic dark matter (LDM) models are
motivated by the fact that DM couplings to quarks are strongly constrained by
gamma ray emission from dwarf galaxies~\cite{GeringerSameth:2011iw,
Ackermann:2013yva}, by direct detection bounds~\cite{Hisano:2011um,
Garny:2012eb, Garny:2013ama}, and by LHC searches~\cite{Goodman:2010yf,
Bai:2010hh, Fox:2011pm, Lin:2013sca, CMS:rwa, ATLAS:2012zim, Aad:2013oja,
Boyd:2013, Richman:2013}.  Additional motivation could be provided by various
cosmic ray anomalies.  For instance, attempts to explain the cosmic ray
positron excess observed by PAMELA~\cite{Adriani:2008zr, Adriani:2013uda},
Fermi-LAT~\cite{FermiLAT:2011ab} and AMS-02~\cite{Aguilar:2013qda} in terms of
DM annihilation typically require a leptophilic DM model~\cite{Cirelli:2008pk,
Donato:2008jk, Nardi:2008ix, Bertone:2008xr, Fox:2008kb, Evoli:2011id,
Kopp:2013eka, Cholis:2013psa, Bergstrom:2013jra, Ibarra:2013zia} in order not
to exceed the measured antiproton flux~\cite{Adriani:2010rc, Bartoli:2012qe}.
Finally, it is intriguing that the possible anomalies in the gamma ray signal
from the Galactic Center~\cite{Hooper:2010mq, Hooper:2011ti, Hooper:2012ft}, in
the gamma ray emission from the Fermi Bubbles~\cite{Su:2010qj, Hooper:2013rwa,
Huang:2013pda, Huang:2013apa}, and in radio signals from filamentary structures
in the inner galaxy~\cite{Linden:2011au} could be explained in leptophilic DM
models. (Note, however, that some of them can also be understood if dark matter
annihilates to $b \bar{b}$ final states.) Direct detection constraints on
leptophilic DM have been studied in~\cite{Kopp:2009et, Essig:2011nj,
Essig:2012yx}.

In the present paper, we derive new constraints on leptophilic DM, and
we translate these constraints into bounds on the cross section for internal
bremsstrahlung.  We also discuss the prospects for probing the parameter space
of leptophilic DM even further with future experiments.
We work in a simplified model which augments the Standard Model (SM) by
a fermionic DM candidate $\chi$ and a charged scalar mediator $\eta$, with
a coupling of the form $\bar\chi \ell \eta + h.c.$, where $\ell$ is a charged
lepton field. This effective scenario can be realized in supersymmetry (SUSY)
(see for instance~\cite{Bringmann:2012vr}),
where $\chi$ could be identified with the lightest neutralino, and $\eta$
would be a slepton.  It also applies to certain radiative neutrino mass models,
whose direct detection phenomenology has been discussed in~\cite{Schmidt:2012yg}.
A simplified framework of the form used here has been employed,
for instance, to explain an anomalous line-like feature at $\sim 135$~GeV
in the Fermi-LAT gamma ray data~\cite{Bringmann:2012vr, Ackermann:2012qk,
Fermi-LAT:2013uma}.  Even though the statistical significance of this
feature is not yet convincing~\cite{Fermi-LAT:2013uma}, and there are
(inconclusive) indications that poorly understood systematic effects may
play a role~\cite{Boyarsky:2012ca, Whiteson:2012hr, Hektor:2012ev,
Finkbeiner:2012ez, Whiteson:2013cs}, it demonstrates the relevance
of internal bremsstrahlung signatures as considered here if
anomalous peaked features are found in future gamma ray observations.

Our starting point is the observation that even in leptophilic models, loop
processes endow the DM with nonzero electromagnetic moments, which in turn
allow it to interact in direct detection experiments. If DM is a Majorana
fermion, only an anapole moment is generated~\cite{Radescu:1985wf,
Kayser:1983wm}, while for Dirac fermions, also a magnetic dipole moment can
exist.  DM with anapole interactions has been studied previously in
\cite{Ho:2012bg, Gresham:2013mua, Gao:2013vfa, DelNobile:2014eta} using an effective field theory
framework, and DM with magnetic dipole moments has been investigated
in~\cite{Heo:2009vt, Masso:2009mu, Schmidt:2012yg, DelNobile:2012tx,
Weiner:2012gm, Barger:2012pf, Gresham:2013mua}.  The importance of loop
processes even for hadrophilic DM has been studied in the context of LHC
searches in~\cite{Haisch:2013uaa}.

Loop processes involving DM particles can also modify electromagnetic
properties of leptons, in particular their anomalous magnetic moments and the
energy levels of dilepton systems such as positronium and muonium.  In the most
general case, also lepton flavor violation could be induced by DM loops.
Finally, if DM couples to electrons, it can be directly produced at LEP or at a
future linear collider, allowing us to derive constraints from searches for
mono-photons plus missing energy.

The paper is organized as follows. In \sect\ref{sec:LDM}, we introduce the
simplified model framework which we will use in the rest of the paper, we
establish its connection to supersymmetric scenarios, and we review the expected
indirect detection (internal  bremsstrahlung) signals from DM annihilation in
this model.  In \sect\ref{sec:dd-limits}, we compute the electromagnetic
form factors of DM and the resulting direct detection cross sections.  We
compare these to LUX~\cite{Akerib:2013tjd} and XENON100~\cite{Aprile:2012nq}
data, and to the expected future sensitivity of XENON1T and LUX-ZEPLIN to
derive constraints. We translate these constraints into limits on the
intensity of possible internal bremsstrahlung signals.  To illustrate the
strength of direct detection limits, we show that for flavor-universal DM couplings to
leptons, the explanation of the aforementioned 135~GeV feature in the Fermi-LAT
data~\cite{Bringmann:2012vr, Ackermann:2012qk, Fermi-LAT:2013uma} in terms of
internal bremsstrahlung is severely constrained. We then investigate in
\sect\ref{sec:collider} the complementary constraints from $e^+e^-$ collider data,
and in \sect\ref{sec:ewpt} the bounds from measurements of the
anomalous magnetic moment of the muon and the electron, from searches for
lepton flavor violation, and from possible future experiments on true muonium
spectroscopy.  We summarize our findings and conclude in
\sect\ref{sec:conclusions}.

\section{\label{sec:LDM} Internal bremsstrahlung in simplified models}

\subsection{The simplest model}

The simplest theoretical models that feature internal bremsstrahlung in DM
annihilation extend the Standard Model by a neutral DM candidate $\chi$ and a
charged mediator $\eta$~\cite{Bringmann:2012vr}.  $\chi$ can be either a
Majorana fermion (as in most supersymmetric theories) or a Dirac fermion (as
for example in supersymmetric theories with preserved
$R$-symmetry~\cite{Chun:2009zx, Buckley:2013sca}).  As explained above, we are
mostly interested in leptophilic models because DM couplings to quarks are
already tightly constrained.  In the simplest case, we thus start with the
interaction Lagrangian
\begin{align} 
  \mathscr{L} \supset
    - y \bar{\chi} P_R \ell \eta - i e \, \eta A^\mu \partial_\mu \eta^* + h.c. \,,
  \label{eq:lagrangian1}
\end{align}
where $A^\mu$ denotes the photon field, $\chi$ is the fermionic DM candidate,
$e$ is the unit electric charge, $\ell$ is a SM lepton field, $P_R = (1 + \gamma^5)/2$ is
the right-handed chiral projection operator, and $y$ is the Yukawa coupling
constant of the DM--lepton interaction.  Unless indicated otherwise, we assume
$\chi$ to be a Majorana fermion. Note that we have omitted couplings to
left-handed leptons here which are more strongly constrained (though not ruled
out) by collider searches and electroweak precision test~\cite{Liu:2013gba}.
We also do not consider the scalar potential for $\eta$ since these terms are
irrelevant to our discussion. Finally, we disregard the vertex
$e^2 \eta^* \eta A^\mu A_\mu$ from the
kinetic term of $\eta$ because it is higher order in the coupling constant
and will thus be phenomenologically negligible.

The simplified model \eqref{eq:lagrangian1} has been studied previously for instance in
\cite{Cao:2009yy}, and it has been shown in \cite{Bringmann:2012vr} that the model could explain the
135~GeV feature in the Fermi-LAT data.  The fit from \cite{Bringmann:2012vr}
results in a preferred DM mass of $m_\chi = 149 \pm 4\,
\text{(stat)}\,{}^{+8}_{-15} \text{(syst)}\ \text{GeV}$ and an annihilation
cross section $\ev{\sigma v_\text{rel}}_{\chi \chi \to \ell \bar{\ell} \gamma}
= (6.2 \pm 1.5\,{}^{+0.9}_{-1.4}) \cdot 10^{-27}\ \text{cm}^3 \text{s}^{-1}$.
(Here, $v_\text{rel}$ is the relative velocity of the two annihilating DM
particles, and the average $\ev{\cdot}$ is taken over $v_\text{rel}$.)

The interactions in eq.~\eqref{eq:lagrangian1} lead to annihilation of DM
particles into pairs of SM leptons via $t$-channel exchange of the charged
scalar $\eta$.  This $2 \to 2$ process can be decomposed into an $s$-wave part
and a $p$-wave part, the latter of which can usually be neglected because it is
suppressed by the square of the small velocity  $v_\text{rel} \sim \text{few}
\times 100 $~km/s of DM particles in the Milky Way.  The $s$-wave contribution
is unsuppressed for Dirac DM, while for Majorana DM, it is helicity-suppressed
by the small mass of the final state lepton~\cite{Bringmann:2012vr}. This can
be understood by noting that DM annihilation through the Yukawa interaction in
\eqn\eqref{eq:lagrangian1} produces two leptons of the same chirality.  For
Majorana DM, however, Pauli blocking in the initial state requires the incoming
DM particles to have opposite spin. Angular momentum conservation therefore requires
a mass insertion on one of the final state lepton lines.  Thus, for Majorana
DM, higher order annihilation processes become important, in particular the $2
\to 3$ process $\chi\chi \to \ell \bar{\ell} \gamma$, with two charged
leptons and a photon in the final state (see \fig\ref{fig:vib-diagrams}). Since
the photon carries away one unit of angular momentum, it can lift the helicity
suppression, see for instance~\cite{Bergstrom:1989jr, Barger:2009xe}.

A helicity suppression of 2-body DM annihilation compared to the
3-body internal bremsstrahlung process exists also in models where the scalar
mediator $\eta$ is replaced by a vector particle~\cite{Barger:2011jg} and in
models with scalar DM and fermionic mediators~\cite{Barger:2011jg,
Toma:2013bka, Giacchino:2013bta}.  We will not consider these possibilities
here, but will instead focus on the scenario from \eqn\eqref{eq:lagrangian1} as
a representative for all internal bremsstrahlung models.

\begin{figure*}
  \hspace{-0.3cm}
  \begin{tabular}{cc}
    \includegraphics[width=0.5\textwidth]{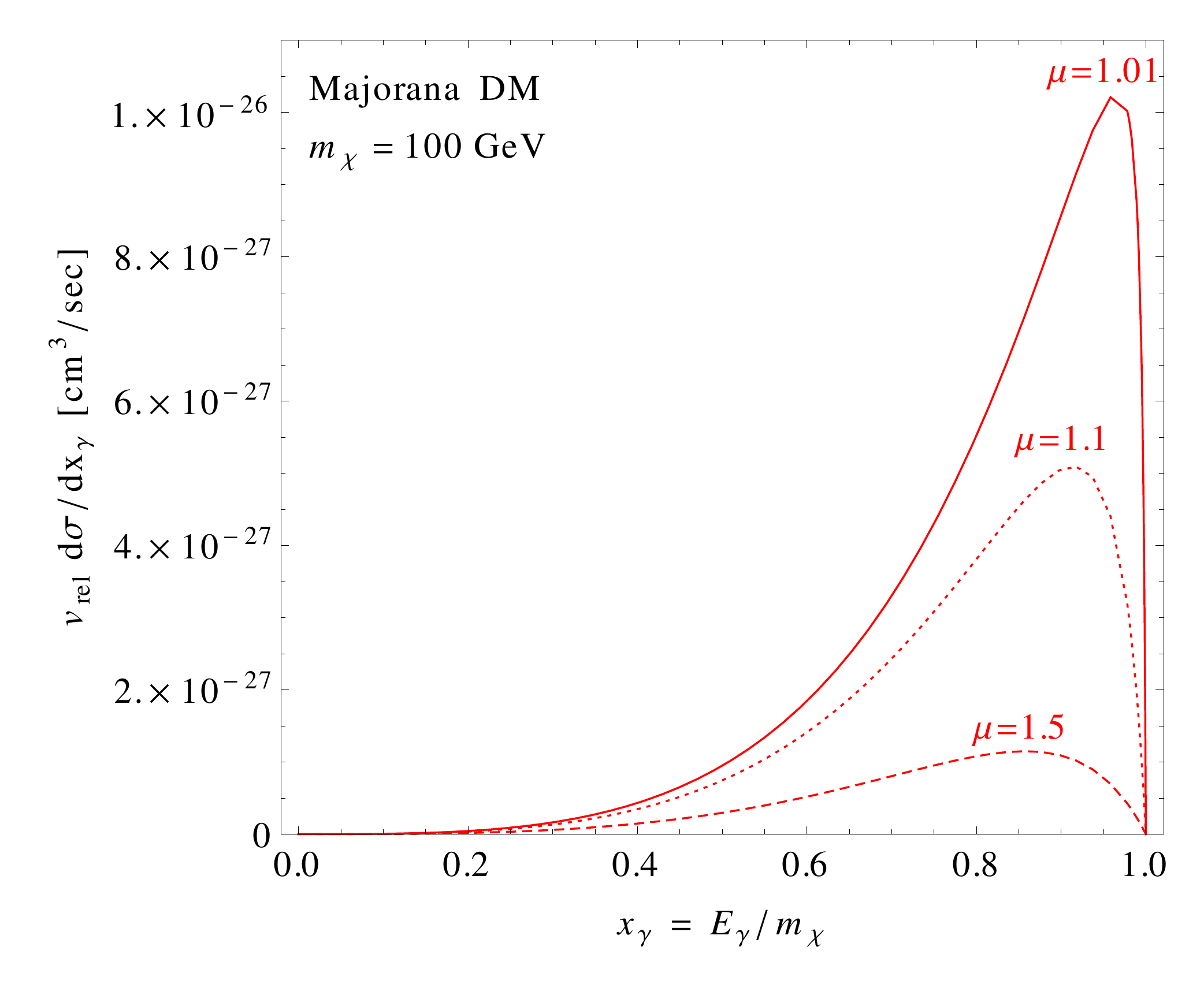} &
    \includegraphics[width=0.5\textwidth]{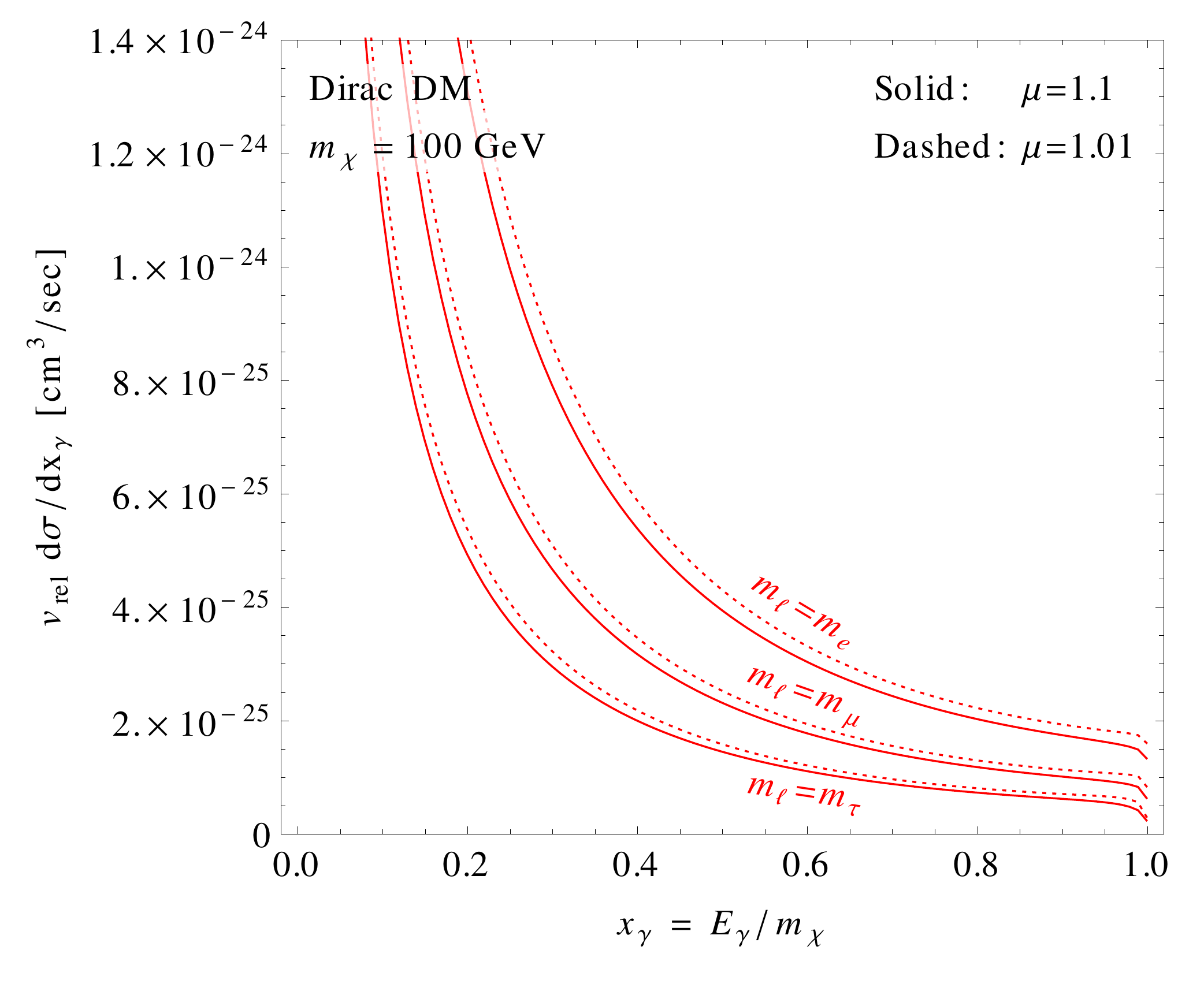} \\
    (a) & (b)
  \end{tabular}
  \caption{The differential cross section \eqn\eqref{eq:vdsigmadx-majorana}
    for the internal bremsstrahlung process (\fig\ref{fig:vib-diagrams}) for different
    values of the degeneracy parameter $\mu = m_\eta^2 / m_\chi^2$.
    (a) is for Majorana DM, (b) is for Dirac DM. We have assumed $y = 1$, $N_\ell = 1$
    and $m_\chi = 100$~GeV.}
  \label{fig:vdsigmadx}
\end{figure*}

If the mediator mass $m_\eta$ and the DM mass $m_\chi$ are nearly degenerate,
the emission of an internal bremsstrahlung photon (first and fourth diagram in
\fig\ref{fig:vib-diagrams}) is strongly peaked if the photon energy $E_\gamma$ gets
close to $m_\chi$. The reason is that, in this case, one of the final state
leptons is very soft, and one of the $\eta$ propagators gets close to the mass
shell. In other words, internal bremsstrahlung with $m_\eta \simeq m_\chi$ and
$E_\gamma \sim m_\chi$ can be viewed as DM annihilation into a lepton and a
photon, with the emission of a soft lepton as a form of initial state
radiation.  While the spectral peak is thus due to internal bremsstrahlung
only, it is important to take into account also the final state radiation diagrams to
guarantee gauge invariance of the process. Note that, in contrast to gamma ray
lines from DM annihilation to $\gamma\gamma$, $\gamma Z$ or $\gamma H$, the
peaked signal from internal bremsstrahlung is not loop-suppressed, hence the
cross section can be sizeable.  The differential three-body cross section for
$\chi\chi \to \ell \bar{\ell} \gamma$ in the case of Majorana DM has the
following form~\cite{Bringmann:2012vr}

\newpage

\begin{align}
  v_\text{rel} \frac{d\sigma_{\chi\chi\to \ell\bar{\ell}\gamma}}{dx} &\simeq
    \frac{y^4 \alpha_{\text{em}} N_\ell}{32 \pi^2 m_\chi^2}
        \big( 1 - x\big) \bigg[ \frac{2x}{(\mu+1)(\mu+1-2x)} \nonumber\\[0.2cm]
  &\hspace{-2.0cm}
      - \frac{x}{(\mu+1-x)^2}
      - \frac{(\mu+1)(\mu+1-2x)}{2(\mu+1-x)^3}
        \log\!\bigg(\! \frac{\mu+1}{\mu+1-2x} \bigg)\!\bigg] \,,
  \label{eq:vdsigmadx-majorana}
\end{align}
with the electromagnetic fine structure constant $\alpha_\text{em}$, the number
of final state lepton flavors $N_\ell$, and with the
definitions $x \equiv E_\gamma / m_\chi$ and $\mu \equiv m_\eta^2 / m_\chi^2$.
In \eqn\eqref{eq:vdsigmadx-majorana}, we have neglected the lepton mass $m_\ell$ and
the DM velocity $v_\text{rel}$.
$v_\text{rel} d\sigma/dx$ is plotted in \fig\ref{fig:vdsigmadx} for different values of $\mu$.
It is clear that, in order to have a distinct peak, a small degeneracy parameter $\mu \lesssim 1.1$
is necessary. Integrating over $x$, we immediately obtain also the full cross
section~\cite{Bringmann:2012vr}
\begin{multline} 
  v_\text{rel} \sigma_{\chi\chi\to \ell\bar{\ell}\gamma} \simeq
    \frac{y^4 \alpha _{\text{em}} N_\ell}{64 \pi ^2 m_{\chi }^2}
      \bigg[ \frac{1}{2\mu} \big(4 \mu^2 - 3 \mu - 1 \big) \log \bigg(\frac{\mu-1}{\mu+1} \bigg) \\
    + \frac{4 \mu + 3}{\mu + 1}
    - (\mu + 1) \bigg\{ \! \log ^2\bigg(\!\frac{\mu +1}{2 \mu }\bigg)
    + 2 \text{Li}_2 \bigg(\!\frac{\mu +1}{2 \mu }\bigg) - \frac{\pi^2}{6}\bigg\} \bigg] \,.
  \label{eq:sigmav-majorana}
\end{multline}
Here, $\text{Li}_2$ is the dilogarithm function.

The approximate expression for the relic density of Majorana
DM in our toy model is~\cite{Bringmann:2012vr}
\begin{align} 
  \Omega_\chi h^2 \simeq
    0.11 \frac{1}{N_\ell} \bigg( \frac{0.35}{y} \bigg)^4 \bigg( \frac{m_\chi}{100\ \text{GeV}} \bigg)^2
    \frac{(1+\mu)^4}{1+\mu^2} \, 
  \label{eq:relic-density}
\end{align}
for $\mu \gtrsim 1.2$. For smaller $\mu$, $\Omega_\chi h^2$ is smaller by an
$\mathcal{O}(1)$ factor due to coannihilations~\cite{Bringmann:2012vr}
(see \cite{Giacchino:2013bta} \fig~VII, for a quantitative estimate
of the effect of co-annihilations). We see that
in the interesting parameter range
$0.1 \lesssim y \lesssim 1$, $m_\chi \gtrsim 100$~GeV, the model
\eqref{eq:lagrangian1} naturally predicts a relic density comparable
to the observed value $0.089 < \Omega_\chi h^2 < 0.1227$.
Here, the quoted upper limit on $\Omega_\chi h^2$ is taken from
Planck~\cite{Ade:2013zuv}, whereas for the lower limit, we conservatively
use the WMAP value~\cite{Hinshaw:2012aka}. We thus account in a qualitative way
for the uncertainty in $\Omega_\chi h^2$ from the yet unresolved tension
between different measurements of the Hubble constant $H_0 = h \cdot
\text{100~km s$^{-1}$ Mpc$^{-1}$}$.

For Dirac DM, $v_\text{rel} \sigma_{\chi\bar\chi\to \ell\bar{\ell}\gamma}$ is not a
well-defined quantity in the limit $m_\ell \to 0$ due to infrared divergences
in the phase space region where the photon is soft or collinear with one of the
leptons.  For $m_\ell \neq 0$, we can evaluate $v_\text{rel}
\sigma_{\chi\bar\chi\to \ell\bar{\ell}\gamma}$ numerically, see
\fig\ref{fig:vdsigmadx}.\footnote{We have checked that the logarithms appearing
in the expression for $v_\text{rel} \sigma_{\chi\bar\chi\to \ell\bar{\ell}\gamma}$
are sufficiently small for a perturbative treatment to be approximately valid.}
We find that the spectrum is entirely dominated by
final state radiation and no internal bremsstrahlung peak is discernable at
$E_\gamma \sim m_\chi$.  This means in particular that no sharp spectral
features are expected for Dirac DM. In the following, we will therefore use the
two-body annihilation cross section
\begin{align}
  v_\text{rel} \sigma_{\chi\bar\chi\to \ell\bar{\ell}} =
    \frac{y^4 N_\ell}{32\pi m_\chi^2} \frac{1}{(1+\mu)^2}
  \label{eq:sigmav-dirac}
\end{align}
as a figure of merit for indirect detection of Dirac DM.

\subsection{Extended models and connection to supersymmetry}

A natural realization of scenario \eqref{eq:lagrangian1} is provided by the
leptonic sector of supersymmetric extensions of the SM. There, the mediator
$\eta$ is the lightest slepton and the DM candidate $\chi$ is the lightest
neutralino, which is given in terms of its bino ($\tilde{B}$), wino
($\tilde{W}^3$) and higgsino ($\tilde{H}_1^0$, $\tilde{H}_2^0$) components as
$\chi = N_{11} \tilde{B} + N_{12} \tilde{W}^3 + N_{13}\tilde{H}_1^0 +
N_{14}\tilde{H}_2^0$. Here, $N_{ij}$ are elements of the neutralino mixing
matrix. The next-to-lightest slepton, as well as the squarks, are assumed to be
much heavier than $m_\eta$.  In the MSSM, the Yukawa coupling $y$ can be
written in terms of the unit electric charge $e$, the Weinberg angle
$\theta_W$, and the neutralino mixing matrix element $N_{11}$
as~\cite{Haber:1984rc}
\begin{align} 
  y &= \sqrt{2} \frac{e}{\cos\theta_W} \, N_{11} \,.
  \label{eq:SUSYCouplings1}
\end{align}
If, instead of \eqn\eqref{eq:lagrangian1}, we were considering couplings to
left handed leptons and their corresponding sleptons, the Yukawa coupling in the MSSM would be
given by  $y = \sqrt{2} ( Q_f - T^3) \,N_{11}\, e / \cos\theta_W
+ \sqrt{2} T^3 g N_{12}$.  Since for conventional mechanisms of supersymmetry breaking,
slepton masses of one chirality tend to be similar, we will also
generalize~\eqref{eq:lagrangian1} to include all three lepton flavors
$\ell^\alpha$ and slepton flavors $\eta^\alpha$ of one chirality, where $\alpha =
e$, $\mu$, $\tau$:
\begin{align}
  \mathscr{L} \supset
    \! \sum_{\alpha = e, \mu, \tau} \Big( -y \bar{\chi} P_R \ell^\alpha \eta^\alpha  
    - i e \, \eta^\alpha A^\mu \partial_\mu \eta^{\alpha *} \Big) + h.c. \,.
  \label{eq:lagrangian2}
\end{align}
Finally, we will also consider a more general model (which cannot be realized
in the MSSM), in which couplings to both left-handed and right-handed fermions
are included, and couplings are allowed to be flavor off-diagonal. The
Lagrangian for this generalized toy model is
\begin{align} 
  \mathscr{L} &\supset
    - \sum_{\alpha, j} y^{\alpha j}_R \bar{\chi} P_R \ell^\alpha \eta^j
    - \sum_{\alpha, j} y^{\alpha j}_L \bar{\chi} P_L \ell^\alpha \eta^j \nonumber\\
    &\quad- i e \, \sum_j \eta^j A^\mu \partial_\mu \eta^{j *} + h.c. \,.
  \label{eq:general-lagrangian}
\end{align}
Here, $(y^{\alpha j}_{L/R})$ are the Yukawa matrices, and
$\eta^j$ are the mass eigenstates of the scalar mediators, of which an arbitrary
nmber could exist.  The index $\alpha$ runs over $e$, $\mu$, $\tau$,
while $j$ runs over all $\eta^j$ mass eigenstates.

Since our main motivation is the possibility of observing internal bremsstrahlung
signals in future gamma ray observations, we will mostly focus on the case where
the mass scale of the mediator(s), $m_\eta$, is similar to the DM mass. The reason
is that in this case the photon spectrum from internal bremsstrahlung
is strongly peaked.  Note that models with $m_\eta \sim m_\chi$ are notoriously
difficult to probe at colliders because the charged leptons produced in slepton decays
are very soft.  In the supersymmetric context, a model with nearly degenerate
neutralino and slepton masses has been studied with a different goal in
\cite{Konishi:2013gda}. We will comment on this model also at the end of
\sect\ref{sec:conclusions}.

\section{\label{sec:dd-limits}Electromagnetic form factors of dark matter
         and direct detection constraints}

\begin{figure}
  \begin{center}
    \includegraphics[width=0.5\textwidth]{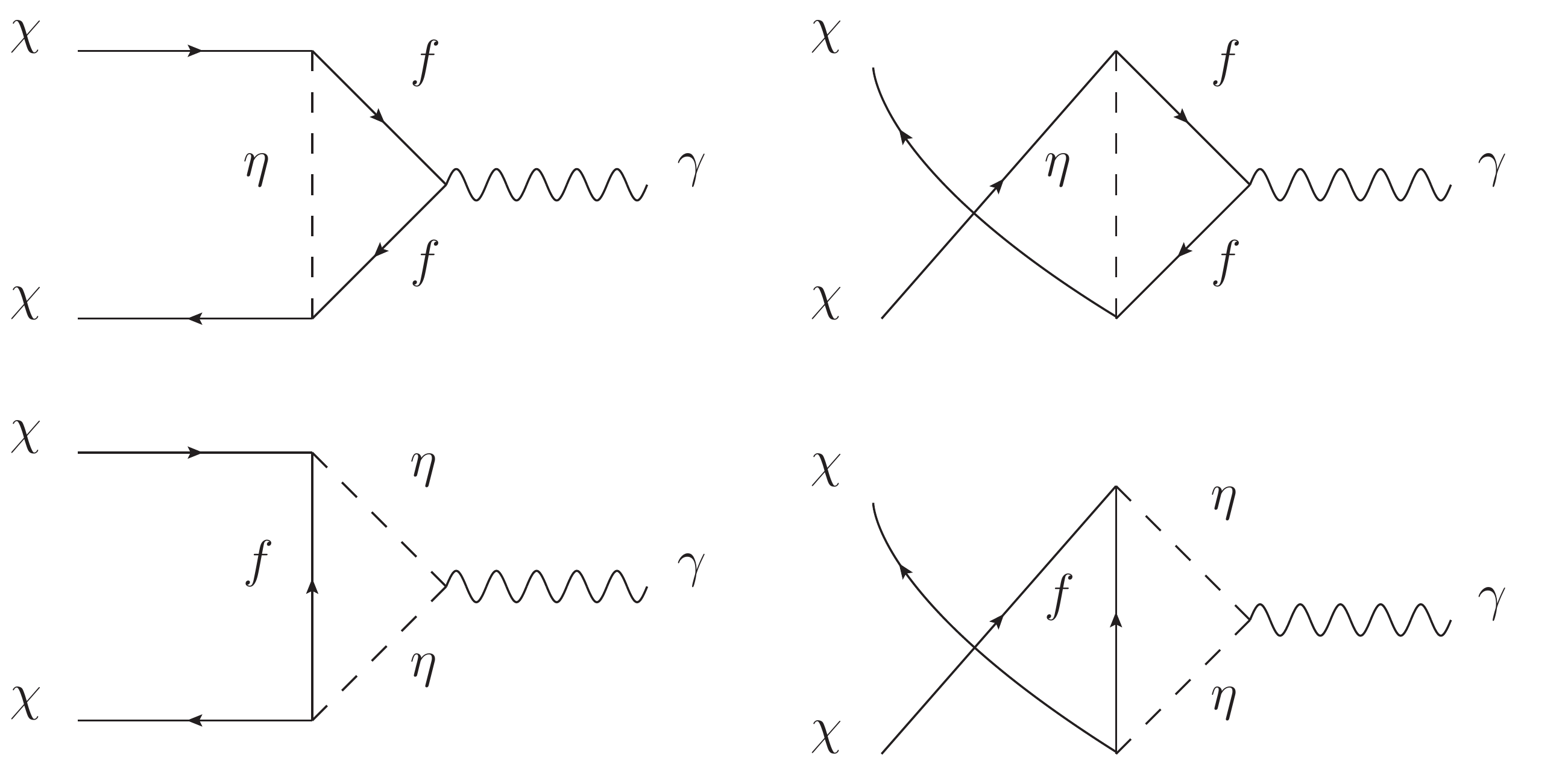}
  \end{center}
  \caption{The one loop diagrams generating the effective dark matter--photon
    coupling for Majorana DM.  For Dirac DM, the two diagrams on the right are absent.}
  \label{fig:em-moment-loops}
\end{figure} 

We now establish the connection between indirect gamma ray signatures of DM in
our toy model and direct laboratory searches on Earth.  Connecting the final
state fermion lines in the internal bremsstrahlung and final state radiation
diagrams from \fig \ref{fig:vib-diagrams}, we obtain an effective vertex
coupling the DM particle to the photon through loops of the form shown in
\fig\ref{fig:em-moment-loops}.  At dimension 5 and 6, the most general form of
this effective interaction for a neutral fermion $\chi$ is~\cite{Fukushima:2013efa}
\begin{multline}
  \mathscr{L}_\text{eff} \supset
    \frac{d_M}{2} \bar\chi \sigma^{\mu\nu} \chi \, F_{\mu\nu}
    + \frac{d_E}{2} \, \bar\chi \sigma^{\mu\nu} \gamma^5 \chi \, F_{\mu\nu} \\
    + \mathcal{A} \, \bar\chi \gamma^\mu \gamma^5 \chi \, \partial^\nu F_{\mu\nu}  \,,
  \label{eq:Leff}
\end{multline}
where $d_M$ is the magnetic dipole moment, $d_E$ is the electric dipole moment, and
$\mathcal{A}$ is the anapole moment.  For Majorana DM,
only the anapole term is nonzero~\cite{Radescu:1985wf, Kayser:1983wm},
as can be seen by using the fact that a Majorana field is invariant under
the charge conjugation operation $\hat{C}$, i.e.\ $\hat{C} \chi \hat{C}
\equiv -i \gamma^2 \chi^* = \chi$.  Applying this identity to the fermion
fields in \eqn\eqref{eq:Leff}, it is straightforward to show that
the magnetic and electric dipole terms vanish.

Note that establishing a similar connection between DM annihilation and
loop-induced electromagnetic form factors is also possible in internal
bremsstrahlung models with scalar DM and fermionic mediators, or with Majorana
DM and vector mediators~\cite{Barger:2011jg, Frandsen:2013bfa}. We have seen
above that these scenarios are phenomenologically as interesting as our model
with Majorana DM and a scalar mediator because internal bremsstrahlung
dominates over DM annihilation to 2-body final states in all of them.  The
connection between gamma ray lines from DM annihilation and direct detection
signals has been made also for models with loop-induced DM annihilation to
photons in~\cite{Frandsen:2012db}.

\subsection{\label{sec:loop} One Loop Contribution to the Electromagnetic Moments}

We will now compute the loop induced electromagnetic interactions for the DM particles
in our toy model \eqn\eqref{eq:lagrangian1}.

\subsubsection{Anapole moment for Majorana fermions}

\begin{figure*}
  \begin{tabular}{cc}
    \includegraphics[width=0.55\textwidth]{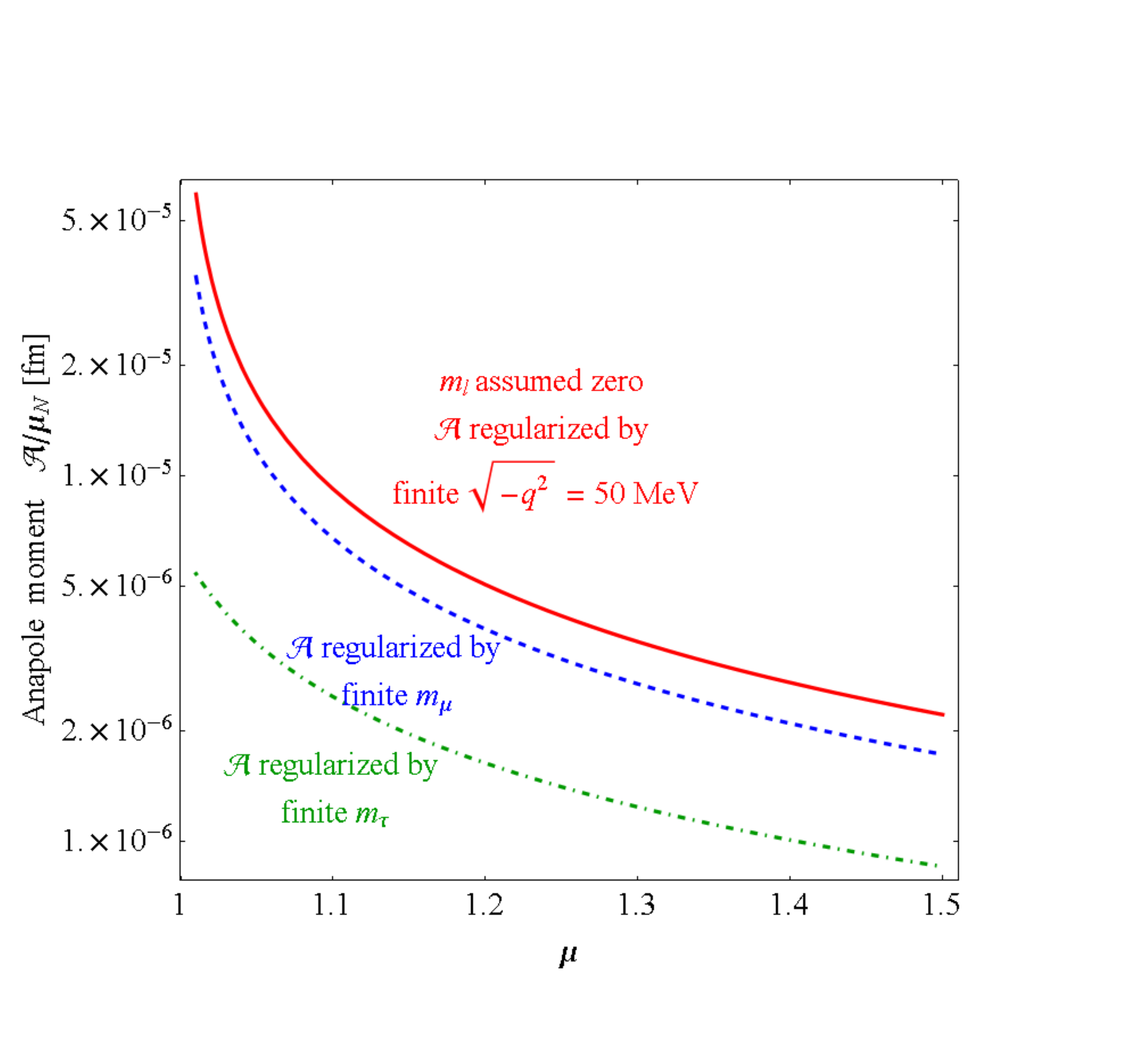} \hspace{-1.0cm} &
    \includegraphics[width=0.55\textwidth]{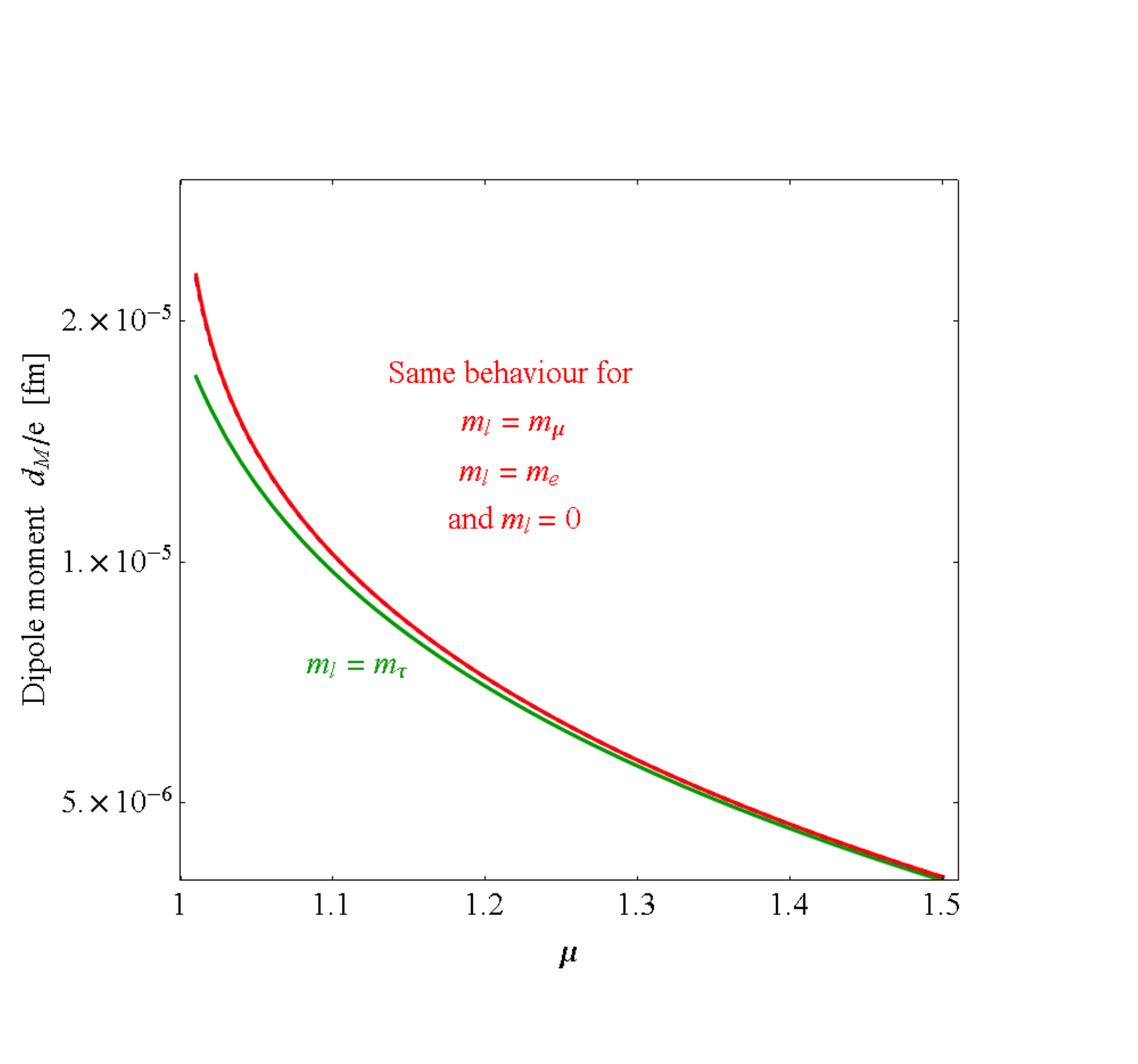} \\[-0.5cm]
    (a) & (b)
  \end{tabular}
  \caption{The (a) anapole moment for Majorana DM and (b) magnetic dipole moment
    for Dirac DM as a function of $\mu = m_\eta^2 / m_\chi^2$. We show results
    for DM couplings to electrons, muons, and tau leptons. Note that for couplings
    to electrons, the divergence in $\mathcal{A}$ is regularized by the momentum
    transfer $q^2$ rather than $m_e$ because in typical DM--nucleus scattering processes,
    $|q^2| \gg m_e^2$.  We have assumed $y=1$ and $m_\chi = 100$~GeV.}
  \label{fig:aN}
\end{figure*}

We begin by evaluating the diagrams in \fig\ref{fig:em-moment-loops} to obtain the
anapole form factor $\mathcal{A}$ in \eqn\eqref{eq:Leff} for Majorana DM.
For negligible 4-momentum transfer $q$ we find
\begin{widetext}
\begin{align}
  \mathcal{A} &= -\frac{e y^2}{96 \pi^2 m_\chi^2} \bigg[
          \frac{3}{2} \log \frac{\mu}{\epsilon}
            - \frac{1 + 3\mu - 3\epsilon}{\sqrt{(\mu-1-\epsilon)^2 - 4\epsilon}}
              \arctanh\bigg( \frac{\sqrt{(\mu-1-\epsilon)^2 - 4\epsilon}}{\mu - 1 + \epsilon} \bigg)
        \bigg] \,,
  & \text{($|q^2| \ll m_\ell^2$)}
  \label{eq:anapole-moment}
\end{align}
with $\mu= m_\eta^2/ m_\chi^2$ , $\epsilon = m_\ell^2 / m_\chi^2$.
Taking into account the behavior of the $\arctanh$ function when its argument approaches 1,
it is easy to see that for $1 \gg \mu - 1 \gg \epsilon \to 0$, the anapole moment diverges
logarithmically as $\mathcal{A} \sim e y^2 / (48 \pi^2 m_\chi^2) \times
\log(\epsilon) / (\mu - 1)$. This behavior can be qualitatively understood by noting
that, if $m_\eta \simeq m_\chi$ and $q^2 \simeq 0$, all three propagators in the loops of
\fig\ref{fig:em-moment-loops} can be close to the mass shell simultaneously.
In the limit $\mu - 1 \ll \epsilon \ll 1$, on the other hand,
the leading term in $\mathcal{A}$ is proportional to $1/\sqrt{\epsilon}$. Note that in
this limit, the expression in \eqn\eqref{eq:anapole-moment} requires
analytic continuation of the $\arctanh$ function into the complex plane.
The dependence of $\mathcal{A}$ on the degeneracy parameter $\mu$ is shown in
\fig\ref{fig:aN} (a) for $y=1$ and $m_\chi = 100$~GeV.

If $|q^2| \gg m_\ell^2$, a case that is relevant for instance in DM--nucleus scattering
through loops containing electrons, the approximation $q^2 \to 0$ underlying
\eqn\eqref{eq:anapole-moment} is not applicable. In this case, it is instead
convenient to set $m_\ell = 0$ and keep only to the leading term in
$\xi \equiv \sqrt{|q^2|}/ m_\chi$, which leads to
\begin{align}
  \mathcal{A} &= -\frac{e y^2}{32 \pi^2 m_\chi^2} \bigg[
                  \frac{-10 + 12 \log\xi - (3 + 9 \mu) \log(\mu -1) -
                        (3 - 9 \mu) \log\mu}{9 (\mu - 1)}
                 \bigg] \,,
  & \text{($|q^2| \gg m_\ell^2$)}\,.
  \label{eq:anapole-moment-ml0}
\end{align}
At very small $\epsilon$ or $\xi$, one may wonder whether a calculation at fixed order
in perturbation theory is still valid. However, in the case of interest to us,
namely $\mu - 1 \gg \epsilon$, the divergent logarithms in
\eqns\eqref{eq:anapole-moment} and \eqref{eq:anapole-moment-ml0} are at most of
order 10 even for DM couplings to electrons.

\subsubsection{Dipole moment for Dirac fermions}

If $\chi$ is a Dirac fermion rather than a Majorana particle, only the two diagrams on
the left in \fig\ref{fig:em-moment-loops} exist. They generate an anapole moment
$\mathcal{A}$ that is half as large as the one for Majorana DM, \eqn\eqref{eq:anapole-moment},
and a magnetic dipole moment $d_M$ given by
\begin{align}
  d_M = \frac{y^2 e}{32 \pi^2 m_\chi} \bigg[
      -1 + \frac{1}{2} (\epsilon - \mu) \log \Big( \frac{\epsilon}{\mu} \Big)
      -\frac{(\mu - 1)(\mu - 2\epsilon) - \epsilon (3 - \epsilon)}
            {\sqrt{(\mu-1)^2 - 2\epsilon(\mu + 1) + \epsilon^2}}
       \arctanh\bigg( \frac{\sqrt{(\mu-1)^2 - 2\epsilon(\mu + 1) + \epsilon^2}}
                           {\mu - 1 + \epsilon} \bigg)
      \bigg] \,
  \label{eq:dipole-moment}
\end{align}
\end{widetext}
for $q^2 \to 0$. The dipole moment will turn out to be numerically much more important
than the anapole moment in scattering processes involving Dirac DM.
If $m_\ell$ is neglected compared to $m_\chi$,
i.e.\ $\epsilon \to 0$, \eqn\eqref{eq:dipole-moment} simplifies to
\begin{align}
  d_M = \frac{y^2 e}{32 \pi^2 m_\chi} \bigg( \mu \log\frac{\mu}{\mu - 1} - 1 \bigg) \,.
\end{align}
Note that, unlike the anapole moment $\mathcal{A}$, the dipole moment $d_M$ is
not divergent for $\epsilon \to 0$.  For $\mu - 1 \ll \epsilon \ll 1$, the
leading term in $d_M$ is proportional to $1/\sqrt{\epsilon}$.
The behavior of $d_M$ as a function of $\mu$ is shown in \fig\ref{fig:aN} (b).

\subsection{\label{sec:dd-signals} Direct detection signals}

In this section we will discuss the experimental limits on dark matter
scattering through anapole and magnetic dipole interactions. This has been done
previously at the effective field theory level for instance in
\refs\cite{Sigurdson:2004zp, Masso:2009mu, Barger:2010gv, Fitzpatrick:2010br, Banks:2010eh,
DelNobile:2012tx, Weiner:2012cb, Ho:2012bg, DelNobile:2013cva,
Gresham:2013mua, DelNobile:2014eta}. Here, we carry out a similar analysis using the latest
LUX~\cite{Akerib:2013tjd} and XENON100~\cite{Aprile:2012nq} data, and we then
translate the resulting constraints into new limits on the expected indirect
detection signals in our toy model. Since the differential DM--nucleus
scattering cross section $d\sigma/dE_r$ (where $E_r$ is the nuclear recoil
energy) for anapole and dipole interactions differs from the conventional
spin-independent or spin-dependent scenarios, we cannot directly use the
published exclusion limits from LUX and XENON100, but instead have to fit the
data at the event level.  We do this by using a framework developed
in \refs\cite{Kopp:2009et, Kopp:2009qt, Kopp:2011yr}, which we have extended by
including LUX data and by implementing anapole and dipole interactions.

\begin{figure*}
  \begin{tabular}{cc}
    \hspace{-0.4cm}
    \includegraphics[width=0.58\textwidth]{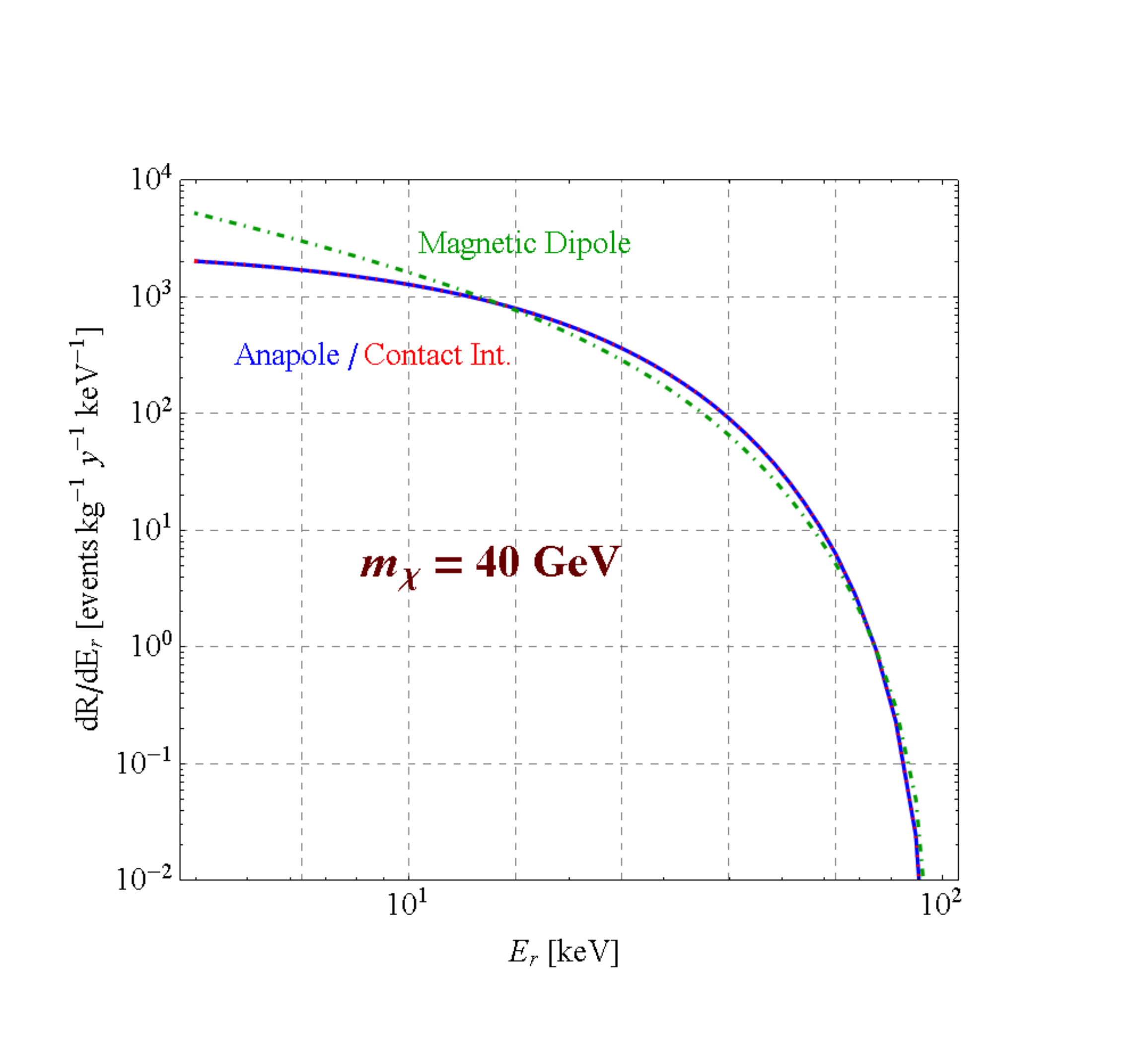} \hspace*{-1.4cm} &
    \includegraphics[width=0.58\textwidth]{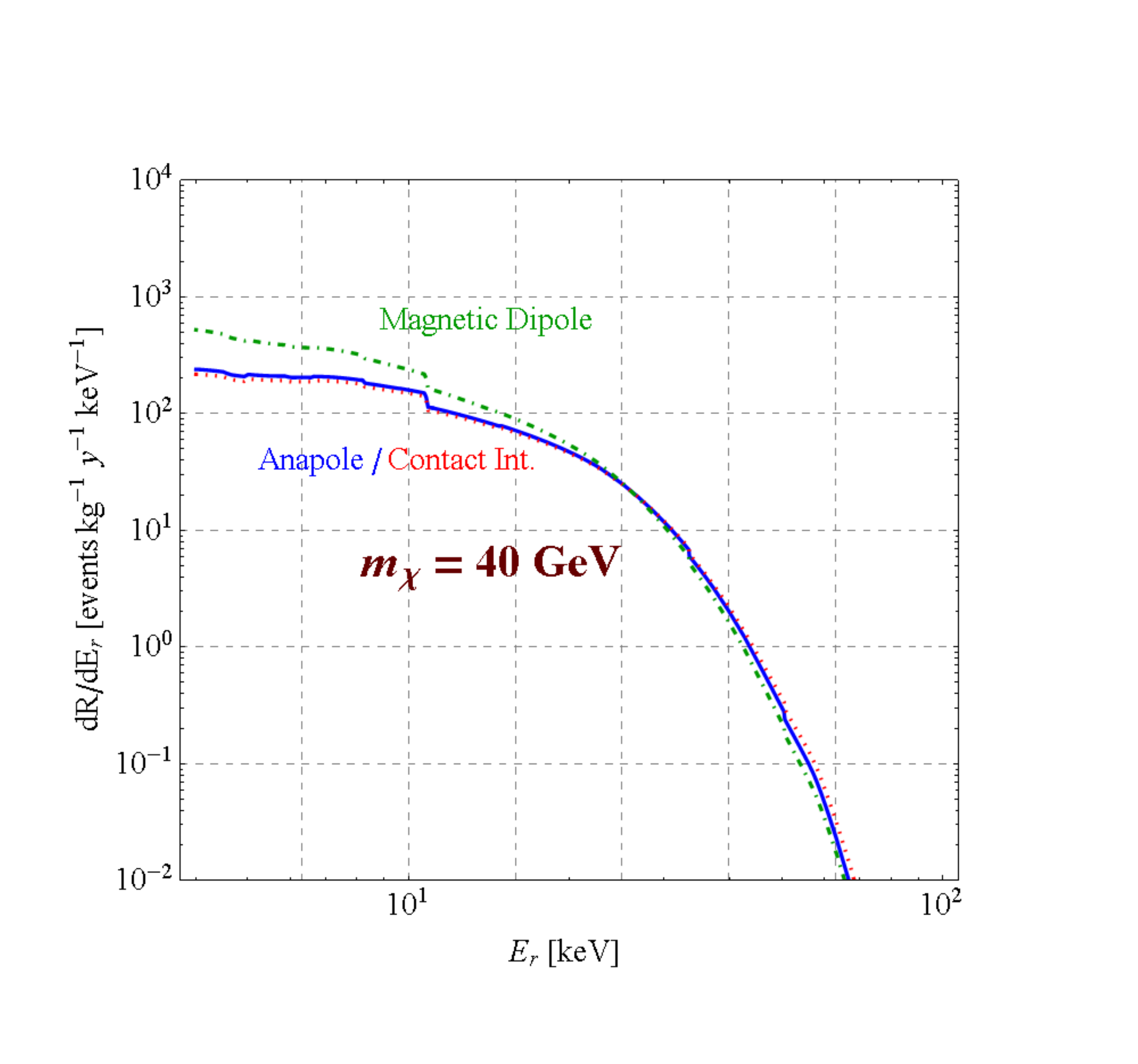} \\[-0.5cm]
    (a) & (b)
  \end{tabular}
  \caption{
      Comparison of the differential DM--nucleus scattering rates on a xenon
      target for anapole interactions (blue), magnetic dipole interactions (green
      dot-dashed) and conventional spin-independent contact interactions (red).
      We show (a) the theoretical rate without nuclear form factor and detector
      effects, and (b) the event rate expected in the XENON100 detector, taking into
      account the detection efficiency, light
      yield and energy resolution as given in~\cite{Aprile:2012nq, Aprile:2011hi}.
      We have used a DM mass of $m_\chi = 40$~GeV
      and coupling constants $\mathcal{A} = 3.8 \times 10^{-3} \mu_N$~fm, $d_M =
      1.9 \times 10^{-4} e$~fm and $\sigma_{\chi p} = 5.0 \times 10^{-41}$~cm$^2$ for
      anapole, dipole and contact interactions, respectively.  (Here,
      $\sigma_{\chi p}$ is the total DM--nucleon cross section.) For the DM
      velocity profile and the nuclear form factor, we have used standard
      assumptions (see text for details).}
  \label{fig:dRdE}
\end{figure*}

The differential cross section for DM--nucleus scattering through an
anapole interaction is (cf.\ also~\cite{Ho:2012bg, DelNobile:2014eta})
\begin{align}
  \frac{d\sigma_{\chi N}^\text{anapole}}{dE_r}
    &= 4 \alpha_\text{em} \mathcal{A}^2 Z^2 [F_Z(E_r)]^2
       \bigg[ 2 m_N - \bigg(1 + \frac{m_N}{m_\chi}\bigg)^2 \frac{E_r}{v^2} \bigg] \nonumber\\
    &\qquad + 4 \mathcal{A}^2 d_A^2 [F_s(E_r)]^2 \bigg(\frac{J+1}{3J}\bigg)
              \frac{2 E_r m_N^2}{\pi v^2} \,,
  \label{eq:dsdE-anapole}
\end{align}
while for dipole interactions we have~\cite{Pospelov:2000bq,
Chang:2010en, Barger:2010gv, Banks:2010eh}
\begin{align}
  \frac{d\sigma_{\chi N}^\text{dipole}}{dE_r}
  &= \frac{\alpha_\text{em} Z^2 [F(E_r)]^2 d_M^2}{E_r}
       \bigg[ 1 - \frac{E_r}{2 m_N v^2} \bigg( 1 + 2\frac{m_N}{m_\chi} \bigg) \bigg] \nonumber\\
  &\qquad + d_M^2 d_A^2 [F_s(E_r)]^2 \bigg(\frac{J+1}{3J}\bigg) \frac{m_N}{\pi v^2} \,.
  \label{eq:dsdE-dipole}
\end{align}
In both equations, the first line corresponds to scattering on the nuclear
charge $Z$, while the second line describes scattering on the nuclear dipole
moment $d_A$.\footnote{Note that the contributions from the nuclear charge and
from the nuclear dipole moment must be separated carefully. For instance, a
naive calculation involving the standard QED vertex for the nucleus would
correctly describe DM--charge scattering, but the contribution from DM--dipole
scattering would be correct only for a truly pointlike nucleus with magnetic
dipole moment $e / (2 m_N)$.  Here, instead, this spurious DM--dipole
scattering term must be subtracted out and replaced by the correct term for
scattering on the dipole moments of extended nuclei (second line of
\eqns\eqref{eq:dsdE-anapole} and~\eqref{eq:dsdE-dipole}).}
The nuclear mass is denoted by $m_N$, and $v$ is the velocity of
the incoming DM particle. We have also included the nuclear charge form factor
$F_Z(E_r)$ and the spin form factor $F_s(E_r)$.  We parametrize $F_Z(E_r)$
as~\cite{Jungman:1995df} $F_Z(E_r) = 3 e^{-\kappa^2 s^2/2} [\sin(\kappa r) -
\kappa r \cos(\kappa r)] / (\kappa r)^3$, where $\kappa = \sqrt{2 m_N E_r}$, $s
= 1$~fm, $r = \sqrt{R^2 - 5 s^2}$, $R = 1.2 A^{1/3}$~fm (with the
nuclear mass number $A$). For $F_s(E_r)$, we use~\cite{Banks:2010eh}
$F_s(E_r) = \sin\kappa R_s / (q R_s)$ for $\kappa R_s < 2.55$ and $\kappa R_s >
4.5$, and $F_s(E_r) = 0.217$ otherwise. Here, $R_s = A^{1/3}$.  Note that
nuclear dipole moments are subdominant in many target materials, including
xenon, which we mostly focus on in this paper. The contribution from the nuclear
dipole moment may be comparable to the contribution from the nuclear charge for
instance in fluorine, sodium and iodine~\cite{Chang:2010en}.  Note that
\eqn\eqref{eq:dsdE-anapole} can be integrated over $E_r$ to yield a total cross
section, while \eqn\eqref{eq:dsdE-dipole} has an infrared divergence, which
makes the total cross section for dipole interactions an ill-defined
quantity.

The differential DM--nucleus scattering rate per unit target mass is given by
\begin{align}
  \frac{dR}{dE_r} &= \frac{\rho_0}{m_\chi m_N}
    \int_{v_\text{min}}^\infty \! d^3v \frac{d\sigma}{dE_r} v \, f_\oplus(\vec{v}) \,,
\end{align}
where $\rho_0 \simeq 0.3$~GeV/cm$^3$ is the local DM density, $v_\text{min} =
\sqrt{m_N E_r / 2} / M_{\chi N}$ is the minimal DM velocity required to yield a
recoil energy $E_r$, $M_{\chi N} = m_\chi m_N / (m_\chi + m_N)$ is the reduced mass of the DM--nucleus
system, and $f_\oplus(\vec{v})$ is the DM velocity distribution in
the rest frame of the detector.  We obtain $f_\oplus(\vec{v})$ by a Galilean
transformation of the DM velocity distribution in the Milky Way rest frame,
$f_\text{MW}$. For the latter, in turn, we assume the conventional
Maxwell-Boltzmann form with a smooth cutoff, $f_\text{MW} \propto
\exp(-\vec{v}^2 / v_0^2) - \exp(-v_\text{esc}^2 / v_0^2)$, with velocity
dispersion $v_0 = 220$~km/s and escape velocity $v_\text{esc} = 550$~km/s.
We expect the dependence of our results on this choice of velocity profile to
be similar to what was found for DM scattering through contact
interactions in the literature, see for instance~\cite{McCabe:2010zh,
Farina:2011pw, Green:2011bv, Fairbairn:2012zs}.

In \fig\ref{fig:dRdE}, we compare the differential reaction rates $dR/dE_r$ for
anapole, dipole and spin-independent contact interactions, both with and
without including nuclear form factor and detector effects. For easier
comparison, all rates are normalized to a total rate of 1~event above 10~keV
per~kg per~day before taking into account nuclear form factor and detector effects.
We see that anapole and contact interactions lead to
similar event spectra, while dipole interactions are strongly enhanced at low
energies due to the $1/E_r$ dependence of the first term in
\eqn\eqref{eq:dsdE-dipole}. The nuclear form factor leads to a suppression of
$dR/dE_r$ at higher energies.  Note that at low energies, the scattering
rate remains sizeable down to few keV even because such low energy events can
occasionally produce a detectable number of photoelectrons due to Poisson statistics.

We conclude that with sufficient statistical power direct detection
experiments could relatively easily distinguish dipole interactions from other
interaction structures, while discriminating between anapole and contact
interaction is challenging.

In the absence of a signal, we next derive limits on the anapole moment
$\mathcal{A}$, the dipole moment $d_M$ and the total DM--nucleon scattering
cross section for contact interactions, $\sigma_{\chi p}$.

\subsection{\label{sec:dd-constraints} Constraints from direct detection data}

\begin{figure*}
  \begin{tabular}{cc}
    \hspace{-0.7cm}
    \includegraphics[width=0.58\textwidth]{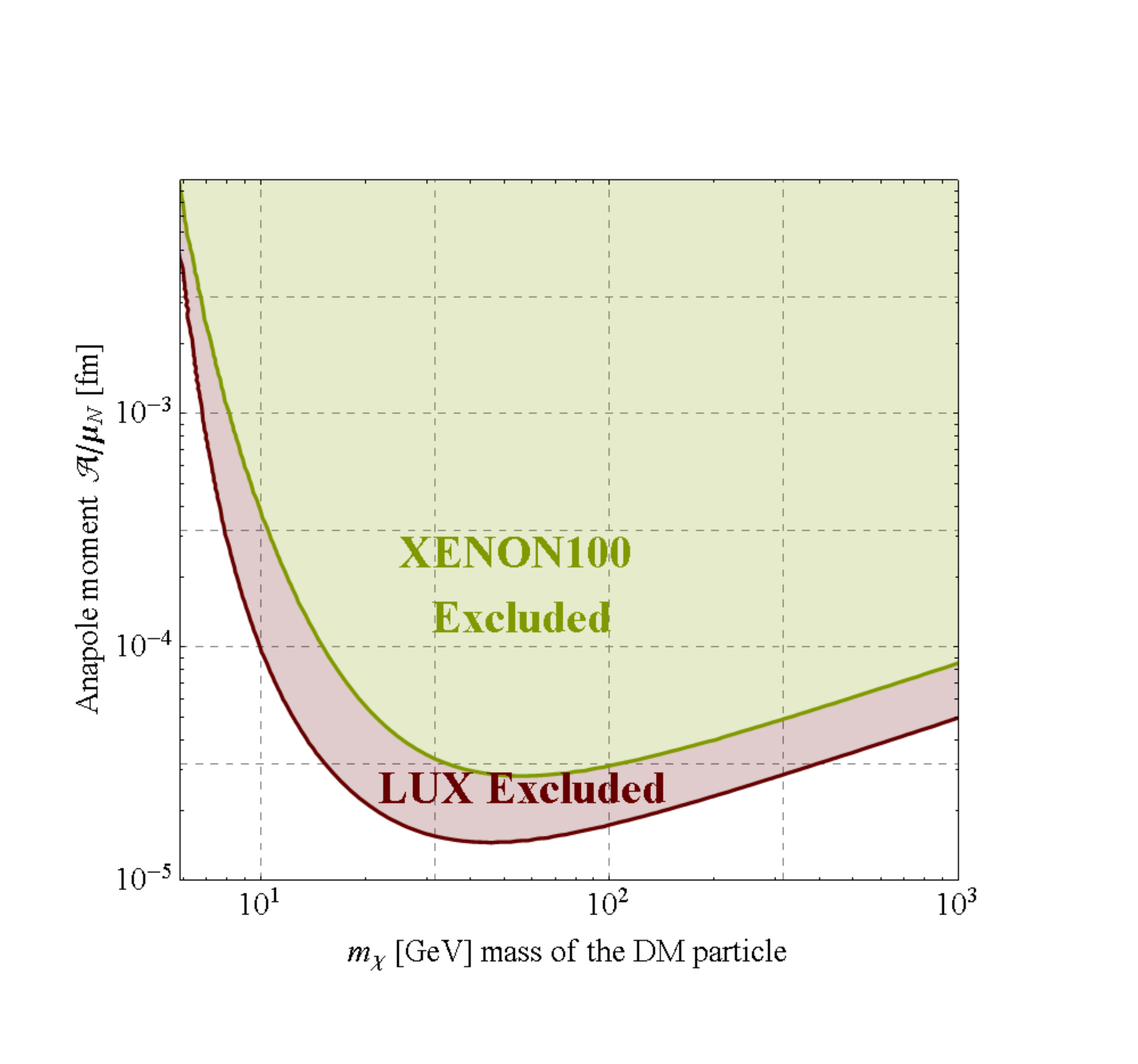} \hspace{-1.5cm} &
    \includegraphics[width=0.58\textwidth]{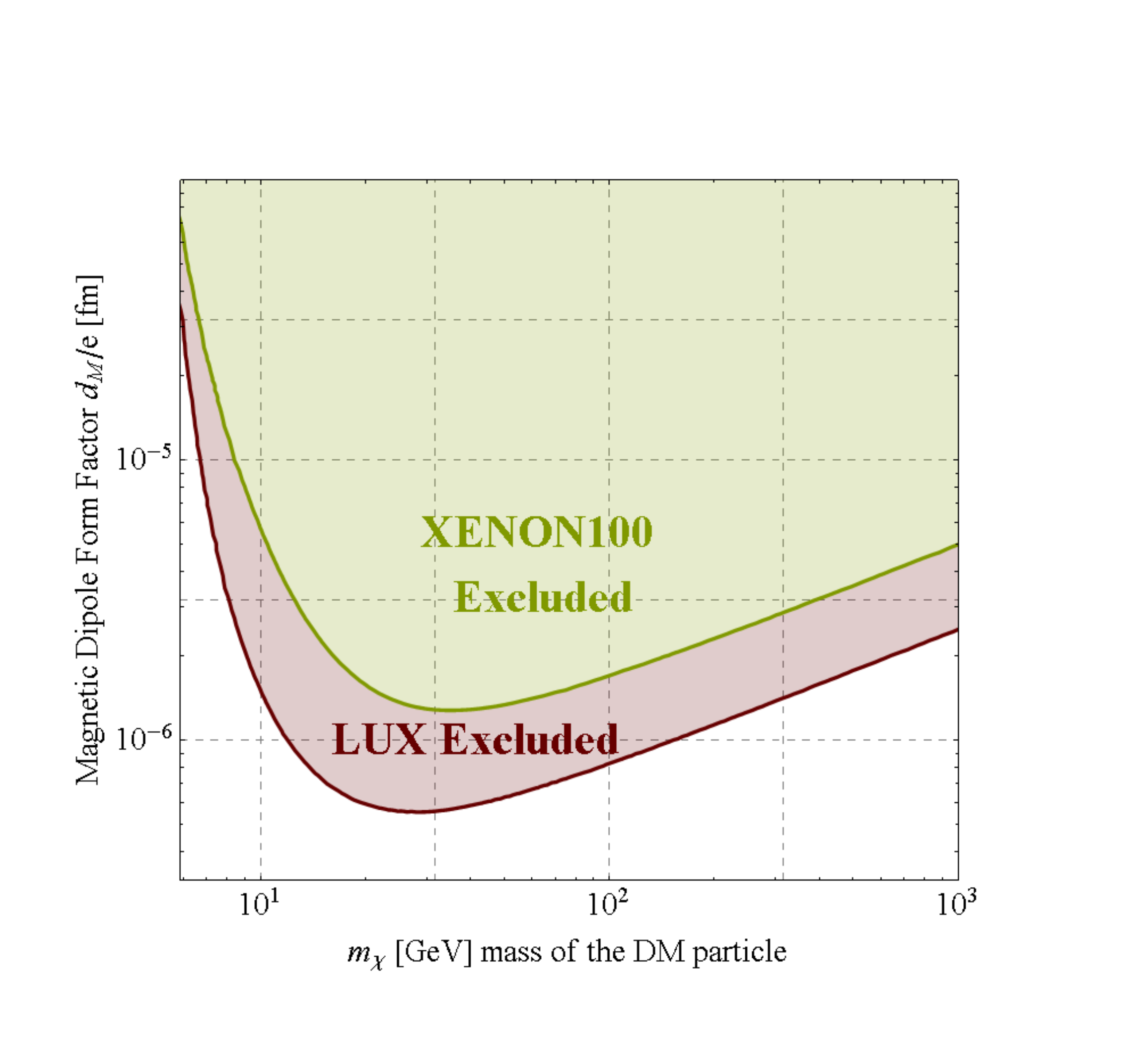} \\[-0.5cm]
    (a) & (b)
  \end{tabular}
  \caption{XENON100 and LUX 90\% CL limits on (a) the anapole moment and (b)
    the magnetic dipole moment of dark matter.}
  \label{fig:limits-em-moments}
\end{figure*}

In \fig\ref{fig:limits-em-moments} we show the constraints on the anapole and magnetic
dipole moments of dark matter from 85.3~days of LUX data~\cite{Akerib:2013tjd}
and from 225~days of XENON100 data~\cite{Aprile:2012nq}.  For the statistical
analysis, we have used Yellin's maximum gap
method~\cite{Yellin:2002xd}.  The code employed to derive limits has been
developed in~\cite{Kopp:2009et, Kopp:2009qt, Kopp:2011yr}, and we have checked
that it reproduces the XENON100 and LUX limits on standard spin-independent DM--nucleus
scattering to very good accuracy.
Note that the qualitative shape of the exclusion
curves is similar to the well-known exclusion limit for scattering through
contact interactions. At low DM mass, the loss in sensitivity is slightly less
steep for dipole interactions due to the enhancement of the scattering rate at
low energies (see \fig\ref{fig:dRdE}).

We now derive our main results by translating the LUX constraint on the anapole
moment from \fig\ref{fig:limits-em-moments}~(a) into a constraint on the
annihilation cross section $\ev{\sigma v_\text{rel}}_{\chi \chi \to \ell \bar{\ell}
\gamma}$ into two charged leptons plus an internal bremsstrahlung photon
using \eqns\eqref{eq:anapole-moment} and \eqref{eq:sigmav-majorana}.
Similarly, we convert the LUX limits on the dipole moment of Dirac DM from
\fig\ref{fig:limits-em-moments}~(b) into bounds on the DM annihilation cross
section into two charged leptons, $\ev{\sigma v_\text{rel}}_{\chi \bar\chi \to
\ell \bar{\ell}}$ using \eqns\eqref{eq:dipole-moment} and
\eqref{eq:sigmav-dirac}.  Note that the total cross section for the 3-body
final state $\ell \bar{\ell} \gamma$ is ill-defined in the Dirac case due to
infrared divergences.  Moreover, annihilation into $\ell \bar{\ell}
\gamma$ is subdominant for Dirac DM.

\begin{figure*}
  \vspace{-0.7cm}
  \begin{tabular}{cc}
    \hspace{-0.5cm}
    \includegraphics[width=0.5\textwidth,clip=true,trim=5mm 5mm 20mm 20mm]{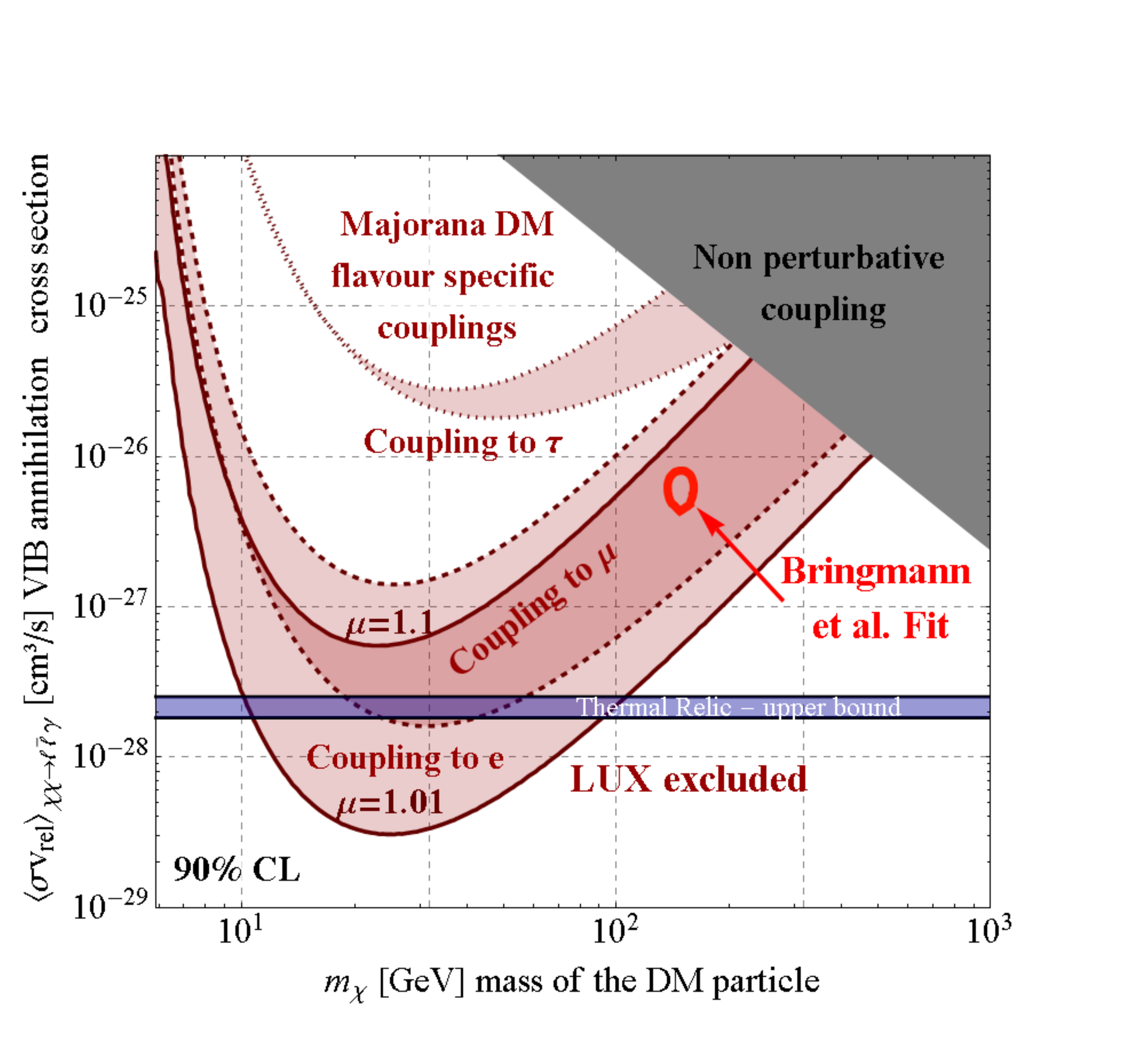} &
    \includegraphics[width=0.5\textwidth,clip=true,trim=5mm 5mm 20mm 20mm]{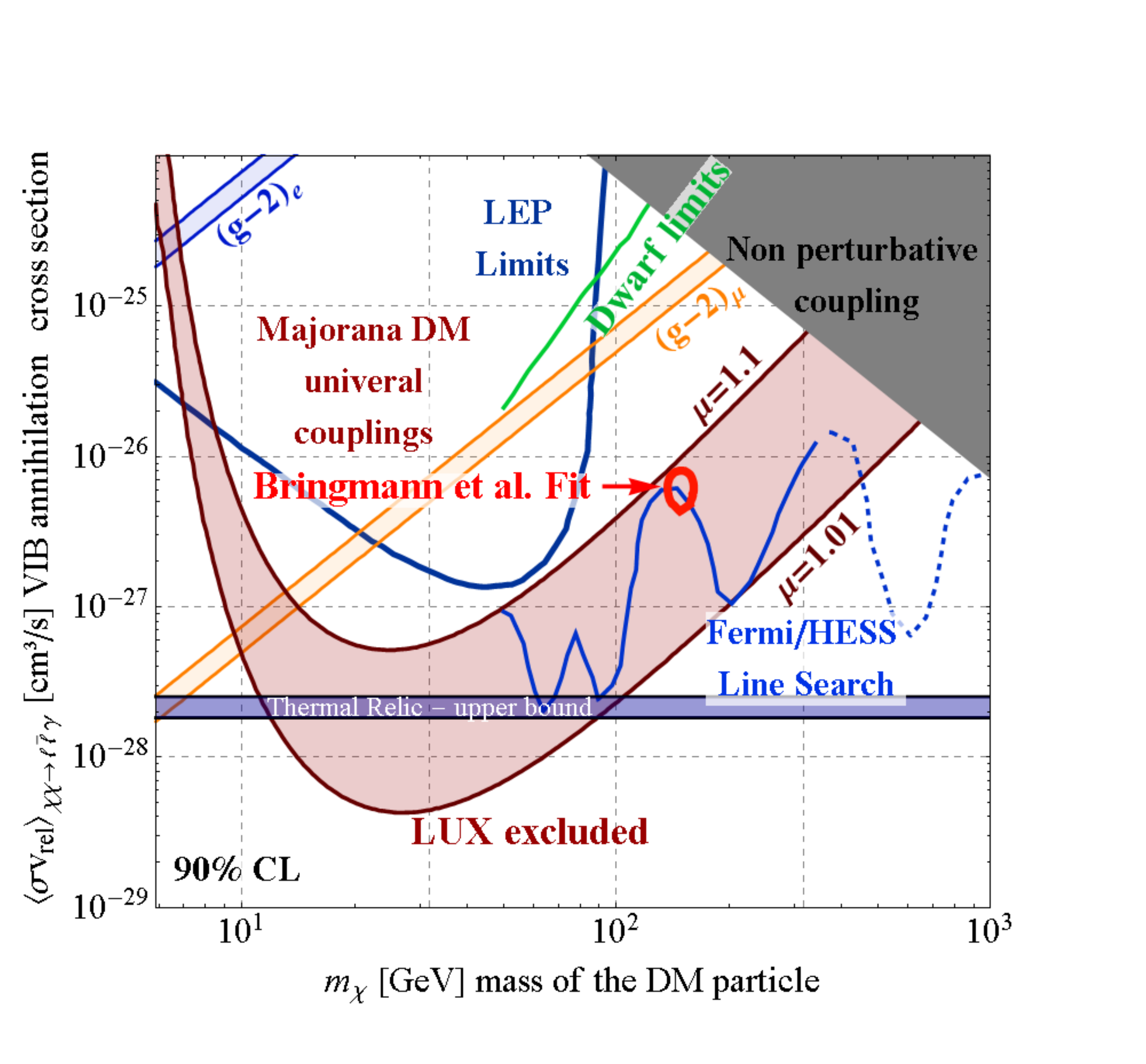} \\
    (a) & (b) \\
    \includegraphics[width=0.5\textwidth,clip=true,trim=5mm 5mm 20mm 20mm]{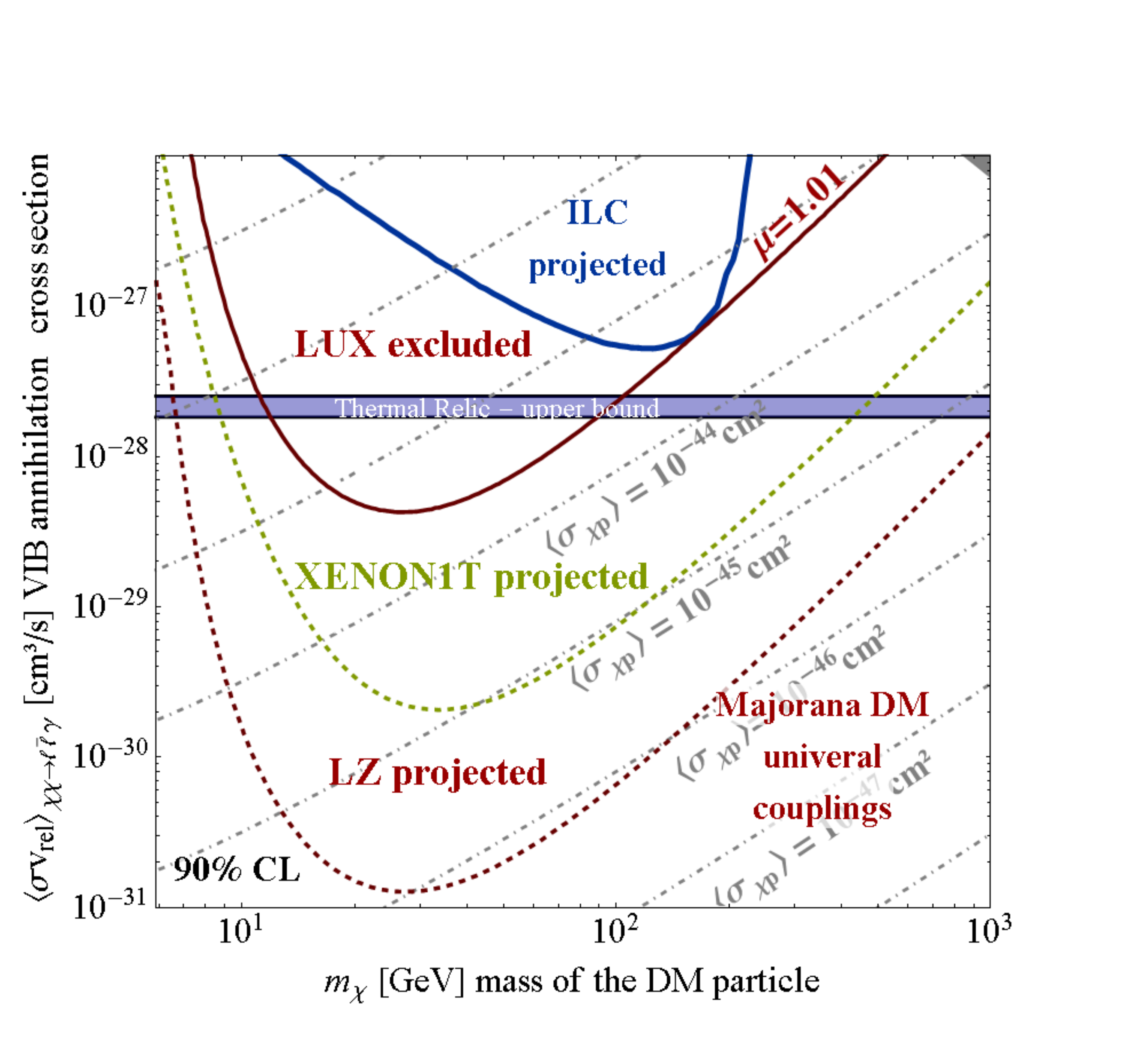} &
    \includegraphics[width=0.5\textwidth,clip=true,trim=5mm 5mm 20mm 20mm]{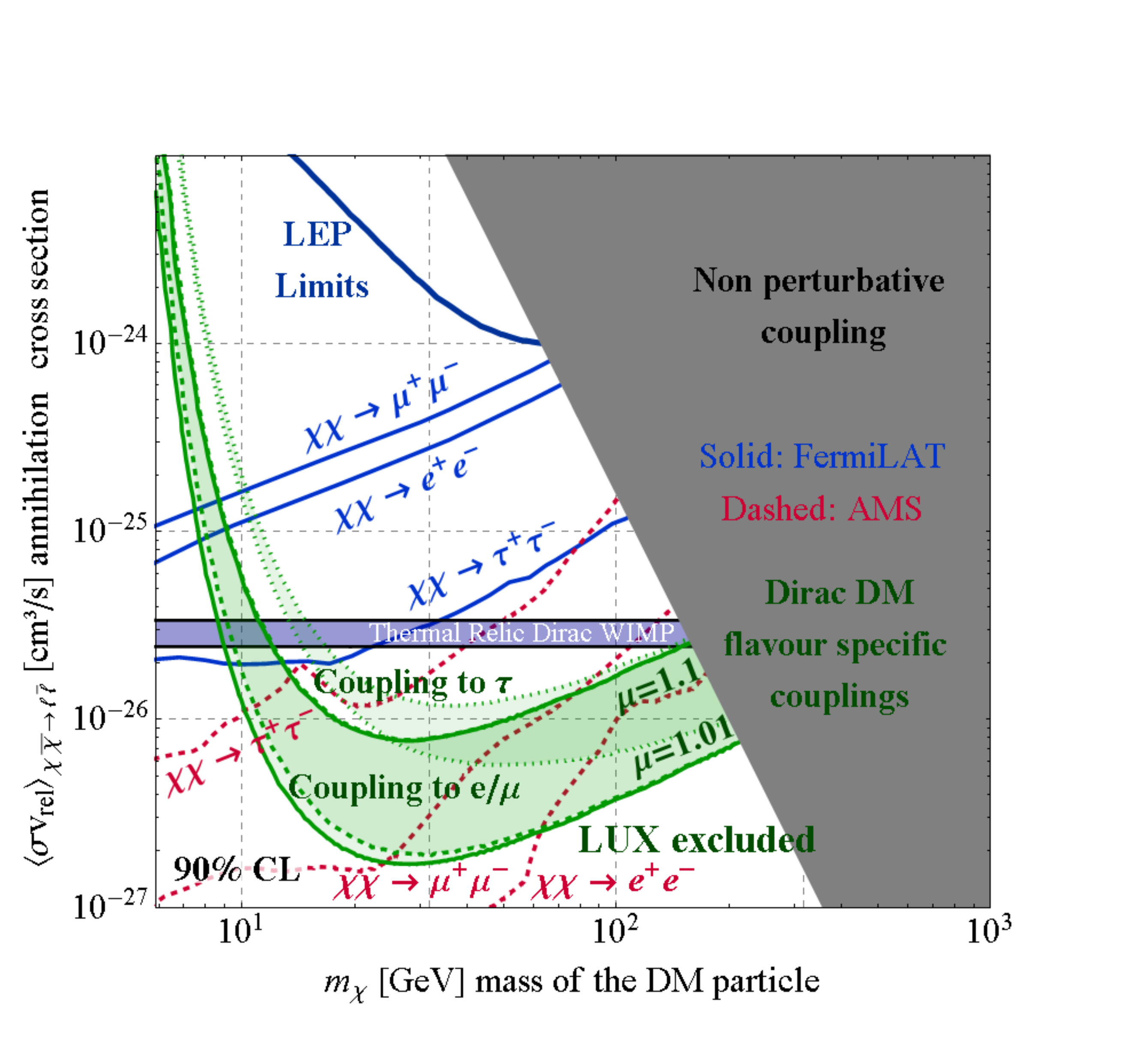} \\
    (c) & (d)
  \end{tabular}
  \caption{LUX 90\% CL limits on the DM annihilation cross section
    in our toy model, \eqn\eqref{eq:lagrangian1}.  In (a) we show
    direct detection constraints induced by anapole interactions for Majorana
    DM coupling only to electrons (thick solid), only to muons (thick dotted),
    and only to taus (thin dotted).  The upper and lower boundaries of the
    colored bands correspond to $\mu \equiv m_\eta^2 / m_\chi^2 = 1.1$ and $\mu
    = 1.01$, respectively. For illustration, we also show the upper limit on the cross
    section required for a thermal relic (neglecting coannihilations and using
    \eqn\eqref{eq:relic-density}), and the tentative best fit region
    from Bringmann et al.~\cite{Bringmann:2012vr}.  The gray region corresponds to $y^2 > 4\pi$
    and thus cannot be reached in our toy model.
    In (b), we compare the LUX bounds on Majorana DM with flavor-universal
    couplings to limits from LEP mono-photon searches (see
    \sect\ref{sec:collider}), $g-2$ measurements (see \sect\ref{sec:g-2}), a
    Fermi-LAT search for continuum gamma rays from dwarf
    galaxies~\cite{Bringmann:2012vr}, and Fermi-LAT (solid) and H.E.S.S.\ (dotted)
    searches for gamma ray lines from the Galactic Center~\cite{Garny:2013ama}.
    In (c), we project the future sensitivities of ton-scale direct detection
    experiments and of a future linear collider for Majorana DM with
    flavor-universal couplings and with $\mu = 1.01$.  For illustration, we
    have also drawn contours of constant velocity-averaged direct detection cross section
    $\ev{\sigma_{\chi p}}$ (see \eqn\eqref{eq:avg-sigma-dd}).
    In (d), we summarize direct detection constraints induced by magnetic
    dipole interactions for Dirac DM with flavor-specific couplings, and we
    compare again to the thermal relic cross section, to LEP mono-photon
    limits, to Fermi-LAT limits from dwarf galaxies~\cite{Ackermann:2013yva}, and to the AMS limits
    from~\cite{Bergstrom:2013jra}. Note that no sharp features are expected
    in the gamma ray spectrum from annihilation of Dirac DM.} 
      \label{fig:limits-sigmav}
\end{figure*}

Our results are shown in \fig\ref{fig:limits-sigmav}~(a), (b) and (c) for
Majorana DM, and in \fig\ref{fig:limits-sigmav} (d) for Dirac DM.
\figs\ref{fig:limits-sigmav}~(a) and (d) are for couplings to only one
lepton species $\ell$, while (b) and (c) are for flavor-universal couplings.

For Majorana DM, \fig\ref{fig:limits-sigmav}~(a) clearly reflects the increase
in the anapole moment for small $\epsilon = m_\ell^2 / m_\chi^2$, which here
translates into stronger limits on the model parameters and on $\ev{\sigma
v_\text{rel}}_{\chi \chi \to \ell \bar{\ell} \gamma}$ for coupling to electrons
than for coupling to $\mu$ or $\tau$.  We also clearly see the effect of
degenerate $m_\eta$ and $m_\chi$: for $\mu = m_\eta^2 / m_\chi^2$ close to
unity, the anapole moment is significantly larger than for well separated
$m_\eta$ and $m_\chi$ (see \eqn\eqref{eq:anapole-moment} and \fig\ref{fig:aN}).
Comparing to the preferred parameter region from the gamma ray line search
in~\cite{Bringmann:2012vr}, we find that this region is still marginally compatible
with direct detection constraints if $\mu=1.1$. For $\mu = 1.01$, it is disfavored
at the $5\sigma$ confidence level if DM has couplings to electrons and at the
$3\sigma$ confidence level if DM couples only to muons.
Comparing to the cross sections required for thermal relic DM (horizontal blue line
in \fig\ref{fig:limits-sigmav}~(a)), we see that direct detection limits are just
starting to probe this region. Note that our estimates for the thermal relic cross
section are based on \eqn\eqref{eq:relic-density}. They do not include the
effect of co-annihilations~\cite{Bringmann:2012vr}, which would move
the thermal relic cross section to smaller values. Note also that our
perturbative calculations become inaccurate close to the gray regions
in \fig\ref{fig:limits-sigmav}, inside of which $y^2$ is larger than $4\pi$. 

Comparing direct detection constraints to limits from gamma ray searches
(\fig\ref{fig:limits-sigmav}~(b)), we find that for flavor-universal couplings
and $\mu$ not too far from unity, direct searches are significantly more
sensitive than continuum gamma ray searches in dwarf
galaxies~\cite{Bringmann:2012vr} and competitive with the bounds from gamma ray
line searches~\cite{Garny:2013ama}. (Note that in \refs\cite{Bringmann:2012vr,
Garny:2013ama} these bounds are shown only for $m_\chi \gtrsim 50$~GeV, even
though in principle, Fermi-LAT and H.E.S.S. are sensitive also to lower DM
masses.) At $m_\chi \lesssim 10$~GeV, direct detection limits are superseded by
constraints from the anomalous magnetic moment $g-2$ of the electron and the
muon (see \sect\ref{sec:g-2}).

Looking into the future, \fig\ref{fig:limits-sigmav}~(c) illustrates that the
sensitivity of direct detection experiments can be expected to improve by more
than two orders of magnitude in the coming years thanks to the planned
XENON1T~\cite{Aprile:2012zx} and LUX-ZEPLIN (LZ)~\cite{Malling:2011va}
experiments.  This will make direct DM searches highly sensitive to thermal
relic DM.  For XENON1T, we have assumed a total exposure of 2\,200~kg~yrs,
while for LZ we use 10\,000~kg~yrs. In both cases, this corresponds to roughly
2~years of data taking. For comparison, we plot in \fig\ref{fig:limits-sigmav}~(c)
also contours of constant $\ev{\sigma_{\text{ana}}}$ (gray dot-dashed curves),
where $\ev{\sigma_{\text{ana}}}$ is the direct detection cross section averaged
over the DM velocity distribution:
\begin{align}
  \ev{\sigma_{\chi N}^\text{anapole}} =
    \int_{v_{min}}^\infty f_\oplus(\vec{v}) \, \sigma_{\chi N}^\text{anapole} \, d^3v \,.
  \label{eq:avg-sigma-dd}
\end{align}
Note that direct detection limits on $\langle\sigma_{\chi N}^\text{anapole}\rangle$ are
more than an order of magnitude weaker than direct detection limits on the
cross section for DM--nucleon scattering through contact interactions.
The reasons are the velocity dependence in $\sigma_{\chi N}^\text{anapole}$
as well as the fact that anapole interactions are proportional to the
nuclear charge rather than the nuclear mass. As discussed in \sect\ref{sec:dd-signals},
couplings to nuclear dipole moments are subdominant for the target material considered here.

For Dirac DM, \fig\ref{fig:limits-sigmav}~(d) shows that the qualitative
picture is similar to Majorana DM, but the dependence on the lepton mass $m_\ell$ is less
strong. Comparing the direct detection limits to constraints from the Fermi-LAT
analysis of gamma ray signals from dwarf galaxies~\cite{Ackermann:2013yva}, we
find that for DM masses $> 10$~GeV, direct detection provides significantly
stronger limits if $m_\chi$ and $m_\eta$ are not too different. In this case,
also thermal production (horizontal blue band in
\fig\ref{fig:limits-sigmav}~(d)) is excluded for $\text{10 GeV}
\lesssim m_\chi \lesssim \text{few} \times 100$~GeV.

\section{\label{sec:collider} Collider searches for leptophilic dark matter}

A set of constraints on leptophilic DM complementary to the limits from direct
detection can be obtained from collider data.  Since tree level production of
DM at hadron colliders~\cite{Cao:2009uw, Beltran:2010ww, Goodman:2010yf,
Bai:2010hh, Goodman:2010ku, Fox:2011pm, Goodman:2011jq} is impossible in the
leptophilic case, the strongest constraints are expected to come from
mono-photon events at LEP~\cite{Fox:2011fx}.  In the future, mono-photon
searches at a linear collider may improve on these
bounds~\cite{Dreiner:2012xm}.

Here, we apply the procedure described in~\cite{Fox:2011fx} to our toy model,
\eqn\eqref{eq:lagrangian1}.  We simulate the process $e^+ e^- \to
\chi\chi\gamma$ in CalcHEP~3.4~\cite{Belyaev:2012qa} including the effect of
initial state radiation and beamstrahlung (with default parameters) on the beam
energy.  We analyze the simulated events in a modified version of
MadAnalysis~1.1.2 (from the MadGraph~4 package)~\cite{Alwall:2007st} that
implements the efficiencies and resolutions of the DELPHI detector at
LEP~\cite{Abdallah:2003np, DELPHI:2008zg}, see \cite{Fox:2011fx} for details.
We have checked that our simulation reproduces the predicted $\gamma \bar\nu
\nu$ background from~\cite{DELPHI:2008zg} to very good accuracy.  To set
limits, we add our signal prediction to the background prediction
from~\cite{DELPHI:2008zg}, and compare to the DELPHI data from \fig 1 of
\cite{DELPHI:2008zg}, which corresponds to an integrated luminosity of
650~pb$^{-1}$.  Following~\cite{Fox:2011fx} we use a simple $\chi^2$ analysis
to set limits on the Yukawa coupling $y$ as a function of $m_\chi$ and
$m_\eta$, and then convert these limits into constraints on $\ev{\sigma
v}_{\chi\chi \to \ell \bar{\ell} \gamma}$, which are shown in
\fig\ref{fig:limits-sigmav} (b) and (d).  Systematic uncertainties are
subdominant compared to statistical uncertainties in DELPHI and are therefore
neglected in our analysis.

We also estimate the sensitivity of a future linear collider with a center of
mass energy $\sqrt{s} = 500$~GeV to leptophilic DM in our toy model. We
simulate the signal and the dominant $\gamma \bar\nu \nu$ background in
CalcHEP~3.4~\cite{Belyaev:2012qa} while for the $\gamma \gamma \bar\nu \nu$
final state (with one photon escaping undetected) and for $\gamma e^+ e^-$
events (with an undetected $e^+ e^-$ pair) we follow~\cite{Dreiner:2012xm}: we
qualitatively include the $\gamma \gamma \bar\nu \nu$ background by simply
increasing the $\gamma \bar\nu \nu$ background by 10\%. For $\gamma e^+ e^-$
events, we reweight the $\gamma \bar\nu \nu$ spectrum by the energy dependent
factor $0.825 \, [1 - E / (0.9\ \text{GeV})]^2$. Negative reweighting factors
are excluded.  The detector response of an ILC detector is modeled according to
the information given in~\cite{Abe:2010aa, Bartels:2011dea, Dreiner:2012xm}. We
assume an energy resolution of $\Delta E / E = 0.011 \oplus 0.166 / \sqrt{E /
\text{GeV}}$, where the notation $\oplus$ means that the different terms
correspond to separate, statistically independent Gaussian distributions. We
restrict our analysis to the photon energy range $\text{10 GeV} < E_\gamma <
\text{220 GeV}$ (divided into 5~GeV bins) to remove events with on-shell $Z$
production, and to the rapidity range $|y| < 2.3$.  The detection efficiency is
given by $0.941 - 0.00129 E_\gamma / \text{GeV}$. We derive limits using a
simple $\chi^2$ analysis, assuming an integrated luminosity of 50~fb$^{-1}$ and
neglecting systematic uncertainties. Our projected ILC limits are included in
\fig\ref{fig:limits-sigmav} (c) and (d).

\section{\label{sec:ewpt} Constraints from Precision Experiments}

\subsection{\label{sec:g-2} Lepton magnetic dipole moments}

\begin{figure}
  \begin{center}
    \includegraphics[width=0.35\textwidth]{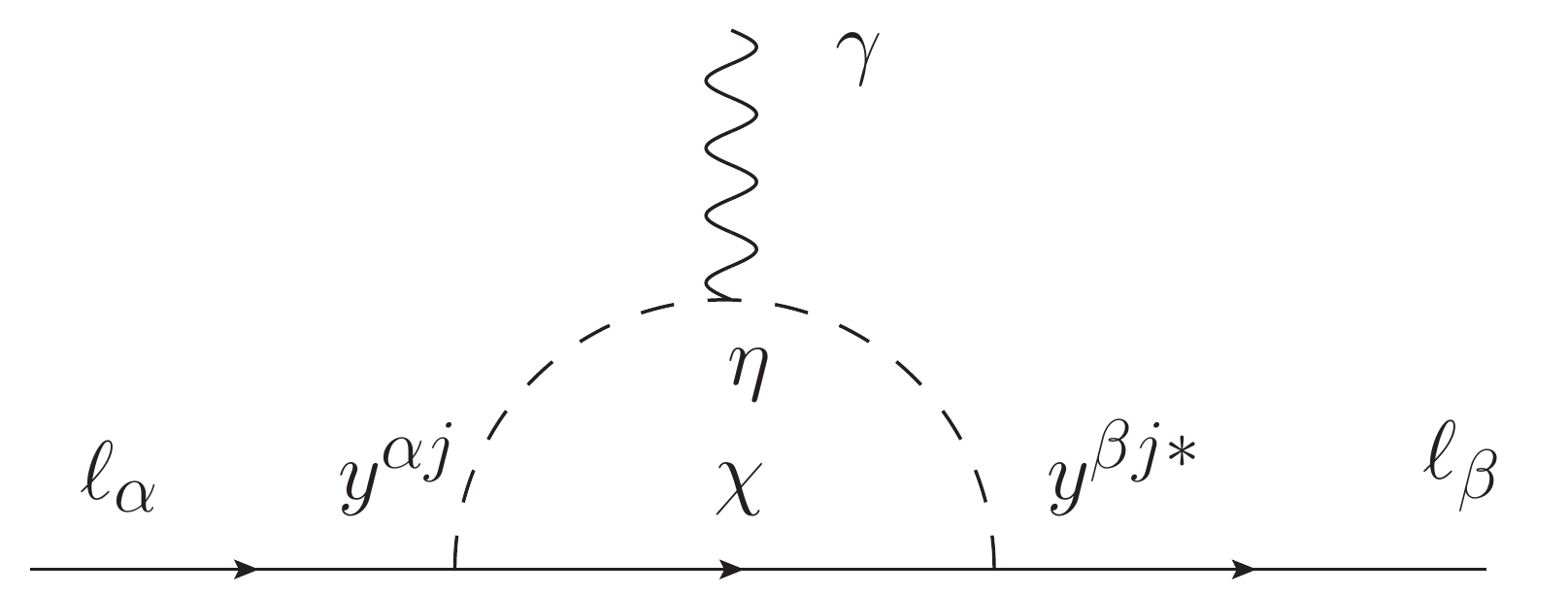}
  \end{center}
  \caption{New physics contribution to the lepton magnetic dipole moment ($\alpha = \beta$)
    and to flavor violating lepton decays ($\alpha \neq \beta$) in our
    simplified model.}
  \label{fig:LFV}
\end{figure}

The extension of the SM by a DM particle and a charged mediator in our toy
model \eqn\eqref{eq:lagrangian1} leads to a new
contribution to the anomalous magnetic moment $(g-2)_\ell$ of leptons $\ell$
via the vertex correction loop shown in \fig\ref{fig:LFV}. This has been used
previously in~\cite{Bringmann:2012vr, Fukushima:2013efa} to constrain DM
annihilation through charged mediators.  In the case of complex Yukawa
couplings, there can also be contributions to electric dipole moments, but we
will not consider this possibility here. In the limit $m_\ell
\ll m_\eta$, $m_\chi$, the anomalous magnetic moment of charged leptons is modified
by~\cite{Bringmann:2012vr}
\begin{align}
  \Delta a_\ell
    &\equiv \Delta \Big( \frac{g-2}{2} \Big)_\ell \nonumber\\
    &= -\frac{y^2 m_\ell^2 }{96\pi^2 m_\chi^2}
      \frac{\mu^3 - 6\mu^2 + 3\mu + 6 \mu\log\mu + 2}{(\mu - 1)^4} \,.
  \label{eq:g-2}
\end{align}
For DM couplings to electrons, we compare \eqn\eqref{eq:g-2} to the difference
between the SM prediction for $a_e$ and the experimentally measured
value, $a_e^\text{exp} - a_e^\text{SM} = (-1.06 \pm 0.82) \times 10^{-12}$
\cite{Aoyama:2012wj} to derive the exclusion bound shown in
\fig\ref{fig:limits-sigmav}~(b) for $\mu = 1.1$ (lower edge of
colored band) and for $\mu = 1.01$ (upper edge of
colored band).

For the $g-2$ of the muon, the difference between the measured best fit value
and the theoretical prediction is $a_\mu^\text{exp} - a_\mu^\text{SM} = [2.87
\pm 0.63\ \text{(exp.)} \pm 0.49\ \text{(theor.)}] \times
10^{-9}$~\cite{Beringer:2012zz}.  We add the experimental and theoretical
uncertainties in quadrature.  To account for the significant discrepancy
between theory and experiment, we artificially inflate the error by linearly
adding an ad-hoc uncertainty given by the central value of the discrepancy,
$2.87 \times 10^{-9}$.  Note that the discrepancy has a sign \emph{opposite} to
the one predicted by \eqn\eqref{eq:g-2}. The resulting constraint on
$\ev{\sigma v_\text{rel}}_{\chi\chi \to \ell\bar\ell\gamma}$ is shown in
\fig\ref{fig:limits-sigmav}~(b).

We see that $g-2$ constraints are competitive with direct and indirect searches only
at DM masses $< 10$~GeV.

\subsection{\label{sec:muonium} Positronium and muonium spectroscopy}

\begin{figure}
  \begin{center}
    \includegraphics[width=0.5\textwidth]{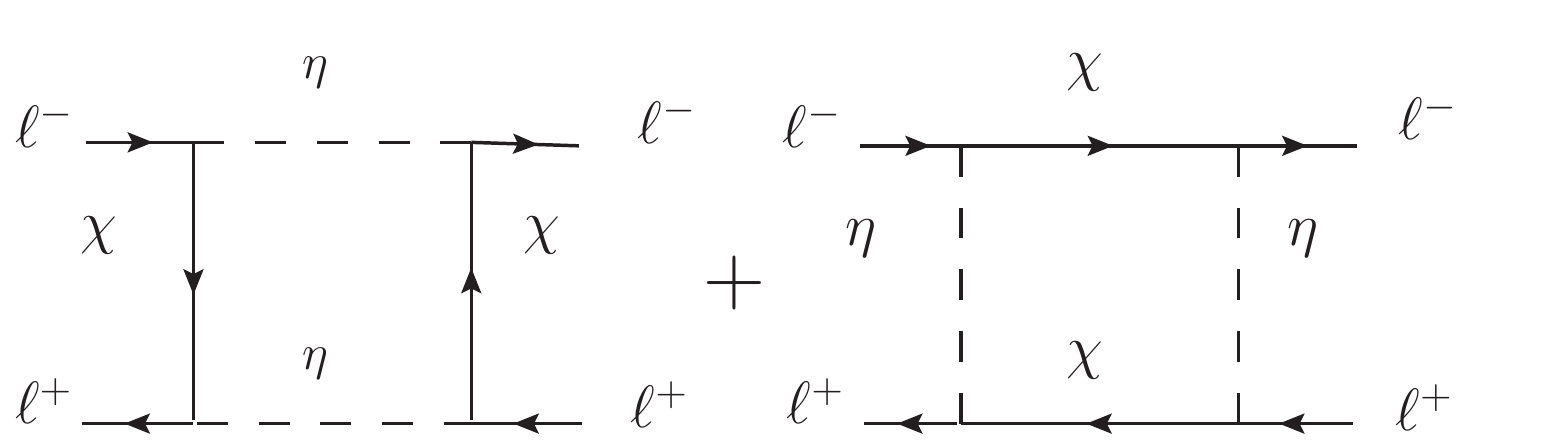}
  \end{center}
  \caption{Diagrams contributing to the hyperfine splitting in $\ell^+ \ell^-$
    systems such as positronium and true muonium.}
  \label{fig:box}
\end{figure} 

Lepton--antilepton bound states such as positronium ($e^+ e^-$) and true muonium
($\mu^+ \mu^-$) are interesting laboratories for precision tests of QED because
they can be studied accurately using spectroscopy, but are theoretically simpler than
atoms. In particular, there are no nuclear effects that need to be taken into account.
In our toy model for leptophilic DM, the box diagrams shown in \fig\ref{fig:box}
lead to an effective contact interaction of the form
\begin{align}
  \mathscr{L}_{\ell^+\ell^-} \equiv \frac{1}{2} C_{\ell^+\ell^-}
    (\bar\ell \gamma^\mu P_R \ell) (\bar\ell \gamma_\mu P_R \ell)
  \label{eq:Leff-dilepton} \\
\intertext{with}
  C_{\ell^+\ell^-} \equiv -\frac{y^4}{64\pi^2 m_\chi^2}
                          \frac{\mu^2 - 2 \mu \log\mu - 1}{(\mu - 1)^3} \,.
\end{align}
This contact interaction contributes to the electrostatic potential between
the $\ell^+$ and $\ell^-$, thus modifying the hyperfine splitting $E_\text{hfs}$
between the energy of the ortho-state (parallel spins,
${}^3\text{S}_1$) and the para-state (antiparallel spins, ${}^1\text{S}_0$).
To obtain the new contribution $\Delta E_\text{hfs}$ to
$E_\text{hfs}$, we first calculate the new term in the Hamilton operator
of the system by plugging explicit expressions for the $\ell^+$ and $\ell^-$
wave functions into \eqref{eq:Leff-dilepton}, integrating over $d^3x$ and adding a
minus sign from the Legendre transform that converts the Lagrangian into the
Hamiltonian as well as a factor 4 from the different ways in which the lepton fields can
be contracted with the incoming and outgoing fermion states.
The lepton wave functions are given by
\begin{align}
  \ell(x) = \frac{(\alpha_\text{em} m_\ell)^{3/2}}{\sqrt{\pi}} \,
    \exp\big[-\alpha_\text{em} m_\ell |\vec{x}| - i E t \big] \, \xi \,,
  \label{eq:lepton-wf}
\end{align}
where $\xi$ is a non-relativistic particle or antiparticle Dirac spinor
normalized to unity. We find that the energy of the ortho-state remains unchanged
while the energy of the para-state is increased. The
splitting between the two states is thus reduced, with
\begin{align}
  \Delta E_\text{hfs} = 
  -\frac{\alpha_\text{em}^3 m_\ell^3}{8\pi}
    \frac{y^4}{64\pi^2 m_\chi^2}
    \frac{\mu^2 - 2 \mu \log\mu - 1}{(\mu - 1)^3} \,.
  \label{eq:DeltaEhfs}
\end{align}
For positronium, this implies
\begin{align}
  \Delta E_\text{hfs}^{e^+ e^-} = 
  -0.17\ \text{Hz} \times y^4 \bigg( \frac{100\ \text{GeV}}{m_\chi} \bigg)^2 \,,
\end{align}
which is an $\mathcal{O}(10^{-12})$ correction to $E_\text{hfs}^{e^+ e^-}
= [203.3941 \pm 0.0016\ \text{(stat)} \pm 0.0011\ \text{(syst)}]
\times 10^{9}\ \text{Hz}$~\cite{Ishida:2013waa}, well below the experimental
precision and the precision of the SM prediction.  The reason for
the low sensitivity is that positronium is a relatively large system, whereas
the contact interaction is effective only at very short distance. The same
is true for $e^\pm \mu^\mp$ bound states.

More promising as a probe for contact interactions of the form of
eq.~\eqref{eq:Leff-dilepton}, and of new physics in the lepton sector in general,
seems to be ``true muonium'', i.e.\ a $\mu^+\mu^-$ bound state. Even though
true muonium has never been directly produced and studied in the laboratory,
precision experiments seem feasible~\cite{Brodsky:2009gx}.
For true muonium, we have
\begin{align}
  \Delta E_\text{hfs}^{\mu^+ \mu^-} = 
  -1.47\ \text{MHz} \times y^4 \bigg( \frac{100\ \text{GeV}}{m_\chi} \bigg)^2 \,,
  \label{eq:Ehfs-muonium}
\end{align}
which is only an $\mathcal{O}(10^{-7})$ correction to the leading term
$E_\text{hfs}^{\mu^+ \mu^-} \simeq 4.23 \times
10^7$~MHz~\cite{Jentschura:1997ma}.  Using \eqn\eqref{eq:relic-density} and
comparing to \eqn\eqref{eq:Ehfs-muonium}, we obtain that to exclude thermal
relic dark matter with $m_\chi = 130$~GeV, $\mu = 1.1$, $E_\text{hfs}^{\mu^+ \mu^-}$
needs to be measured with an accuracy of 0.2~MHz.

\subsection{\label{sec:dd-lfv} Lepton Flavor Violation}

Even though in the simplest versions of our toy model motivated by supersymmetry,
\eqns\eqref{eq:lagrangian1} and \eqref{eq:lagrangian2}, DM couplings to leptons
are flavor diagonal, we now consider also the general Lagrangian
\eqn\eqref{eq:general-lagrangian} including flavor off-diagonal couplings.
We derive constraints on these couplings from searches for the rare decays
$\mu \to e \gamma$, $\tau \to e \gamma$ and $\tau \to \mu \gamma$, which
are mediated by the diagram shown in \fig\ref{fig:LFV}. Computing this diagram,
we obtain for the decay rate
\begin{align}
  \Gamma_{\ell_\alpha \to \ell_\beta \gamma} &=
  \frac{\alpha_\text{em} m_\alpha^3}{1024 \pi^4 m_\chi^2} \big( |c_L|^2 + |c_R|^2 \big)
  \label{eq:LFV-rate}
\end{align}
where
\begin{align}
  c_L &\equiv \sum_j y_L^{\alpha j} y_R^{\beta j *} J(\mu_j) 
              + \frac{m_\alpha}{m_\chi} \sum_j y_R^{\alpha j} y_R^{\beta j *} I(\mu_j) \,, \\
  c_R &\equiv \sum_j y_R^{\alpha j} y_L^{\beta j *} J(\mu_j)
              + \frac{m_\alpha}{m_\chi} \sum_j y_L^{\alpha j} y_L^{\beta j *} I(\mu_j)
\end{align}
are Wilson coefficients in the effective Lagrangian
\begin{align}
  \mathscr{L}_{\mu\to e\gamma}  &\equiv  \frac{e}{32\pi^2 m_\chi} \Big[
      c_L \bar\ell_\beta \sigma^{\mu\nu} P_L \ell_\alpha
    + c_R \bar\ell_\beta \sigma^{\mu\nu} P_R \ell_\alpha
  \Big] F_{\mu\nu} \,,
  \label{eq:Leff-meg}
\end{align}
and the loop functions $J(\mu)$, $I(\mu)$ are given by
\begin{align}
  J(\mu) &\equiv \frac{\mu^2 - 2 \mu \log\mu - 1}{2(\mu - 1)^3} \,, \\
  I(\mu) &\equiv \frac{\mu^3 - 6 \mu^2 + 3\mu + 6\mu \log\mu + 2}{12(\mu - 1)^4} \,.
\end{align}
We have used the definition $\mu_j \equiv m_{\eta_j}^2 / m_\chi^2$, where
$m_{\eta_j}$ are the masses of the charged mediators (see
\eqn\eqref{eq:general-lagrangian}).

We consider for illustrative purposes the special case where only three
charged mediator $\eta_1$, $\eta_2$, $\eta_3$ exist, and where $y_L = 0$.
This can be realized in supersymmetry if all left-handed sleptons are too
heavy to be phenomenologically relevant.  We obtain in this special
case for the branching ratios $\BR_{\ell_\alpha \to \ell_\beta \gamma}
\simeq \Gamma_{\ell_\alpha \to \ell_\beta \gamma} / \Gamma_\text{SM}$
(with the SM width $\Gamma_\text{SM}$)
\begin{align}
  \BR_{\mu \to e \gamma} &\simeq 0.032 \, \bigg( \frac{\text{100 GeV}}{m_\chi} \bigg)^4
    \Big[ \sum_j y_R^{\mu j} y_R^{e j *} I(\mu_j) \Big]^2 \,, \\
  \BR_{\tau \to \mu \gamma} &\simeq 0.0057 \, \bigg( \frac{\text{100 GeV}}{m_\chi} \bigg)^4
    \Big[ \sum_j y_R^{\tau j} y_R^{\mu j *} I(\mu_j) \Big]^2 \,.
\end{align}
The expression for $\BR_{\tau \to e \gamma}$ is identical to the one for
$\BR_{\tau \to \mu \gamma}$, with the replacements $y_R^{\mu j *} \to y_R^{e j
*}$.  With the current experimental limits $\BR_{\mu \to e \gamma} < 5.7 \times
10^{-13}$~\cite{Adam:2013mnn}, $\BR_{\tau \to \mu \gamma} < 4.4 \times
10^{-8}$~\cite{Aubert:2009ag} and $\BR_{\tau \to e \gamma} < 3.3 \times
10^{-8}$~\cite{Aubert:2009ag}, and using $m_\chi = 100$~GeV, we then obtain the
following constraints on the elements of $y_R$ at $\mu = 1.1$: \\
\begin{center}
  \begin{ruledtabular}
  \begin{tabular}{lcl}
    Process &  Coupling & Limit \\ \hline
    $\mu \to e \gamma$    & $\big[\sum_j (y_R^{\mu j}  y_R^{e j *})^2\big]^{1/2}$
      & $< 1.0 \times 10^{-4}$   \\
    $\tau \to \mu \gamma$ & $\big[\sum_j (y_R^{\tau j} y_R^{\mu j *})^2\big]^{1/2}$
      & $<$ 7.0 $ 10^{-2}$ \\
    $\tau \to e \gamma$   & $\big[\sum_j (y_R^{\tau j} y_R^{e j *})^2\big]^{1/2}$
      & $<$ 6.1 $ 10^{-2} $\\
  \end{tabular}
  \end{ruledtabular}
\end{center}
We have seen in \eqn\eqref{eq:relic-density} that in our simplified model setup,
at least one of the Yukawa couplings should be of order 0.1--1 to avoid DM
overproduction.  The above constraints show that flavor off-diagonal Yukawa
couplings are therefore always subdominant. This justifies our neglecting them
in the preceding sections.

We have also studied the decay $\mu \to 3e$, which constrains a different
combination of Yukawa couplings because it also receives contributions
from box diagrams similar to \fig\ref{fig:box}.  If we assume that flavor-diagonal
Yukawa couplings are $\mathcal{O}(1)$, we obtain limits on the flavor off-diagonal
couplings that are about a factor of 8 weaker than the limit from
$\mu \to e\gamma$. To arrive at this estimate, we have used 
\reference~\cite{Kuno:1999jp} to express $\BR(\mu \to 3e)$ in terms of the
Wilson coefficients of the effective operators in \eqns\eqref{eq:Leff-meg}
and \eqref{eq:Leff-dilepton}. We have then compared the predicted branching
ratio to the current experimental limit from~\cite{Bellgardt:1987du, Beringer:2012zz}.
Note that planned searches for $\mu \to 3e$ will improve the limit on
$\BR(\mu \to 3e)$ by up to four orders of magnitude~\cite{Blondel:2013ia}.

\section{\label{sec:conclusions}Conclusions}

In this paper, we have studied leptophilic dark matter models in which DM
annihilation proceeds through a charged mediator and can therefore be
accompanied by emission of a virtual internal bremsstrahlung photon.  Such
models are of great interest for indirect dark matter searches because internal
bremsstrahlung can lead to spectral peaks in the gamma ray sky, a feature which
is easily distinguishable from the large astrophysical gamma ray flux.
Leptophilic DM models are also well motivated theoretically: they
can be realized for instance in supersymmetric scenarios or radiative neutrino
mass models, and in most cases, their parameter space is relatively
unconstrained.

Here, we have established a connection between internal bremsstrahlung signals
and loop-induced electromagnetic form factors of DM particles in leptophilic
models.  In particular, upon connecting the charged lepton lines in the
internal bremsstrahlung diagrams in \fig\ref{fig:vib-diagrams} to a loop, one
immediately obtains the electromagnetic vertex corrections in
\fig\ref{fig:em-moment-loops}.  For Majorana DM, these lead to an anapole
moment, while for Dirac DM, both anapole and magnetic dipole moments are generated,
with the dipole moment being dominant in DM scattering processes.
Interactions of the anapole and dipole moments with atomic nuclei then allow us
to constrain the internal bremsstrahlung cross section using DM--nucleus
scattering data from direct detection experiments. We have carried out this
analysis for the most recent LUX and XENON100 data, and have found that direct
detection constraints can be competitive with internal bremsstrahlung
searches. This is true in particular if the mass splitting between the DM
particle $\chi$ and the charged mediator $\eta$ is very small---the case which is
also most interesting for internal bremsstrahlung searches due to the peaked
gamma ray spectrum.

If DM is a Majorana fermion that couples universally to all charged leptons,
direct detection limits are of the same order as limits from gamma ray line
searches, and better than continuum gamma ray constraints from dwarf galaxies
(see \fig\ref{fig:limits-sigmav} (b)).  Specifically, for small mass
splitting $m_\eta^2 / m_\chi^2 \lesssim 1.1$, LUX constrains the internal
bremsstrahlung cross section $\ev{\sigma v_\text{rel}}_{\chi\chi \to \ell \bar{\ell} \gamma}$
to be below $\text{few} \times 10^{-28}$~cm$^3$/s at $m_\chi \sim 20$~GeV.
At DM masses of order 100~GeV, which have been invoked previously to explain
a bump in Fermi-LAT gamma ray data~\cite{Bringmann:2012vr}, LUX
constraints imply that this interpretation is disfavored if DM couples
to electrons or muons and if $m_\eta$ and $m_\chi$ differ by $\lesssim \text{few \%}$.
If the last condition is significantly violated, however, the expected bump in
the gamma ray spectrum becomes relatively broad, making line searches less
sensitive.  If $m_\eta / m_\chi \gg 1$, also direct searches for the charged
mediator $\eta$ at colliders will impose important constraints, disfavoring
$m_\eta \lesssim \text{few} \times 100$~GeV~\cite{Bringmann:2012vr,Liu:2013gba}.
These constraints are ineffective if $m_\eta \sim m_\chi$ because the leptons
from $\eta$ decay will be very soft in this case and thus hard to detect.

We note an interesting connection between our results and the scenario studied
by Konishi et al.~\cite{Konishi:2013gda} to solve the cosmological lithium-7
problem in the Constrained Minimal Supersymmetric Standard Model (CMSSM) with
sleptons that are nearly mass degenerate with the lightest neutralino. For the
preferred mass range from~\cite{Konishi:2013gda}, $\text{300~GeV} \lesssim m_\chi
\lesssim \text{500~GeV}$, this scenario would predict
$\ev{\sigma v_\text{rel}}_{\chi\chi \to \ell \bar{\ell} \gamma} \sim
10^{-28}$~cm$^3$/s, well within the region testable by next generation
direct detection experiments.

If DM is a Dirac fermion and the masses of $\chi$ and $\eta$ are of the same
order of magnitude, but still differ by $\gtrsim 10\%$ so that coannihilations
are not yet relevant), direct detection constraints disfavor thermal relic
production of DM for $m_\chi$ between 10--20~GeV and up to a few hundred GeV
(see \fig\ref{fig:limits-sigmav} (d)).  For $m_\chi > 20$~GeV, direct detection
limits are also significantly stronger than astrophysical limits from gamma ray
line searches and from continuum gamma rays searches in dwarf galaxies.

In the future, we expect the XENON1T and LUX-ZEPLIN experiments to improve
these direct detection limits by about two orders of magnitude. These
experiments will thus test the thermal relic hypothesis for DM masses of order
$\text{10~GeV} \lesssim m_\chi \lesssim \text{few} \times \text{100~GeV}$,
except for scenarios with a per cent level degeneracy between $m_\chi$ and
$m_\eta$, where coannihilations dominate in the early Universe. If a signal is
detected, the spectrum of recoil events can be used to discriminate between
anapole and dipole interaction and hence between Majorana and Dirac DM.

We have also studied constraints on our simplified model from low energy
precision experiments. We confirm that bounds from the anomalous magnetic
moment $g-2$ of the electron and the muon are weaker than the direct detection
constraints at $m_\chi \gtrsim 10$~GeV.  Searches for the lepton flavor
violating decays $\tau \to \mu\gamma$, $\tau \to e\gamma$, $\mu \to e\gamma$
and $\mu \to 3e$
are very powerful in setting bounds on DM annihilation into flavor violating
final states.  Finally, we have studied the possibility of obtaining
constraints from a future measurement of the hyperfine splitting in true
muonium (a $\mu^+\mu^-$ bound state).  We have found such a measurement to be
challenging for heavy DM ($m_\chi \sim 100$~GeV), where excluding
thermal relic DM would require a measurement with a relative accuracy better than
$10^{-7}$ (see \eqn\eqref{eq:Ehfs-muonium}).  For lighter DM ($m_\chi \lesssim
10$~GeV), however, requirements are weaker and an interesting measurement may
be possible.

In summary, our results show that direct dark matter searches are powerful
tools to search for leptophilic DM even though DM--nucleus scattering occurs
only at the loop level in this case.  They are complementary to, and sometimes
significantly superior to, indirect searches and precision experiments.
Particularly in a scenario where a peak is observed in the cosmic gamma ray
spectrum, but no other indirect hints for DM are found, virtual internal
bremsstrahlung in a leptophilic DM model provides an attractive explanation.
Our results show how this scenario can be confirmed in direct detection
experiments by looking for the electromagnetic moment interactions of DM with
nuclei.  This illustrates once again that the search for Dark Matter is an
interdisciplinary task, and that only a combination of different search
strategies can yield optimal results.

\section*{Acknowledgments}

We would like to thank E.~del~Nobile, B.~Kayser, T.~Marrod\'{a}n Undagoitia, H.~Patel,
T.~Plehn, P.~Schichtel, D. Schmeier, J. Tattersall, and C.~Weniger
for very helpful discussions.  It is also a pleasure to thank Jonathan Schuster
for his unusual but creative contributions to this work.  JS acknowledges
support from the IMPRS for Precision Tests of Fundamental Symmetries. JK would
like to thank the Aspen Center for Physics (supported by NSF grant 1066293) for
kind hospitality during part of this work.


\begin{thebibliography}{124}
\expandafter\ifx\csname natexlab\endcsname\relax\def\natexlab#1{#1}\fi
\expandafter\ifx\csname bibnamefont\endcsname\relax
  \def\bibnamefont#1{#1}\fi
\expandafter\ifx\csname bibfnamefont\endcsname\relax
  \def\bibfnamefont#1{#1}\fi
\expandafter\ifx\csname citenamefont\endcsname\relax
  \def\citenamefont#1{#1}\fi
\expandafter\ifx\csname url\endcsname\relax
  \def\url#1{\texttt{#1}}\fi
\expandafter\ifx\csname urlprefix\endcsname\relax\def\urlprefix{URL }\fi
\providecommand{\bibinfo}[2]{#2}
\providecommand{\eprint}[2][]{\url{#2}}

\bibitem[{\citenamefont{Aharonian et~al.}(2012)\citenamefont{Aharonian,
  Khangulyan, and Malyshev}}]{Aharonian:2012cs}
\bibinfo{author}{\bibfnamefont{F.}~\bibnamefont{Aharonian}},
  \bibinfo{author}{\bibfnamefont{D.}~\bibnamefont{Khangulyan}},
  \bibnamefont{and} \bibinfo{author}{\bibfnamefont{D.}~\bibnamefont{Malyshev}}
  (\bibinfo{year}{2012}), \eprint{1207.0458}.

\bibitem[{\citenamefont{Bringmann et~al.}(2008)\citenamefont{Bringmann,
  Bergstrom, and Edsjo}}]{Bringmann:2007nk}
\bibinfo{author}{\bibfnamefont{T.}~\bibnamefont{Bringmann}},
  \bibinfo{author}{\bibfnamefont{L.}~\bibnamefont{Bergstrom}},
  \bibnamefont{and} \bibinfo{author}{\bibfnamefont{J.}~\bibnamefont{Edsjo}},
  \bibinfo{journal}{JHEP} \textbf{\bibinfo{volume}{0801}}, \bibinfo{pages}{049}
  (\bibinfo{year}{2008}), \eprint{0710.3169}.

\bibitem[{\citenamefont{Bell et~al.}(2011)\citenamefont{Bell, Dent, Jacques,
  and Weiler}}]{Bell:2010ei}
\bibinfo{author}{\bibfnamefont{N.~F.} \bibnamefont{Bell}},
  \bibinfo{author}{\bibfnamefont{J.~B.} \bibnamefont{Dent}},
  \bibinfo{author}{\bibfnamefont{T.~D.} \bibnamefont{Jacques}},
  \bibnamefont{and} \bibinfo{author}{\bibfnamefont{T.~J.}
  \bibnamefont{Weiler}}, \bibinfo{journal}{Phys.Rev.}
  \textbf{\bibinfo{volume}{D83}}, \bibinfo{pages}{013001}
  (\bibinfo{year}{2011}), \eprint{1009.2584}.

\bibitem[{\citenamefont{Bringmann et~al.}(2012)\citenamefont{Bringmann, Huang,
  Ibarra, Vogl, and Weniger}}]{Bringmann:2012vr}
\bibinfo{author}{\bibfnamefont{T.}~\bibnamefont{Bringmann}},
  \bibinfo{author}{\bibfnamefont{X.}~\bibnamefont{Huang}},
  \bibinfo{author}{\bibfnamefont{A.}~\bibnamefont{Ibarra}},
  \bibinfo{author}{\bibfnamefont{S.}~\bibnamefont{Vogl}}, \bibnamefont{and}
  \bibinfo{author}{\bibfnamefont{C.}~\bibnamefont{Weniger}}
  (\bibinfo{year}{2012}), \eprint{1203.1312}.

\bibitem[{\citenamefont{Geringer-Sameth and
  Koushiappas}(2011)}]{GeringerSameth:2011iw}
\bibinfo{author}{\bibfnamefont{A.}~\bibnamefont{Geringer-Sameth}}
  \bibnamefont{and} \bibinfo{author}{\bibfnamefont{S.~M.}
  \bibnamefont{Koushiappas}}, \bibinfo{journal}{Phys.Rev.Lett.}
  \textbf{\bibinfo{volume}{107}}, \bibinfo{pages}{241303}
  (\bibinfo{year}{2011}), \eprint{1108.2914}.

\bibitem[{\citenamefont{Ackermann et~al.}(2013)}]{Ackermann:2013yva}
\bibinfo{author}{\bibfnamefont{M.}~\bibnamefont{Ackermann}}
  \bibnamefont{et~al.} (\bibinfo{collaboration}{Fermi-LAT Collaboration})
  (\bibinfo{year}{2013}), \eprint{1310.0828}.

\bibitem[{\citenamefont{Hisano et~al.}(2011)\citenamefont{Hisano, Ishiwata, and
  Nagata}}]{Hisano:2011um}
\bibinfo{author}{\bibfnamefont{J.}~\bibnamefont{Hisano}},
  \bibinfo{author}{\bibfnamefont{K.}~\bibnamefont{Ishiwata}}, \bibnamefont{and}
  \bibinfo{author}{\bibfnamefont{N.}~\bibnamefont{Nagata}},
  \bibinfo{journal}{Phys.Lett.} \textbf{\bibinfo{volume}{B706}},
  \bibinfo{pages}{208} (\bibinfo{year}{2011}), \eprint{1110.3719}.

\bibitem[{\citenamefont{Garny et~al.}(2012)\citenamefont{Garny, Ibarra, Pato,
  and Vogl}}]{Garny:2012eb}
\bibinfo{author}{\bibfnamefont{M.}~\bibnamefont{Garny}},
  \bibinfo{author}{\bibfnamefont{A.}~\bibnamefont{Ibarra}},
  \bibinfo{author}{\bibfnamefont{M.}~\bibnamefont{Pato}}, \bibnamefont{and}
  \bibinfo{author}{\bibfnamefont{S.}~\bibnamefont{Vogl}}
  (\bibinfo{year}{2012}), \eprint{1207.1431}.

\bibitem[{\citenamefont{Garny et~al.}(2013)\citenamefont{Garny, Ibarra, Pato,
  and Vogl}}]{Garny:2013ama}
\bibinfo{author}{\bibfnamefont{M.}~\bibnamefont{Garny}},
  \bibinfo{author}{\bibfnamefont{A.}~\bibnamefont{Ibarra}},
  \bibinfo{author}{\bibfnamefont{M.}~\bibnamefont{Pato}}, \bibnamefont{and}
  \bibinfo{author}{\bibfnamefont{S.}~\bibnamefont{Vogl}}
  (\bibinfo{year}{2013}), \eprint{1306.6342}.

\bibitem[{\citenamefont{Goodman et~al.}(2011)\citenamefont{Goodman, Ibe,
  Rajaraman, Shepherd, Tait et~al.}}]{Goodman:2010yf}
\bibinfo{author}{\bibfnamefont{J.}~\bibnamefont{Goodman}},
  \bibinfo{author}{\bibfnamefont{M.}~\bibnamefont{Ibe}},
  \bibinfo{author}{\bibfnamefont{A.}~\bibnamefont{Rajaraman}},
  \bibinfo{author}{\bibfnamefont{W.}~\bibnamefont{Shepherd}},
  \bibinfo{author}{\bibfnamefont{T.~M.} \bibnamefont{Tait}},
  \bibnamefont{et~al.}, \bibinfo{journal}{Phys.Lett.}
  \textbf{\bibinfo{volume}{B695}}, \bibinfo{pages}{185} (\bibinfo{year}{2011}),
  \eprint{1005.1286}.

\bibitem[{\citenamefont{Bai et~al.}(2010)\citenamefont{Bai, Fox, and
  Harnik}}]{Bai:2010hh}
\bibinfo{author}{\bibfnamefont{Y.}~\bibnamefont{Bai}},
  \bibinfo{author}{\bibfnamefont{P.~J.} \bibnamefont{Fox}}, \bibnamefont{and}
  \bibinfo{author}{\bibfnamefont{R.}~\bibnamefont{Harnik}},
  \bibinfo{journal}{JHEP} \textbf{\bibinfo{volume}{1012}}, \bibinfo{pages}{048}
  (\bibinfo{year}{2010}), \eprint{1005.3797}.

\bibitem[{\citenamefont{Fox et~al.}(2012)\citenamefont{Fox, Harnik, Kopp, and
  Tsai}}]{Fox:2011pm}
\bibinfo{author}{\bibfnamefont{P.~J.} \bibnamefont{Fox}},
  \bibinfo{author}{\bibfnamefont{R.}~\bibnamefont{Harnik}},
  \bibinfo{author}{\bibfnamefont{J.}~\bibnamefont{Kopp}}, \bibnamefont{and}
  \bibinfo{author}{\bibfnamefont{Y.}~\bibnamefont{Tsai}},
  \bibinfo{journal}{Phys.Rev.} \textbf{\bibinfo{volume}{D85}},
  \bibinfo{pages}{056011} (\bibinfo{year}{2012}), \eprint{1109.4398}.

\bibitem[{\citenamefont{Lin et~al.}(2013)\citenamefont{Lin, Kolb, and
  Wang}}]{Lin:2013sca}
\bibinfo{author}{\bibfnamefont{T.}~\bibnamefont{Lin}},
  \bibinfo{author}{\bibfnamefont{E.~W.} \bibnamefont{Kolb}}, \bibnamefont{and}
  \bibinfo{author}{\bibfnamefont{L.-T.} \bibnamefont{Wang}}
  (\bibinfo{year}{2013}), \eprint{1303.6638}.

\bibitem[{\citenamefont{{The CMS collaboration}}(2013)}]{CMS:rwa}
\bibinfo{author}{\bibnamefont{{The CMS collaboration}}} (\bibinfo{year}{2013}),
  \bibinfo{note}{{CMS-PAS-EXO-12-048}}.

\bibitem[{\citenamefont{{The ATLAS collaboration}}(2012)}]{ATLAS:2012zim}
\bibinfo{author}{\bibnamefont{{The ATLAS collaboration}}}
  (\bibinfo{year}{2012}), \bibinfo{note}{{ATLAS-CONF-2012-147,
  ATLAS-COM-CONF-2012-190}}.

\bibitem[{\citenamefont{Aad et~al.}(2013)}]{Aad:2013oja}
\bibinfo{author}{\bibfnamefont{G.}~\bibnamefont{Aad}} \bibnamefont{et~al.}
  (\bibinfo{collaboration}{ATLAS Collaboration}) (\bibinfo{year}{2013}),
  \eprint{1309.4017}.

\bibitem[{\citenamefont{Boyd}(2013)}]{Boyd:2013}
\bibinfo{author}{\bibfnamefont{J.}~\bibnamefont{Boyd}}
  (\bibinfo{collaboration}{ATLAS}), \emph{\bibinfo{title}{{Overview of SUSY
  results from the ATLAS experiment}}} (\bibinfo{year}{2013}),
  \bibinfo{note}{{talk given at the SUSY 2013 conference, slides available from
  http://susy2013.ictp.it}}.

\bibitem[{\citenamefont{Richman}(2013)}]{Richman:2013}
\bibinfo{author}{\bibfnamefont{J.~D.} \bibnamefont{Richman}}
  (\bibinfo{collaboration}{CMS}), \emph{\bibinfo{title}{{Searches for
  Supersymmetry in the CMS Experiment}}} (\bibinfo{year}{2013}),
  \bibinfo{note}{{talk given at the SUSY 2013 conference, slides available from
  http://susy2013.ictp.it}}.

\bibitem[{\citenamefont{Adriani et~al.}(2009)}]{Adriani:2008zr}
\bibinfo{author}{\bibfnamefont{O.}~\bibnamefont{Adriani}} \bibnamefont{et~al.}
  (\bibinfo{collaboration}{PAMELA Collaboration}), \bibinfo{journal}{Nature}
  \textbf{\bibinfo{volume}{458}}, \bibinfo{pages}{607} (\bibinfo{year}{2009}),
  \eprint{0810.4995}.

\bibitem[{\citenamefont{Adriani et~al.}(2013)}]{Adriani:2013uda}
\bibinfo{author}{\bibfnamefont{O.}~\bibnamefont{Adriani}} \bibnamefont{et~al.}
  (\bibinfo{collaboration}{PAMELA Collaboration}) (\bibinfo{year}{2013}),
  \eprint{1308.0133}.

\bibitem[{\citenamefont{Ackermann
  et~al.}(2012{\natexlab{a}})}]{FermiLAT:2011ab}
\bibinfo{author}{\bibfnamefont{M.}~\bibnamefont{Ackermann}}
  \bibnamefont{et~al.} (\bibinfo{collaboration}{Fermi LAT Collaboration}),
  \bibinfo{journal}{Phys.Rev.Lett.} \textbf{\bibinfo{volume}{108}},
  \bibinfo{pages}{011103} (\bibinfo{year}{2012}{\natexlab{a}}),
  \eprint{1109.0521}.

\bibitem[{\citenamefont{Aguilar et~al.}(2013)}]{Aguilar:2013qda}
\bibinfo{author}{\bibfnamefont{M.}~\bibnamefont{Aguilar}} \bibnamefont{et~al.}
  (\bibinfo{collaboration}{AMS Collaboration}),
  \bibinfo{journal}{Phys.Rev.Lett.} \textbf{\bibinfo{volume}{110}},
  \bibinfo{pages}{141102} (\bibinfo{year}{2013}).

\bibitem[{\citenamefont{Cirelli et~al.}(2009)\citenamefont{Cirelli, Kadastik,
  Raidal, and Strumia}}]{Cirelli:2008pk}
\bibinfo{author}{\bibfnamefont{M.}~\bibnamefont{Cirelli}},
  \bibinfo{author}{\bibfnamefont{M.}~\bibnamefont{Kadastik}},
  \bibinfo{author}{\bibfnamefont{M.}~\bibnamefont{Raidal}}, \bibnamefont{and}
  \bibinfo{author}{\bibfnamefont{A.}~\bibnamefont{Strumia}},
  \bibinfo{journal}{Nucl.Phys.} \textbf{\bibinfo{volume}{B813}},
  \bibinfo{pages}{1} (\bibinfo{year}{2009}), \eprint{0809.2409}.

\bibitem[{\citenamefont{Donato et~al.}(2009)\citenamefont{Donato, Maurin, Brun,
  Delahaye, and Salati}}]{Donato:2008jk}
\bibinfo{author}{\bibfnamefont{F.}~\bibnamefont{Donato}},
  \bibinfo{author}{\bibfnamefont{D.}~\bibnamefont{Maurin}},
  \bibinfo{author}{\bibfnamefont{P.}~\bibnamefont{Brun}},
  \bibinfo{author}{\bibfnamefont{T.}~\bibnamefont{Delahaye}}, \bibnamefont{and}
  \bibinfo{author}{\bibfnamefont{P.}~\bibnamefont{Salati}},
  \bibinfo{journal}{Phys.Rev.Lett.} \textbf{\bibinfo{volume}{102}},
  \bibinfo{pages}{071301} (\bibinfo{year}{2009}), \eprint{0810.5292}.

\bibitem[{\citenamefont{Nardi et~al.}(2009)\citenamefont{Nardi, Sannino, and
  Strumia}}]{Nardi:2008ix}
\bibinfo{author}{\bibfnamefont{E.}~\bibnamefont{Nardi}},
  \bibinfo{author}{\bibfnamefont{F.}~\bibnamefont{Sannino}}, \bibnamefont{and}
  \bibinfo{author}{\bibfnamefont{A.}~\bibnamefont{Strumia}},
  \bibinfo{journal}{JCAP} \textbf{\bibinfo{volume}{0901}}, \bibinfo{pages}{043}
  (\bibinfo{year}{2009}), \eprint{0811.4153}.

\bibitem[{\citenamefont{Bertone et~al.}(2009)\citenamefont{Bertone, Cirelli,
  Strumia, and Taoso}}]{Bertone:2008xr}
\bibinfo{author}{\bibfnamefont{G.}~\bibnamefont{Bertone}},
  \bibinfo{author}{\bibfnamefont{M.}~\bibnamefont{Cirelli}},
  \bibinfo{author}{\bibfnamefont{A.}~\bibnamefont{Strumia}}, \bibnamefont{and}
  \bibinfo{author}{\bibfnamefont{M.}~\bibnamefont{Taoso}},
  \bibinfo{journal}{JCAP} \textbf{\bibinfo{volume}{0903}}, \bibinfo{pages}{009}
  (\bibinfo{year}{2009}), \eprint{0811.3744}.

\bibitem[{\citenamefont{Fox and Poppitz}(2009)}]{Fox:2008kb}
\bibinfo{author}{\bibfnamefont{P.~J.} \bibnamefont{Fox}} \bibnamefont{and}
  \bibinfo{author}{\bibfnamefont{E.}~\bibnamefont{Poppitz}},
  \bibinfo{journal}{Phys.Rev.} \textbf{\bibinfo{volume}{D79}},
  \bibinfo{pages}{083528} (\bibinfo{year}{2009}), \eprint{0811.0399}.

\bibitem[{\citenamefont{Evoli et~al.}(2011)\citenamefont{Evoli, Cholis, Grasso,
  Maccione, and Ullio}}]{Evoli:2011id}
\bibinfo{author}{\bibfnamefont{C.}~\bibnamefont{Evoli}},
  \bibinfo{author}{\bibfnamefont{I.}~\bibnamefont{Cholis}},
  \bibinfo{author}{\bibfnamefont{D.}~\bibnamefont{Grasso}},
  \bibinfo{author}{\bibfnamefont{L.}~\bibnamefont{Maccione}}, \bibnamefont{and}
  \bibinfo{author}{\bibfnamefont{P.}~\bibnamefont{Ullio}}
  (\bibinfo{year}{2011}), \eprint{1108.0664}.

\bibitem[{\citenamefont{Kopp}(2013)}]{Kopp:2013eka}
\bibinfo{author}{\bibfnamefont{J.}~\bibnamefont{Kopp}} (\bibinfo{year}{2013}),
  \eprint{1304.1184}.

\bibitem[{\citenamefont{Cholis and Hooper}(2013)}]{Cholis:2013psa}
\bibinfo{author}{\bibfnamefont{I.}~\bibnamefont{Cholis}} \bibnamefont{and}
  \bibinfo{author}{\bibfnamefont{D.}~\bibnamefont{Hooper}}
  (\bibinfo{year}{2013}), \eprint{1304.1840}.

\bibitem[{\citenamefont{Bergstrom et~al.}(2013)\citenamefont{Bergstrom,
  Bringmann, Cholis, Hooper, and Weniger}}]{Bergstrom:2013jra}
\bibinfo{author}{\bibfnamefont{L.}~\bibnamefont{Bergstrom}},
  \bibinfo{author}{\bibfnamefont{T.}~\bibnamefont{Bringmann}},
  \bibinfo{author}{\bibfnamefont{I.}~\bibnamefont{Cholis}},
  \bibinfo{author}{\bibfnamefont{D.}~\bibnamefont{Hooper}}, \bibnamefont{and}
  \bibinfo{author}{\bibfnamefont{C.}~\bibnamefont{Weniger}},
  \bibinfo{journal}{Phys.Rev.Lett.} \textbf{\bibinfo{volume}{111}},
  \bibinfo{pages}{171101} (\bibinfo{year}{2013}), \eprint{1306.3983}.

\bibitem[{\citenamefont{Ibarra et~al.}(2013)\citenamefont{Ibarra,
  Lamperstorfer, and Silk}}]{Ibarra:2013zia}
\bibinfo{author}{\bibfnamefont{A.}~\bibnamefont{Ibarra}},
  \bibinfo{author}{\bibfnamefont{A.~S.} \bibnamefont{Lamperstorfer}},
  \bibnamefont{and} \bibinfo{author}{\bibfnamefont{J.}~\bibnamefont{Silk}}
  (\bibinfo{year}{2013}), \eprint{1309.2570}.

\bibitem[{\citenamefont{Adriani et~al.}(2010)}]{Adriani:2010rc}
\bibinfo{author}{\bibfnamefont{O.}~\bibnamefont{Adriani}} \bibnamefont{et~al.}
  (\bibinfo{collaboration}{PAMELA Collaboration}),
  \bibinfo{journal}{Phys.Rev.Lett.} \textbf{\bibinfo{volume}{105}},
  \bibinfo{pages}{121101} (\bibinfo{year}{2010}), \eprint{1007.0821}.

\bibitem[{\citenamefont{Bartoli et~al.}(2012)}]{Bartoli:2012qe}
\bibinfo{author}{\bibfnamefont{B.}~\bibnamefont{Bartoli}} \bibnamefont{et~al.}
  (\bibinfo{collaboration}{ARGO-YBJ Collaboration}),
  \bibinfo{journal}{Phys.Rev.} \textbf{\bibinfo{volume}{D85}},
  \bibinfo{pages}{022002} (\bibinfo{year}{2012}), \eprint{1201.3848}.

\bibitem[{\citenamefont{Hooper and Goodenough}(2011)}]{Hooper:2010mq}
\bibinfo{author}{\bibfnamefont{D.}~\bibnamefont{Hooper}} \bibnamefont{and}
  \bibinfo{author}{\bibfnamefont{L.}~\bibnamefont{Goodenough}},
  \bibinfo{journal}{Phys.Lett.} \textbf{\bibinfo{volume}{B697}},
  \bibinfo{pages}{412} (\bibinfo{year}{2011}), \eprint{1010.2752}.

\bibitem[{\citenamefont{Hooper and Linden}(2011)}]{Hooper:2011ti}
\bibinfo{author}{\bibfnamefont{D.}~\bibnamefont{Hooper}} \bibnamefont{and}
  \bibinfo{author}{\bibfnamefont{T.}~\bibnamefont{Linden}},
  \bibinfo{journal}{Phys.Rev.} \textbf{\bibinfo{volume}{D84}},
  \bibinfo{pages}{123005} (\bibinfo{year}{2011}), \eprint{1110.0006}.

\bibitem[{\citenamefont{Hooper}(2012)}]{Hooper:2012ft}
\bibinfo{author}{\bibfnamefont{D.}~\bibnamefont{Hooper}}
  (\bibinfo{year}{2012}), \eprint{1201.1303}.

\bibitem[{\citenamefont{Su et~al.}(2010)\citenamefont{Su, Slatyer, and
  Finkbeiner}}]{Su:2010qj}
\bibinfo{author}{\bibfnamefont{M.}~\bibnamefont{Su}},
  \bibinfo{author}{\bibfnamefont{T.~R.} \bibnamefont{Slatyer}},
  \bibnamefont{and} \bibinfo{author}{\bibfnamefont{D.~P.}
  \bibnamefont{Finkbeiner}}, \bibinfo{journal}{Astrophys.J.}
  \textbf{\bibinfo{volume}{724}}, \bibinfo{pages}{1044} (\bibinfo{year}{2010}),
  \eprint{1005.5480}.

\bibitem[{\citenamefont{Hooper and Slatyer}(2013)}]{Hooper:2013rwa}
\bibinfo{author}{\bibfnamefont{D.}~\bibnamefont{Hooper}} \bibnamefont{and}
  \bibinfo{author}{\bibfnamefont{T.~R.} \bibnamefont{Slatyer}}
  (\bibinfo{year}{2013}), \eprint{1302.6589}.

\bibitem[{\citenamefont{Huang et~al.}(2013{\natexlab{a}})\citenamefont{Huang,
  Urbano, and Xue}}]{Huang:2013pda}
\bibinfo{author}{\bibfnamefont{W.-C.} \bibnamefont{Huang}},
  \bibinfo{author}{\bibfnamefont{A.}~\bibnamefont{Urbano}}, \bibnamefont{and}
  \bibinfo{author}{\bibfnamefont{W.}~\bibnamefont{Xue}}
  (\bibinfo{year}{2013}{\natexlab{a}}), \eprint{1307.6862}.

\bibitem[{\citenamefont{Huang et~al.}(2013{\natexlab{b}})\citenamefont{Huang,
  Urbano, and Xue}}]{Huang:2013apa}
\bibinfo{author}{\bibfnamefont{W.-C.} \bibnamefont{Huang}},
  \bibinfo{author}{\bibfnamefont{A.}~\bibnamefont{Urbano}}, \bibnamefont{and}
  \bibinfo{author}{\bibfnamefont{W.}~\bibnamefont{Xue}}
  (\bibinfo{year}{2013}{\natexlab{b}}), \eprint{1310.7609}.

\bibitem[{\citenamefont{Linden et~al.}(2011)\citenamefont{Linden, Hooper, and
  Yusef-Zadeh}}]{Linden:2011au}
\bibinfo{author}{\bibfnamefont{T.}~\bibnamefont{Linden}},
  \bibinfo{author}{\bibfnamefont{D.}~\bibnamefont{Hooper}}, \bibnamefont{and}
  \bibinfo{author}{\bibfnamefont{F.}~\bibnamefont{Yusef-Zadeh}},
  \bibinfo{journal}{Astrophys.J.} \textbf{\bibinfo{volume}{741}},
  \bibinfo{pages}{95} (\bibinfo{year}{2011}), \eprint{1106.5493}.

\bibitem[{\citenamefont{Kopp et~al.}(2009)\citenamefont{Kopp, Niro, Schwetz,
  and Zupan}}]{Kopp:2009et}
\bibinfo{author}{\bibfnamefont{J.}~\bibnamefont{Kopp}},
  \bibinfo{author}{\bibfnamefont{V.}~\bibnamefont{Niro}},
  \bibinfo{author}{\bibfnamefont{T.}~\bibnamefont{Schwetz}}, \bibnamefont{and}
  \bibinfo{author}{\bibfnamefont{J.}~\bibnamefont{Zupan}},
  \bibinfo{journal}{Phys. Rev.} \textbf{\bibinfo{volume}{D80}},
  \bibinfo{pages}{083502} (\bibinfo{year}{2009}), \eprint{0907.3159}.

\bibitem[{\citenamefont{Essig et~al.}(2012{\natexlab{a}})\citenamefont{Essig,
  Mardon, and Volansky}}]{Essig:2011nj}
\bibinfo{author}{\bibfnamefont{R.}~\bibnamefont{Essig}},
  \bibinfo{author}{\bibfnamefont{J.}~\bibnamefont{Mardon}}, \bibnamefont{and}
  \bibinfo{author}{\bibfnamefont{T.}~\bibnamefont{Volansky}},
  \bibinfo{journal}{Phys.Rev.} \textbf{\bibinfo{volume}{D85}},
  \bibinfo{pages}{076007} (\bibinfo{year}{2012}{\natexlab{a}}),
  \eprint{1108.5383}.

\bibitem[{\citenamefont{Essig et~al.}(2012{\natexlab{b}})\citenamefont{Essig,
  Manalaysay, Mardon, Sorensen, and Volansky}}]{Essig:2012yx}
\bibinfo{author}{\bibfnamefont{R.}~\bibnamefont{Essig}},
  \bibinfo{author}{\bibfnamefont{A.}~\bibnamefont{Manalaysay}},
  \bibinfo{author}{\bibfnamefont{J.}~\bibnamefont{Mardon}},
  \bibinfo{author}{\bibfnamefont{P.}~\bibnamefont{Sorensen}}, \bibnamefont{and}
  \bibinfo{author}{\bibfnamefont{T.}~\bibnamefont{Volansky}}
  (\bibinfo{year}{2012}{\natexlab{b}}), \eprint{1206.2644}.

\bibitem[{\citenamefont{Schmidt et~al.}(2012)\citenamefont{Schmidt, Schwetz,
  and Toma}}]{Schmidt:2012yg}
\bibinfo{author}{\bibfnamefont{D.}~\bibnamefont{Schmidt}},
  \bibinfo{author}{\bibfnamefont{T.}~\bibnamefont{Schwetz}}, \bibnamefont{and}
  \bibinfo{author}{\bibfnamefont{T.}~\bibnamefont{Toma}},
  \bibinfo{journal}{Phys.Rev.} \textbf{\bibinfo{volume}{D85}},
  \bibinfo{pages}{073009} (\bibinfo{year}{2012}), \eprint{1201.0906}.

\bibitem[{\citenamefont{Ackermann
  et~al.}(2012{\natexlab{b}})}]{Ackermann:2012qk}
\bibinfo{author}{\bibfnamefont{M.}~\bibnamefont{Ackermann}}
  \bibnamefont{et~al.} (\bibinfo{collaboration}{LAT Collaboration})
  (\bibinfo{year}{2012}{\natexlab{b}}), \eprint{1205.2739}.

\bibitem[{\citenamefont{{The Fermi-LAT
  collaboration}}(2013)}]{Fermi-LAT:2013uma}
\bibinfo{author}{\bibnamefont{{The Fermi-LAT collaboration}}}
  (\bibinfo{year}{2013}), \eprint{1305.5597}.

\bibitem[{\citenamefont{Boyarsky et~al.}(2012)\citenamefont{Boyarsky, Malyshev,
  and Ruchayskiy}}]{Boyarsky:2012ca}
\bibinfo{author}{\bibfnamefont{A.}~\bibnamefont{Boyarsky}},
  \bibinfo{author}{\bibfnamefont{D.}~\bibnamefont{Malyshev}}, \bibnamefont{and}
  \bibinfo{author}{\bibfnamefont{O.}~\bibnamefont{Ruchayskiy}}
  (\bibinfo{year}{2012}), \eprint{1205.4700}.

\bibitem[{\citenamefont{Whiteson}(2012)}]{Whiteson:2012hr}
\bibinfo{author}{\bibfnamefont{D.}~\bibnamefont{Whiteson}}
  (\bibinfo{year}{2012}), \eprint{1208.3677}.

\bibitem[{\citenamefont{Hektor et~al.}(2012)\citenamefont{Hektor, Raidal, and
  Tempel}}]{Hektor:2012ev}
\bibinfo{author}{\bibfnamefont{A.}~\bibnamefont{Hektor}},
  \bibinfo{author}{\bibfnamefont{M.}~\bibnamefont{Raidal}}, \bibnamefont{and}
  \bibinfo{author}{\bibfnamefont{E.}~\bibnamefont{Tempel}}
  (\bibinfo{year}{2012}), \eprint{1209.4548}.

\bibitem[{\citenamefont{Finkbeiner et~al.}(2012)\citenamefont{Finkbeiner, Su,
  and Weniger}}]{Finkbeiner:2012ez}
\bibinfo{author}{\bibfnamefont{D.~P.} \bibnamefont{Finkbeiner}},
  \bibinfo{author}{\bibfnamefont{M.}~\bibnamefont{Su}}, \bibnamefont{and}
  \bibinfo{author}{\bibfnamefont{C.}~\bibnamefont{Weniger}}
  (\bibinfo{year}{2012}), \eprint{1209.4562}.

\bibitem[{\citenamefont{Whiteson}(2013)}]{Whiteson:2013cs}
\bibinfo{author}{\bibfnamefont{D.}~\bibnamefont{Whiteson}}
  (\bibinfo{year}{2013}), \eprint{1302.0427}.

\bibitem[{\citenamefont{Radescu}(1985)}]{Radescu:1985wf}
\bibinfo{author}{\bibfnamefont{E.}~\bibnamefont{Radescu}},
  \bibinfo{journal}{Phys.Rev.} \textbf{\bibinfo{volume}{D32}},
  \bibinfo{pages}{1266} (\bibinfo{year}{1985}).

\bibitem[{\citenamefont{Kayser and Goldhaber}(1983)}]{Kayser:1983wm}
\bibinfo{author}{\bibfnamefont{B.}~\bibnamefont{Kayser}} \bibnamefont{and}
  \bibinfo{author}{\bibfnamefont{A.~S.} \bibnamefont{Goldhaber}},
  \bibinfo{journal}{Phys.Rev.} \textbf{\bibinfo{volume}{D28}},
  \bibinfo{pages}{2341} (\bibinfo{year}{1983}).

\bibitem[{\citenamefont{Ho and Scherrer}(2012)}]{Ho:2012bg}
\bibinfo{author}{\bibfnamefont{C.~M.} \bibnamefont{Ho}} \bibnamefont{and}
  \bibinfo{author}{\bibfnamefont{R.~J.} \bibnamefont{Scherrer}}
  (\bibinfo{year}{2012}), \eprint{1211.0503}.

\bibitem[{\citenamefont{Gresham and Zurek}(2013)}]{Gresham:2013mua}
\bibinfo{author}{\bibfnamefont{M.~I.} \bibnamefont{Gresham}} \bibnamefont{and}
  \bibinfo{author}{\bibfnamefont{K.~M.} \bibnamefont{Zurek}}
  (\bibinfo{year}{2013}), \eprint{1311.2082}.

\bibitem[{\citenamefont{Gao et~al.}(2013)\citenamefont{Gao, Ho, and
  Scherrer}}]{Gao:2013vfa}
\bibinfo{author}{\bibfnamefont{Y.}~\bibnamefont{Gao}},
  \bibinfo{author}{\bibfnamefont{C.~M.} \bibnamefont{Ho}}, \bibnamefont{and}
  \bibinfo{author}{\bibfnamefont{R.~J.} \bibnamefont{Scherrer}}
  (\bibinfo{year}{2013}), \eprint{1311.5630}.

\bibitem[{\citenamefont{Del~Nobile et~al.}(2014)\citenamefont{Del~Nobile,
  Gelmini, Gondolo, and Huh}}]{DelNobile:2014eta}
\bibinfo{author}{\bibfnamefont{E.}~\bibnamefont{Del~Nobile}},
  \bibinfo{author}{\bibfnamefont{G.~B.} \bibnamefont{Gelmini}},
  \bibinfo{author}{\bibfnamefont{P.}~\bibnamefont{Gondolo}}, \bibnamefont{and}
  \bibinfo{author}{\bibfnamefont{J.-H.} \bibnamefont{Huh}}
  (\bibinfo{year}{2014}), \eprint{1401.4508}.

\bibitem[{\citenamefont{Heo}(2010)}]{Heo:2009vt}
\bibinfo{author}{\bibfnamefont{J.~H.} \bibnamefont{Heo}},
  \bibinfo{journal}{Phys.Lett.} \textbf{\bibinfo{volume}{B693}},
  \bibinfo{pages}{255} (\bibinfo{year}{2010}), \eprint{0901.3815}.

\bibitem[{\citenamefont{Masso et~al.}(2009)\citenamefont{Masso, Mohanty, and
  Rao}}]{Masso:2009mu}
\bibinfo{author}{\bibfnamefont{E.}~\bibnamefont{Masso}},
  \bibinfo{author}{\bibfnamefont{S.}~\bibnamefont{Mohanty}}, \bibnamefont{and}
  \bibinfo{author}{\bibfnamefont{S.}~\bibnamefont{Rao}},
  \bibinfo{journal}{Phys. Rev.} \textbf{\bibinfo{volume}{D80}},
  \bibinfo{pages}{036009} (\bibinfo{year}{2009}), \eprint{0906.1979}.

\bibitem[{\citenamefont{Del~Nobile et~al.}(2012)\citenamefont{Del~Nobile,
  Kouvaris, Panci, Sannino, and Virkajarvi}}]{DelNobile:2012tx}
\bibinfo{author}{\bibfnamefont{E.}~\bibnamefont{Del~Nobile}},
  \bibinfo{author}{\bibfnamefont{C.}~\bibnamefont{Kouvaris}},
  \bibinfo{author}{\bibfnamefont{P.}~\bibnamefont{Panci}},
  \bibinfo{author}{\bibfnamefont{F.}~\bibnamefont{Sannino}}, \bibnamefont{and}
  \bibinfo{author}{\bibfnamefont{J.}~\bibnamefont{Virkajarvi}}
  (\bibinfo{year}{2012}), \eprint{1203.6652}.

\bibitem[{\citenamefont{Weiner and Yavin}(2012{\natexlab{a}})}]{Weiner:2012gm}
\bibinfo{author}{\bibfnamefont{N.}~\bibnamefont{Weiner}} \bibnamefont{and}
  \bibinfo{author}{\bibfnamefont{I.}~\bibnamefont{Yavin}}
  (\bibinfo{year}{2012}{\natexlab{a}}), \eprint{1209.1093}.

\bibitem[{\citenamefont{Barger et~al.}(2012{\natexlab{a}})\citenamefont{Barger,
  Keung, Marfatia, and Tseng}}]{Barger:2012pf}
\bibinfo{author}{\bibfnamefont{V.}~\bibnamefont{Barger}},
  \bibinfo{author}{\bibfnamefont{W.-Y.} \bibnamefont{Keung}},
  \bibinfo{author}{\bibfnamefont{D.}~\bibnamefont{Marfatia}}, \bibnamefont{and}
  \bibinfo{author}{\bibfnamefont{P.-Y.} \bibnamefont{Tseng}}
  (\bibinfo{year}{2012}{\natexlab{a}}), \eprint{1206.0640}.

\bibitem[{\citenamefont{Haisch and Kahlhoefer}(2013)}]{Haisch:2013uaa}
\bibinfo{author}{\bibfnamefont{U.}~\bibnamefont{Haisch}} \bibnamefont{and}
  \bibinfo{author}{\bibfnamefont{F.}~\bibnamefont{Kahlhoefer}}
  (\bibinfo{year}{2013}), \eprint{1302.4454}.

\bibitem[{\citenamefont{Akerib et~al.}(2013)}]{Akerib:2013tjd}
\bibinfo{author}{\bibfnamefont{D.}~\bibnamefont{Akerib}} \bibnamefont{et~al.}
  (\bibinfo{collaboration}{LUX Collaboration}) (\bibinfo{year}{2013}),
  \eprint{1310.8214}.

\bibitem[{\citenamefont{Aprile et~al.}(2012)}]{Aprile:2012nq}
\bibinfo{author}{\bibfnamefont{E.}~\bibnamefont{Aprile}} \bibnamefont{et~al.}
  (\bibinfo{collaboration}{XENON100 Collaboration}) (\bibinfo{year}{2012}),
  \eprint{1207.5988}.

\bibitem[{\citenamefont{Chun et~al.}(2010)\citenamefont{Chun, Park, and
  Scopel}}]{Chun:2009zx}
\bibinfo{author}{\bibfnamefont{E.~J.} \bibnamefont{Chun}},
  \bibinfo{author}{\bibfnamefont{J.-C.} \bibnamefont{Park}}, \bibnamefont{and}
  \bibinfo{author}{\bibfnamefont{S.}~\bibnamefont{Scopel}},
  \bibinfo{journal}{JCAP} \textbf{\bibinfo{volume}{1002}}, \bibinfo{pages}{015}
  (\bibinfo{year}{2010}), \eprint{arXiv:0911.5273}.

\bibitem[{\citenamefont{Buckley et~al.}(2013)\citenamefont{Buckley, Hooper, and
  Kumar}}]{Buckley:2013sca}
\bibinfo{author}{\bibfnamefont{M.~R.} \bibnamefont{Buckley}},
  \bibinfo{author}{\bibfnamefont{D.}~\bibnamefont{Hooper}}, \bibnamefont{and}
  \bibinfo{author}{\bibfnamefont{J.}~\bibnamefont{Kumar}}
  (\bibinfo{year}{2013}), \eprint{1307.3561}.

\bibitem[{\citenamefont{Liu et~al.}(2013)\citenamefont{Liu, Shuve, Weiner, and
  Yavin}}]{Liu:2013gba}
\bibinfo{author}{\bibfnamefont{J.}~\bibnamefont{Liu}},
  \bibinfo{author}{\bibfnamefont{B.}~\bibnamefont{Shuve}},
  \bibinfo{author}{\bibfnamefont{N.}~\bibnamefont{Weiner}}, \bibnamefont{and}
  \bibinfo{author}{\bibfnamefont{I.}~\bibnamefont{Yavin}}
  (\bibinfo{year}{2013}), \eprint{1303.4404}.

\bibitem[{\citenamefont{Cao et~al.}(2009)\citenamefont{Cao, Ma, and
  Shaughnessy}}]{Cao:2009yy}
\bibinfo{author}{\bibfnamefont{Q.-H.} \bibnamefont{Cao}},
  \bibinfo{author}{\bibfnamefont{E.}~\bibnamefont{Ma}}, \bibnamefont{and}
  \bibinfo{author}{\bibfnamefont{G.}~\bibnamefont{Shaughnessy}},
  \bibinfo{journal}{Phys.Lett.} \textbf{\bibinfo{volume}{B673}},
  \bibinfo{pages}{152} (\bibinfo{year}{2009}), \eprint{0901.1334}.

\bibitem[{\citenamefont{Bergstrom}(1989)}]{Bergstrom:1989jr}
\bibinfo{author}{\bibfnamefont{L.}~\bibnamefont{Bergstrom}},
  \bibinfo{journal}{Phys.Lett.} \textbf{\bibinfo{volume}{B225}},
  \bibinfo{pages}{372} (\bibinfo{year}{1989}).

\bibitem[{\citenamefont{Barger et~al.}(2009)\citenamefont{Barger, Gao, Keung,
  and Marfatia}}]{Barger:2009xe}
\bibinfo{author}{\bibfnamefont{V.}~\bibnamefont{Barger}},
  \bibinfo{author}{\bibfnamefont{Y.}~\bibnamefont{Gao}},
  \bibinfo{author}{\bibfnamefont{W.~Y.} \bibnamefont{Keung}}, \bibnamefont{and}
  \bibinfo{author}{\bibfnamefont{D.}~\bibnamefont{Marfatia}},
  \bibinfo{journal}{Phys.Rev.} \textbf{\bibinfo{volume}{D80}},
  \bibinfo{pages}{063537} (\bibinfo{year}{2009}), \eprint{0906.3009}.

\bibitem[{\citenamefont{Barger et~al.}(2012{\natexlab{b}})\citenamefont{Barger,
  Keung, and Marfatia}}]{Barger:2011jg}
\bibinfo{author}{\bibfnamefont{V.}~\bibnamefont{Barger}},
  \bibinfo{author}{\bibfnamefont{W.-Y.} \bibnamefont{Keung}}, \bibnamefont{and}
  \bibinfo{author}{\bibfnamefont{D.}~\bibnamefont{Marfatia}},
  \bibinfo{journal}{Phys.Lett.} \textbf{\bibinfo{volume}{B707}},
  \bibinfo{pages}{385} (\bibinfo{year}{2012}{\natexlab{b}}),
  \eprint{1111.4523}.

\bibitem[{\citenamefont{Toma}(2013)}]{Toma:2013bka}
\bibinfo{author}{\bibfnamefont{T.}~\bibnamefont{Toma}},
  \bibinfo{journal}{Phys.Rev.Lett.} \textbf{\bibinfo{volume}{111}},
  \bibinfo{pages}{091301} (\bibinfo{year}{2013}), \eprint{1307.6181}.

\bibitem[{\citenamefont{Giacchino et~al.}(2013)\citenamefont{Giacchino,
  Lopez-Honorez, and Tytgat}}]{Giacchino:2013bta}
\bibinfo{author}{\bibfnamefont{F.}~\bibnamefont{Giacchino}},
  \bibinfo{author}{\bibfnamefont{L.}~\bibnamefont{Lopez-Honorez}},
  \bibnamefont{and} \bibinfo{author}{\bibfnamefont{M.~H.}
  \bibnamefont{Tytgat}}, \bibinfo{journal}{JCAP}
  \textbf{\bibinfo{volume}{1310}}, \bibinfo{pages}{025} (\bibinfo{year}{2013}),
  \eprint{1307.6480}.

\bibitem[{\citenamefont{Ade et~al.}(2013)}]{Ade:2013zuv}
\bibinfo{author}{\bibfnamefont{P.}~\bibnamefont{Ade}} \bibnamefont{et~al.}
  (\bibinfo{collaboration}{Planck Collaboration}) (\bibinfo{year}{2013}),
  \eprint{1303.5076}.

\bibitem[{\citenamefont{Hinshaw et~al.}(2013)}]{Hinshaw:2012aka}
\bibinfo{author}{\bibfnamefont{G.}~\bibnamefont{Hinshaw}} \bibnamefont{et~al.}
  (\bibinfo{collaboration}{WMAP}), \bibinfo{journal}{Astrophys.J.Suppl.}
  \textbf{\bibinfo{volume}{208}}, \bibinfo{pages}{19} (\bibinfo{year}{2013}),
  \eprint{1212.5226}.

\bibitem[{\citenamefont{Haber and Kane}(1985)}]{Haber:1984rc}
\bibinfo{author}{\bibfnamefont{H.~E.} \bibnamefont{Haber}} \bibnamefont{and}
  \bibinfo{author}{\bibfnamefont{G.~L.} \bibnamefont{Kane}},
  \bibinfo{journal}{Phys.Rept.} \textbf{\bibinfo{volume}{117}},
  \bibinfo{pages}{75} (\bibinfo{year}{1985}).

\bibitem[{\citenamefont{Konishi et~al.}(2013)\citenamefont{Konishi, Ohta, Sato,
  Shimomura, Sugai et~al.}}]{Konishi:2013gda}
\bibinfo{author}{\bibfnamefont{Y.}~\bibnamefont{Konishi}},
  \bibinfo{author}{\bibfnamefont{S.}~\bibnamefont{Ohta}},
  \bibinfo{author}{\bibfnamefont{J.}~\bibnamefont{Sato}},
  \bibinfo{author}{\bibfnamefont{T.}~\bibnamefont{Shimomura}},
  \bibinfo{author}{\bibfnamefont{K.}~\bibnamefont{Sugai}}, \bibnamefont{et~al.}
  (\bibinfo{year}{2013}), \eprint{1309.2067}.

\bibitem[{\citenamefont{Fukushima and Kumar}(2013)}]{Fukushima:2013efa}
\bibinfo{author}{\bibfnamefont{K.}~\bibnamefont{Fukushima}} \bibnamefont{and}
  \bibinfo{author}{\bibfnamefont{J.}~\bibnamefont{Kumar}}
  (\bibinfo{year}{2013}), \eprint{1307.7120}.

\bibitem[{\citenamefont{Frandsen et~al.}(2013)\citenamefont{Frandsen, Sannino,
  Shoemaker, and Svendsen}}]{Frandsen:2013bfa}
\bibinfo{author}{\bibfnamefont{M.~T.} \bibnamefont{Frandsen}},
  \bibinfo{author}{\bibfnamefont{F.}~\bibnamefont{Sannino}},
  \bibinfo{author}{\bibfnamefont{I.~M.} \bibnamefont{Shoemaker}},
  \bibnamefont{and} \bibinfo{author}{\bibfnamefont{O.}~\bibnamefont{Svendsen}}
  (\bibinfo{year}{2013}), \eprint{1312.3326}.

\bibitem[{\citenamefont{Frandsen et~al.}(2012)\citenamefont{Frandsen, Haisch,
  Kahlhoefer, Mertsch, and Schmidt-Hoberg}}]{Frandsen:2012db}
\bibinfo{author}{\bibfnamefont{M.~T.} \bibnamefont{Frandsen}},
  \bibinfo{author}{\bibfnamefont{U.}~\bibnamefont{Haisch}},
  \bibinfo{author}{\bibfnamefont{F.}~\bibnamefont{Kahlhoefer}},
  \bibinfo{author}{\bibfnamefont{P.}~\bibnamefont{Mertsch}}, \bibnamefont{and}
  \bibinfo{author}{\bibfnamefont{K.}~\bibnamefont{Schmidt-Hoberg}},
  \bibinfo{journal}{JCAP} \textbf{\bibinfo{volume}{1210}}, \bibinfo{pages}{033}
  (\bibinfo{year}{2012}), \eprint{1207.3971}.

\bibitem[{\citenamefont{Sigurdson et~al.}(2004)\citenamefont{Sigurdson, Doran,
  Kurylov, Caldwell, and Kamionkowski}}]{Sigurdson:2004zp}
\bibinfo{author}{\bibfnamefont{K.}~\bibnamefont{Sigurdson}},
  \bibinfo{author}{\bibfnamefont{M.}~\bibnamefont{Doran}},
  \bibinfo{author}{\bibfnamefont{A.}~\bibnamefont{Kurylov}},
  \bibinfo{author}{\bibfnamefont{R.~R.} \bibnamefont{Caldwell}},
  \bibnamefont{and}
  \bibinfo{author}{\bibfnamefont{M.}~\bibnamefont{Kamionkowski}},
  \bibinfo{journal}{Phys.Rev.} \textbf{\bibinfo{volume}{D70}},
  \bibinfo{pages}{083501} (\bibinfo{year}{2004}), \eprint{astro-ph/0406355}.

\bibitem[{\citenamefont{Barger et~al.}(2011)\citenamefont{Barger, Keung, and
  Marfatia}}]{Barger:2010gv}
\bibinfo{author}{\bibfnamefont{V.}~\bibnamefont{Barger}},
  \bibinfo{author}{\bibfnamefont{W.-Y.} \bibnamefont{Keung}}, \bibnamefont{and}
  \bibinfo{author}{\bibfnamefont{D.}~\bibnamefont{Marfatia}},
  \bibinfo{journal}{Phys.Lett.} \textbf{\bibinfo{volume}{B696}},
  \bibinfo{pages}{74} (\bibinfo{year}{2011}), \eprint{1007.4345}.

\bibitem[{\citenamefont{Fitzpatrick and Zurek}(2010)}]{Fitzpatrick:2010br}
\bibinfo{author}{\bibfnamefont{A.}~\bibnamefont{Fitzpatrick}} \bibnamefont{and}
  \bibinfo{author}{\bibfnamefont{K.~M.} \bibnamefont{Zurek}},
  \bibinfo{journal}{Phys.Rev.} \textbf{\bibinfo{volume}{D82}},
  \bibinfo{pages}{075004} (\bibinfo{year}{2010}), \eprint{1007.5325}.

\bibitem[{\citenamefont{Banks et~al.}(2010)\citenamefont{Banks, Fortin, and
  Thomas}}]{Banks:2010eh}
\bibinfo{author}{\bibfnamefont{T.}~\bibnamefont{Banks}},
  \bibinfo{author}{\bibfnamefont{J.-F.} \bibnamefont{Fortin}},
  \bibnamefont{and} \bibinfo{author}{\bibfnamefont{S.}~\bibnamefont{Thomas}}
  (\bibinfo{year}{2010}), \eprint{1007.5515}.

\bibitem[{\citenamefont{Weiner and Yavin}(2012{\natexlab{b}})}]{Weiner:2012cb}
\bibinfo{author}{\bibfnamefont{N.}~\bibnamefont{Weiner}} \bibnamefont{and}
  \bibinfo{author}{\bibfnamefont{I.}~\bibnamefont{Yavin}}
  (\bibinfo{year}{2012}{\natexlab{b}}), \eprint{1206.2910}.

\bibitem[{\citenamefont{Del~Nobile et~al.}(2013)\citenamefont{Del~Nobile,
  Gelmini, Gondolo, and Huh}}]{DelNobile:2013cva}
\bibinfo{author}{\bibfnamefont{E.}~\bibnamefont{Del~Nobile}},
  \bibinfo{author}{\bibfnamefont{G.}~\bibnamefont{Gelmini}},
  \bibinfo{author}{\bibfnamefont{P.}~\bibnamefont{Gondolo}}, \bibnamefont{and}
  \bibinfo{author}{\bibfnamefont{J.-H.} \bibnamefont{Huh}}
  (\bibinfo{year}{2013}), \eprint{1306.5273}.

\bibitem[{\citenamefont{Kopp et~al.}(2010)\citenamefont{Kopp, Schwetz, and
  Zupan}}]{Kopp:2009qt}
\bibinfo{author}{\bibfnamefont{J.}~\bibnamefont{Kopp}},
  \bibinfo{author}{\bibfnamefont{T.}~\bibnamefont{Schwetz}}, \bibnamefont{and}
  \bibinfo{author}{\bibfnamefont{J.}~\bibnamefont{Zupan}},
  \bibinfo{journal}{JCAP} \textbf{\bibinfo{volume}{1002}}, \bibinfo{pages}{014}
  (\bibinfo{year}{2010}), \eprint{0912.4264}.

\bibitem[{\citenamefont{Kopp et~al.}(2012)\citenamefont{Kopp, Schwetz, and
  Zupan}}]{Kopp:2011yr}
\bibinfo{author}{\bibfnamefont{J.}~\bibnamefont{Kopp}},
  \bibinfo{author}{\bibfnamefont{T.}~\bibnamefont{Schwetz}}, \bibnamefont{and}
  \bibinfo{author}{\bibfnamefont{J.}~\bibnamefont{Zupan}},
  \bibinfo{journal}{JCAP} \textbf{\bibinfo{volume}{1203}}, \bibinfo{pages}{001}
  (\bibinfo{year}{2012}), \eprint{1110.2721}.

\bibitem[{\citenamefont{Aprile et~al.}(2011)}]{Aprile:2011hi}
\bibinfo{author}{\bibfnamefont{E.}~\bibnamefont{Aprile}} \bibnamefont{et~al.}
  (\bibinfo{collaboration}{XENON100 Collaboration}),
  \bibinfo{journal}{Phys.Rev.Lett.}  (\bibinfo{year}{2011}),
  \eprint{1104.2549}.

\bibitem[{\citenamefont{Pospelov and ter Veldhuis}(2000)}]{Pospelov:2000bq}
\bibinfo{author}{\bibfnamefont{M.}~\bibnamefont{Pospelov}} \bibnamefont{and}
  \bibinfo{author}{\bibfnamefont{T.}~\bibnamefont{ter Veldhuis}},
  \bibinfo{journal}{Phys.Lett.} \textbf{\bibinfo{volume}{B480}},
  \bibinfo{pages}{181} (\bibinfo{year}{2000}), \eprint{hep-ph/0003010}.

\bibitem[{\citenamefont{Chang et~al.}(2010)\citenamefont{Chang, Weiner, and
  Yavin}}]{Chang:2010en}
\bibinfo{author}{\bibfnamefont{S.}~\bibnamefont{Chang}},
  \bibinfo{author}{\bibfnamefont{N.}~\bibnamefont{Weiner}}, \bibnamefont{and}
  \bibinfo{author}{\bibfnamefont{I.}~\bibnamefont{Yavin}},
  \bibinfo{journal}{Phys.Rev.} \textbf{\bibinfo{volume}{D82}},
  \bibinfo{pages}{125011} (\bibinfo{year}{2010}), \eprint{1007.4200}.

\bibitem[{\citenamefont{Jungman et~al.}(1996)\citenamefont{Jungman,
  Kamionkowski, and Griest}}]{Jungman:1995df}
\bibinfo{author}{\bibfnamefont{G.}~\bibnamefont{Jungman}},
  \bibinfo{author}{\bibfnamefont{M.}~\bibnamefont{Kamionkowski}},
  \bibnamefont{and} \bibinfo{author}{\bibfnamefont{K.}~\bibnamefont{Griest}},
  \bibinfo{journal}{Phys. Rept.} \textbf{\bibinfo{volume}{267}},
  \bibinfo{pages}{195} (\bibinfo{year}{1996}), \eprint{hep-ph/9506380}.

\bibitem[{\citenamefont{McCabe}(2010)}]{McCabe:2010zh}
\bibinfo{author}{\bibfnamefont{C.}~\bibnamefont{McCabe}},
  \bibinfo{journal}{Phys.Rev.} \textbf{\bibinfo{volume}{D82}},
  \bibinfo{pages}{023530} (\bibinfo{year}{2010}), \eprint{1005.0579}.

\bibitem[{\citenamefont{Farina et~al.}(2011)\citenamefont{Farina, Pappadopulo,
  Strumia, and Volansky}}]{Farina:2011pw}
\bibinfo{author}{\bibfnamefont{M.}~\bibnamefont{Farina}},
  \bibinfo{author}{\bibfnamefont{D.}~\bibnamefont{Pappadopulo}},
  \bibinfo{author}{\bibfnamefont{A.}~\bibnamefont{Strumia}}, \bibnamefont{and}
  \bibinfo{author}{\bibfnamefont{T.}~\bibnamefont{Volansky}},
  \bibinfo{journal}{JCAP} \textbf{\bibinfo{volume}{1111}}, \bibinfo{pages}{010}
  (\bibinfo{year}{2011}), \eprint{1107.0715}.

\bibitem[{\citenamefont{Green}(2012)}]{Green:2011bv}
\bibinfo{author}{\bibfnamefont{A.~M.} \bibnamefont{Green}},
  \bibinfo{journal}{Mod.Phys.Lett.} \textbf{\bibinfo{volume}{A27}},
  \bibinfo{pages}{1230004} (\bibinfo{year}{2012}), \eprint{1112.0524}.

\bibitem[{\citenamefont{Fairbairn et~al.}(2012)\citenamefont{Fairbairn, Douce,
  and Swift}}]{Fairbairn:2012zs}
\bibinfo{author}{\bibfnamefont{M.}~\bibnamefont{Fairbairn}},
  \bibinfo{author}{\bibfnamefont{T.}~\bibnamefont{Douce}}, \bibnamefont{and}
  \bibinfo{author}{\bibfnamefont{J.}~\bibnamefont{Swift}}
  (\bibinfo{year}{2012}), \eprint{1206.2693}.

\bibitem[{\citenamefont{Yellin}(2002)}]{Yellin:2002xd}
\bibinfo{author}{\bibfnamefont{S.}~\bibnamefont{Yellin}},
  \bibinfo{journal}{Phys. Rev.} \textbf{\bibinfo{volume}{D66}},
  \bibinfo{pages}{032005} (\bibinfo{year}{2002}), \eprint{physics/0203002}.

\bibitem[{\citenamefont{Aprile}(2012)}]{Aprile:2012zx}
\bibinfo{author}{\bibfnamefont{E.}~\bibnamefont{Aprile}}
  (\bibinfo{collaboration}{XENON1T collaboration}) (\bibinfo{year}{2012}),
  \eprint{1206.6288}.

\bibitem[{\citenamefont{Malling et~al.}(2011)\citenamefont{Malling, Akerib,
  Araujo, Bai, Bedikian et~al.}}]{Malling:2011va}
\bibinfo{author}{\bibfnamefont{D.}~\bibnamefont{Malling}},
  \bibinfo{author}{\bibfnamefont{D.}~\bibnamefont{Akerib}},
  \bibinfo{author}{\bibfnamefont{H.}~\bibnamefont{Araujo}},
  \bibinfo{author}{\bibfnamefont{X.}~\bibnamefont{Bai}},
  \bibinfo{author}{\bibfnamefont{S.}~\bibnamefont{Bedikian}},
  \bibnamefont{et~al.} (\bibinfo{year}{2011}), \bibinfo{note}{see also
  http://www.hep.ucl.ac.uk/darkMatter/LZ.shtml}, \eprint{1110.0103}.

\bibitem[{\citenamefont{Cao et~al.}(2011)\citenamefont{Cao, Chen, Li, and
  Zhang}}]{Cao:2009uw}
\bibinfo{author}{\bibfnamefont{Q.-H.} \bibnamefont{Cao}},
  \bibinfo{author}{\bibfnamefont{C.-R.} \bibnamefont{Chen}},
  \bibinfo{author}{\bibfnamefont{C.~S.} \bibnamefont{Li}}, \bibnamefont{and}
  \bibinfo{author}{\bibfnamefont{H.}~\bibnamefont{Zhang}},
  \bibinfo{journal}{JHEP} \textbf{\bibinfo{volume}{1108}}, \bibinfo{pages}{018}
  (\bibinfo{year}{2011}), \eprint{0912.4511}.

\bibitem[{\citenamefont{Beltran et~al.}(2010)\citenamefont{Beltran, Hooper,
  Kolb, Krusberg, and Tait}}]{Beltran:2010ww}
\bibinfo{author}{\bibfnamefont{M.}~\bibnamefont{Beltran}},
  \bibinfo{author}{\bibfnamefont{D.}~\bibnamefont{Hooper}},
  \bibinfo{author}{\bibfnamefont{E.~W.} \bibnamefont{Kolb}},
  \bibinfo{author}{\bibfnamefont{Z.~A.} \bibnamefont{Krusberg}},
  \bibnamefont{and} \bibinfo{author}{\bibfnamefont{T.~M.} \bibnamefont{Tait}},
  \bibinfo{journal}{JHEP} \textbf{\bibinfo{volume}{1009}}, \bibinfo{pages}{037}
  (\bibinfo{year}{2010}), \eprint{1002.4137}.

\bibitem[{\citenamefont{Goodman et~al.}(2010)\citenamefont{Goodman, Ibe,
  Rajaraman, Shepherd, Tait et~al.}}]{Goodman:2010ku}
\bibinfo{author}{\bibfnamefont{J.}~\bibnamefont{Goodman}},
  \bibinfo{author}{\bibfnamefont{M.}~\bibnamefont{Ibe}},
  \bibinfo{author}{\bibfnamefont{A.}~\bibnamefont{Rajaraman}},
  \bibinfo{author}{\bibfnamefont{W.}~\bibnamefont{Shepherd}},
  \bibinfo{author}{\bibfnamefont{T.~M.} \bibnamefont{Tait}},
  \bibnamefont{et~al.}, \bibinfo{journal}{Phys.Rev.}
  \textbf{\bibinfo{volume}{D82}}, \bibinfo{pages}{116010}
  (\bibinfo{year}{2010}), \eprint{1008.1783}.

\bibitem[{\citenamefont{Goodman and Shepherd}(2011)}]{Goodman:2011jq}
\bibinfo{author}{\bibfnamefont{J.}~\bibnamefont{Goodman}} \bibnamefont{and}
  \bibinfo{author}{\bibfnamefont{W.}~\bibnamefont{Shepherd}}
  (\bibinfo{year}{2011}), \eprint{1111.2359}.

\bibitem[{\citenamefont{Fox et~al.}(2011)\citenamefont{Fox, Harnik, Kopp, and
  Tsai}}]{Fox:2011fx}
\bibinfo{author}{\bibfnamefont{P.~J.} \bibnamefont{Fox}},
  \bibinfo{author}{\bibfnamefont{R.}~\bibnamefont{Harnik}},
  \bibinfo{author}{\bibfnamefont{J.}~\bibnamefont{Kopp}}, \bibnamefont{and}
  \bibinfo{author}{\bibfnamefont{Y.}~\bibnamefont{Tsai}},
  \bibinfo{journal}{Phys.Rev.} \textbf{\bibinfo{volume}{D84}},
  \bibinfo{pages}{014028} (\bibinfo{year}{2011}), \eprint{1103.0240}.

\bibitem[{\citenamefont{Dreiner et~al.}(2012)\citenamefont{Dreiner, Huck,
  Kramer, Schmeier, and Tattersall}}]{Dreiner:2012xm}
\bibinfo{author}{\bibfnamefont{H.}~\bibnamefont{Dreiner}},
  \bibinfo{author}{\bibfnamefont{M.}~\bibnamefont{Huck}},
  \bibinfo{author}{\bibfnamefont{M.}~\bibnamefont{Kramer}},
  \bibinfo{author}{\bibfnamefont{D.}~\bibnamefont{Schmeier}}, \bibnamefont{and}
  \bibinfo{author}{\bibfnamefont{J.}~\bibnamefont{Tattersall}}
  (\bibinfo{year}{2012}), \eprint{1211.2254}.

\bibitem[{\citenamefont{Belyaev et~al.}(2012)\citenamefont{Belyaev,
  Christensen, and Pukhov}}]{Belyaev:2012qa}
\bibinfo{author}{\bibfnamefont{A.}~\bibnamefont{Belyaev}},
  \bibinfo{author}{\bibfnamefont{N.~D.} \bibnamefont{Christensen}},
  \bibnamefont{and} \bibinfo{author}{\bibfnamefont{A.}~\bibnamefont{Pukhov}}
  (\bibinfo{year}{2012}), \eprint{1207.6082}.

\bibitem[{\citenamefont{Alwall et~al.}(2007)\citenamefont{Alwall, Demin,
  de~Visscher, Frederix, Herquet et~al.}}]{Alwall:2007st}
\bibinfo{author}{\bibfnamefont{J.}~\bibnamefont{Alwall}},
  \bibinfo{author}{\bibfnamefont{P.}~\bibnamefont{Demin}},
  \bibinfo{author}{\bibfnamefont{S.}~\bibnamefont{de~Visscher}},
  \bibinfo{author}{\bibfnamefont{R.}~\bibnamefont{Frederix}},
  \bibinfo{author}{\bibfnamefont{M.}~\bibnamefont{Herquet}},
  \bibnamefont{et~al.}, \bibinfo{journal}{JHEP}
  \textbf{\bibinfo{volume}{0709}}, \bibinfo{pages}{028} (\bibinfo{year}{2007}),
  \eprint{0706.2334}.

\bibitem[{\citenamefont{Abdallah et~al.}(2005)}]{Abdallah:2003np}
\bibinfo{author}{\bibfnamefont{J.}~\bibnamefont{Abdallah}} \bibnamefont{et~al.}
  (\bibinfo{collaboration}{DELPHI Collaboration}),
  \bibinfo{journal}{Eur.Phys.J.} \textbf{\bibinfo{volume}{C38}},
  \bibinfo{pages}{395} (\bibinfo{year}{2005}), \eprint{hep-ex/0406019}.

\bibitem[{\citenamefont{Abdallah et~al.}(2009)}]{DELPHI:2008zg}
\bibinfo{author}{\bibfnamefont{J.}~\bibnamefont{Abdallah}} \bibnamefont{et~al.}
  (\bibinfo{collaboration}{DELPHI Collaboration}),
  \bibinfo{journal}{Eur.Phys.J.} \textbf{\bibinfo{volume}{C60}},
  \bibinfo{pages}{17} (\bibinfo{year}{2009}), \eprint{arXiv:0901.4486}.

\bibitem[{\citenamefont{Abe et~al.}(2010)}]{Abe:2010aa}
\bibinfo{author}{\bibfnamefont{T.}~\bibnamefont{Abe}} \bibnamefont{et~al.}
  (\bibinfo{collaboration}{ILD Concept Group - Linear Collider Collaboration})
  (\bibinfo{year}{2010}), \eprint{1006.3396}.

\bibitem[{\citenamefont{Bartels}(2011)}]{Bartels:2011dea}
\bibinfo{author}{\bibfnamefont{C.}~\bibnamefont{Bartels}}
  (\bibinfo{year}{2011}),
  \urlprefix\url{http://www-library.desy.de/cgi-bin/showprep.pl?thesis11-034}.

\bibitem[{\citenamefont{Aoyama et~al.}(2012)\citenamefont{Aoyama, Hayakawa,
  Kinoshita, and Nio}}]{Aoyama:2012wj}
\bibinfo{author}{\bibfnamefont{T.}~\bibnamefont{Aoyama}},
  \bibinfo{author}{\bibfnamefont{M.}~\bibnamefont{Hayakawa}},
  \bibinfo{author}{\bibfnamefont{T.}~\bibnamefont{Kinoshita}},
  \bibnamefont{and} \bibinfo{author}{\bibfnamefont{M.}~\bibnamefont{Nio}},
  \bibinfo{journal}{Phys.Rev.Lett.} \textbf{\bibinfo{volume}{109}},
  \bibinfo{pages}{111807} (\bibinfo{year}{2012}), \eprint{1205.5368}.

\bibitem[{\citenamefont{Beringer et~al.}(2012)}]{Beringer:2012zz}
\bibinfo{author}{\bibfnamefont{J.}~\bibnamefont{Beringer}} \bibnamefont{et~al.}
  (\bibinfo{collaboration}{Particle Data Group}), \bibinfo{journal}{Phys.Rev.}
  \textbf{\bibinfo{volume}{D86}}, \bibinfo{pages}{010001}
  (\bibinfo{year}{2012}).

\bibitem[{\citenamefont{Ishida et~al.}(2013)\citenamefont{Ishida, Namba, Asai,
  Kobayashi, Saito et~al.}}]{Ishida:2013waa}
\bibinfo{author}{\bibfnamefont{A.}~\bibnamefont{Ishida}},
  \bibinfo{author}{\bibfnamefont{T.}~\bibnamefont{Namba}},
  \bibinfo{author}{\bibfnamefont{S.}~\bibnamefont{Asai}},
  \bibinfo{author}{\bibfnamefont{T.}~\bibnamefont{Kobayashi}},
  \bibinfo{author}{\bibfnamefont{H.}~\bibnamefont{Saito}}, \bibnamefont{et~al.}
  (\bibinfo{year}{2013}), \eprint{1310.6923}.

\bibitem[{\citenamefont{Brodsky and Lebed}(2009)}]{Brodsky:2009gx}
\bibinfo{author}{\bibfnamefont{S.~J.} \bibnamefont{Brodsky}} \bibnamefont{and}
  \bibinfo{author}{\bibfnamefont{R.~F.} \bibnamefont{Lebed}},
  \bibinfo{journal}{Phys.Rev.Lett.} \textbf{\bibinfo{volume}{102}},
  \bibinfo{pages}{213401} (\bibinfo{year}{2009}), \eprint{0904.2225}.

\bibitem[{\citenamefont{Jentschura et~al.}(1997)\citenamefont{Jentschura, Soff,
  Ivanov, and Karshenboim}}]{Jentschura:1997ma}
\bibinfo{author}{\bibfnamefont{U.}~\bibnamefont{Jentschura}},
  \bibinfo{author}{\bibfnamefont{G.}~\bibnamefont{Soff}},
  \bibinfo{author}{\bibfnamefont{V.}~\bibnamefont{Ivanov}}, \bibnamefont{and}
  \bibinfo{author}{\bibfnamefont{S.~G.} \bibnamefont{Karshenboim}}
  (\bibinfo{year}{1997}), \eprint{hep-ph/9706401}.

\bibitem[{\citenamefont{Adam et~al.}(2013)}]{Adam:2013mnn}
\bibinfo{author}{\bibfnamefont{J.}~\bibnamefont{Adam}} \bibnamefont{et~al.}
  (\bibinfo{collaboration}{MEG Collaboration}) (\bibinfo{year}{2013}),
  \eprint{1303.0754}.

\bibitem[{\citenamefont{Aubert et~al.}(2010)}]{Aubert:2009ag}
\bibinfo{author}{\bibfnamefont{B.}~\bibnamefont{Aubert}} \bibnamefont{et~al.}
  (\bibinfo{collaboration}{BaBar Collaboration}),
  \bibinfo{journal}{Phys.Rev.Lett.} \textbf{\bibinfo{volume}{104}},
  \bibinfo{pages}{021802} (\bibinfo{year}{2010}), \eprint{0908.2381}.

\bibitem[{\citenamefont{Kuno and Okada}(2001)}]{Kuno:1999jp}
\bibinfo{author}{\bibfnamefont{Y.}~\bibnamefont{Kuno}} \bibnamefont{and}
  \bibinfo{author}{\bibfnamefont{Y.}~\bibnamefont{Okada}},
  \bibinfo{journal}{Rev.Mod.Phys.} \textbf{\bibinfo{volume}{73}},
  \bibinfo{pages}{151} (\bibinfo{year}{2001}), \eprint{hep-ph/9909265}.

\bibitem[{\citenamefont{Bellgardt et~al.}(1988)}]{Bellgardt:1987du}
\bibinfo{author}{\bibfnamefont{U.}~\bibnamefont{Bellgardt}}
  \bibnamefont{et~al.} (\bibinfo{collaboration}{SINDRUM Collaboration}),
  \bibinfo{journal}{Nucl.Phys.} \textbf{\bibinfo{volume}{B299}},
  \bibinfo{pages}{1} (\bibinfo{year}{1988}).

\bibitem[{\citenamefont{Blondel et~al.}(2013)\citenamefont{Blondel, Bravar,
  Pohl, Bachmann, Berger et~al.}}]{Blondel:2013ia}
\bibinfo{author}{\bibfnamefont{A.}~\bibnamefont{Blondel}},
  \bibinfo{author}{\bibfnamefont{A.}~\bibnamefont{Bravar}},
  \bibinfo{author}{\bibfnamefont{M.}~\bibnamefont{Pohl}},
  \bibinfo{author}{\bibfnamefont{S.}~\bibnamefont{Bachmann}},
  \bibinfo{author}{\bibfnamefont{N.}~\bibnamefont{Berger}},
  \bibnamefont{et~al.} (\bibinfo{year}{2013}), \eprint{1301.6113}.

\end{thebibliography}

\end{document}